\documentclass[aps,amsmath,amssymb,nofootinbib,superscriptaddress,showpacs,floatfix,prb,twocolumn]{revtex4-1}

\usepackage{latexsym,multirow,hyperref}
\usepackage{graphicx,cleveref}
\usepackage{times,psfrag,subfigure}
\usepackage{amsmath}
\usepackage{dsfont}
\usepackage{dcolumn}
\usepackage{bm,bbm}       
\usepackage{color}
\usepackage{soul}
\usepackage{latexsym,amsmath,amssymb,bm,euscript}
\usepackage{threeparttable}
\usepackage{hyperref}

\def\sgn{\mathop{\textrm{sgn}}}

\newcommand{\beq}{\begin{equation}}
\newcommand{\eeq}{\end{equation}}
\newcommand{\beqarray}{\begin{eqnarray}}
\newcommand{\eeqarray}{\end{eqnarray}}

\newcommand{\ket}[1]{| #1 \rangle}
\newcommand{\braket}[2]{\left \langle #1 | #2 \right\rangle}

\newcommand{\bee}{\begin{equation}}
\newcommand{\ee}{\end{equation}}
\newcommand{\bma}{\begin{pmatrix}}
\newcommand{\ema}{\end{pmatrix}}
\newcommand{\balig}{\begin{align}}
\newcommand{\ealig}{\end{align}}

\newcommand{\bZ}{\mathbb{Z}}
\newcommand{\bI}{\mathbbm{1}}
\newcommand{\ba}{\begin{eqnarray}}
\newcommand{\ea}{\end{eqnarray}}
\newcommand\tr{{\mbox{Tr\,}}}
\newcommand{\ignore}[1]{}

\newcommand{\mH}{\mathcal{H}}

\newcommand{\six}{\sigma_x}
\newcommand{\siy}{\sigma_y}
\newcommand{\siz}{\sigma_z}

\newcommand{\mK}{\mathcal{K}}
\newcommand{\dI}{\mathds{1}}

\begin{document}

\allowdisplaybreaks

\title{Classification of reflection symmetry protected topological semimetals and nodal superconductors}

\date{\today}

\author{Ching-Kai Chiu}
\email{chiu7@phas.ubc.ca}
   %\homepage[Introduction Video: http://www.youtube.com/watch$?v=9_0TBJ7pYpw$]{http://www.youtube.com/watch?v=9_0TBJ7pYpw}
\affiliation{Department of Physics and Astronomy, University of British Columbia, Vancouver, BC, Canada V6T 1Z1} 
\affiliation{Quantum Matter Institute, University of British Columbia, Vancouver BC, Canada V6T 1Z4}

\author{Andreas P. Schnyder}
\email{a.schnyder@fkf.mpg.de}
\affiliation{Max-Planck-Institut f\"ur Festk\"orperforschung,
  Heisenbergstrasse 1, D-70569 Stuttgart, Germany}

\begin{abstract}
While the topological classification of insulators, semimetals, and superconductors in terms of nonspatial symmetries is well understood,
less is known about topological states protected by crystalline symmetries, such as mirror reflections and rotations.
In this work, we systematically classify topological semimetals and nodal superconductors 
that are protected, not only by  nonspatial (i.e., global) symmetries, but also by  a crystal reflection symmetry.
We find that the classification crucially depends on (i) the codimension of the Fermi surface (nodal line or point) of the semimetal (superconductor), (ii) whether the mirror symmetry commutes  or anticommutes
with the nonspatial symmetries and
(iii) how the Fermi surfaces (nodal lines or points)  transform
under the mirror reflection and nonspatial symmetries. 
The classification is derived 
by examining all possible symmetry-allowed mass terms that can be added to 
the Bloch or Bogoliubov-de Gennes Hamiltonian in a given symmetry class
and by explicitly deriving topological invariants.
We discuss several examples of reflection symmetry protected topological semimetals and nodal superconductors, including topological crystalline semimetals with mirror $\bZ_2$ numbers and  topological crystalline nodal superconductors with mirror winding numbers.  \\
\\
\href{http://www.youtube.com/watch?v=9_0TBJ7pYpw}{Introduction video: http://www.youtube.com/watch?v$=$9\_0TBJ7pYpw}
\end{abstract}

\date{\today}

\pacs{03.65.Vf,74.50.+r, 73.20.Fz, 73.20.-r:}

\maketitle

\section{Introduction}

Inspired by the recent experimental discovery of two- and three-dimensional topological insulators,\cite{konigJPSJ08,hasan:rmp,hasanAnnuRev,QiRMP11,bernevig06} a multitude of novel topological states protected by different symmetries has been predicted over the last few years.\cite{QiRMP11,Kitaev,SchnyderAIP,Ryu2010ten,Schnyder2008gf} One of the main hallmarks of these topological materials is the appearance of
protected zero-energy surface states, which arise as a consequence of the nontrivial topological characteristics of the bulk wave functions.
For fully gapped topological phases protected by general nonspatial symmetries a complete classification, the tenfold way, has
been obtained for arbitrary dimensions.\cite{Kitaev,SchnyderAIP,Ryu2010ten,Schnyder2008gf}
This scheme classifies fully gapped noninteracting systems in terms of nonspatial symmetries, i.e., symmetries that act locally in 
position space, namely time-reversal symmetry (TRS), particle-hole symmetry (PHS), and chiral or sublattice symmetry (SLS).

However, over the last few years it has become apparent that besides nonspatial symmetries, also crystalline symmetries, i.e., symmetries that act nonlocally in position space, can  lead to nontrivial topological properties of bulk insulating states. \cite{teoPRB08,Fu2011uq,chiuPRB13,morimotoPRB13,slaberNatPhys13,uenoPRL13,zhangPRL13,benalcazar2013,fangPRB12,fangPRB13,Jadaun2013,Teo_hughes_2013,Turner:2012bh,Turner:2010qf,HughesPRB11,luLee2014,Sato_Crystalline_arxiv14,koshino_morimoto_sato_ArXiv} 
 A prime example of a topological material protected by a crystalline symmetry is the topological crystalline insulator SnTe.\cite{Tanaka:2012fk,Hsieh:2012fk,Xu2012,Dziawa2012uq_short}
This band insulator exhibits Dirac-cone surface states that are protected by a mirror reflection symmetry of the crystal. 
Other than reflection symmetry, inversion\cite{Turner:2012bh,Turner:2010qf,HughesPRB11,Sato_Crystalline_arxiv14,luLee2014}
and rotation\cite{benalcazar2013,fangPRB13,koshino_morimoto_sato_ArXiv} can also give rise to topologically nontrivial quantum states of matter.
In fact, it is expected that for any given discrete space group symmetry there is a distinct topological classification
of band insulators and fully gapped superconductors, and that each of these
space-group-symmetry protected topological states can be characterized in terms of an associated crystalline topological number.

Parallel to these developments, the concept of topological band theory has been extended to semimetals with Fermi points or Fermi lines,
and nodal superconductors with point nodes or line nodes.\cite{ryuHatsugai2002,Sato_Crystalline_arxiv14,matsuuraNJP13,ZhaoWangPRL13,ZhaoWangPRB14,HoravaPRL05,beriPRB10,Volovik:book,volovikLectNotes13,Manes_crytal_semimetal,lauTimm2013,ranLee2009,hiddensymmetry,Sato_Blount} 
 Although a global topological number cannot be defined for these gapless systems, it is nevertheless possible to determine their topological characteristics
and the stability of their Fermi points or Fermi lines in terms of momentum-dependent topological numbers. Notable examples of gapless topological states
include Weyl semimetals,\cite{BurkovBalentsPRB11,burkovBalenstPRL11,WanVishwanathSavrasovPRB11,Weyl_semimetal_Fang,Weyl_Delplace,Lu:2013Weyl,WeylChern,WeylDas,Weyl_Gilbert} Weyl superconductors,\cite{biswasPRB2013,fischerPRB14,MengBalentsPRB12,WeylSC_Jay} and nodal noncentrosymmetric superconductors.\cite{SatoPRB06,SchnyderRyuFlat,BrydonSchnyderTimmFlat,nodalSCDungHai,Brydon10,tanakaReview12,brydonNJP13,schnyderPRL13}
Similar to fully gapped topological materials, the topological characteristics of gapless topological states manifest themselves at the surface in the form of
either linearly dispersing boundary modes (i.e., Dirac or Majorana states) or dispersionless states, forming two-dimensional surface flat-bands 
or one-dimensional surface arcs. 
While a complete topological classification of semimetals and nodal superconductors
in terms of nonspatial symmetries has been established recently,\cite{Sato_Crystalline_arxiv14,matsuuraNJP13,ZhaoWangPRL13,ZhaoWangPRB14} the characterization  of gapless topological materials protected
by crystalline symmetries has remained an open problem.

In this paper, we present a complete classification of topological semimetals and nodal superconductors  protected 
by crystal reflection symmetries and possibly one or two nonspatial  (i.e., global) symmetries. We find that the topological 
classification of these reflection symmetry protected gapless states sensitively depends
on (i) the codimension of the Fermi surface, (ii) whether the reflection symmetry commutes or anticommutes with the 
nonspatial symmetries, and (iii) whether the
Fermi points or Fermi lines are left invariant by the mirror symmetry or the nonspatial symmetries.
The outcome of this classification scheme is summarized in Tables~\ref{reflection_table_full} and \ref{table_reflection_off_off},
which constitute the main results of this paper. 
Similar to the ten-fold classification in terms of nonspatial symmetries,\cite{Kitaev,SchnyderAIP,Ryu2010ten,Schnyder2008gf} these tables exhibit 
two-fold and eight-fold Bott periodicities as a function of spatial dimension. 
Two complementary methods are used to derive these classification tables.
The first approach is based on classifying  
all possible symmetry-allowed mass terms that can be added to
the Bloch or Bogoliubov-de Gennes (BdG) Hamiltonian in a given symmetry class.
The second method relies on the explicit derivation of  
 different types of topological invariants that guarantee the  stability of the Fermi surfaces (superconducting nodes).
In order to illustrate the new topological phases predicted by these classification schemes, we
 discuss several specific examples of reflection symmetry protected topological semimetals and nodal superconductors, see Sec.~\ref{sec:examples}. 
 
The remainder of this article is organized as follows. In Sec.~\ref{sec:II} we briefly review
the classification of gapless topological materials in terms of nonspatial symmetries.
This is followed by the derivation of the topological classification of reflection symmetry protected
semimetals and nodal superconductors
in Sec.~\ref{classReflecGapless}, which is the principal result of this paper. 
We present some explicit examples of topological semimetals and nodal superconductors protected by reflection symmetries 
in Sec.~\ref{sec:examples} 
and conclude with a brief summary in Sec.~\ref{sec:Conclu}.
Some technical details have been relegated to  appendices.

\section{Gapless topological materials protected by nonspatial symmetries}
\label{sec:II}

Since the classification of reflection symmetry protected topological semimetals and nodal superconductors is closely 
related to the topological classification of gapless states protected by global symmetries, we first briefly review the ten-fold classification
of gapless topological materials (cf.~Appendix~\ref{appendixA}).
 This brief review also aims to clarify some open questions which recently arose in the literature.\cite{Sato_Crystalline_arxiv14,matsuuraNJP13,ZhaoWangPRL13,ZhaoWangPRB14} 
The ten-fold scheme classifies gapless fermionic systems in terms of three fundamental global symmetries, i.e., 
antiunitary time-reversal and particle-hole symmetry, as well as chiral (i.e., sublattice) symmetry.\cite{altlandZirnbauerPRB10,Zirnbauer1996fk} 
In momentum space, TRS and PHS of the Bloch or BdG Hamiltonian $H ( {\bf k} )$ are implemented by  antiunitary operators 
$T$ and $C$, which act on $H ( {\bf k} )$ as
\begin{equation} \label{TRSopPHSop}
T^{-1}  H ( -{\bf k}) T 
= + H ( {\bf k} )
\; \textrm{and} \;
C^{-1} H (-{\bf k} ) C 
=  - H ( {\bf k} ),
\end{equation}
respectively. 
Both $T$ and $C$ can square either to $+1$ or $-1$, depending on the type of the symmetry (see last three columns of Table~\ref{original table}).
Chiral symmetry, on the other hand, is implemented by
\begin{eqnarray} \label{chiralOPP}
S^{-1} H ({\bf k} ) S 
&=& 
- H ( {\bf k} ),
\end{eqnarray}
where $S$ is a unitary operator.

\subsection{Ten-fold classification of gapless topological materials}

%%%%%%%%%%% begin Figure
\begin{figure}[thb]
 \begin{center} 
\includegraphics[clip,width=0.98\columnwidth]{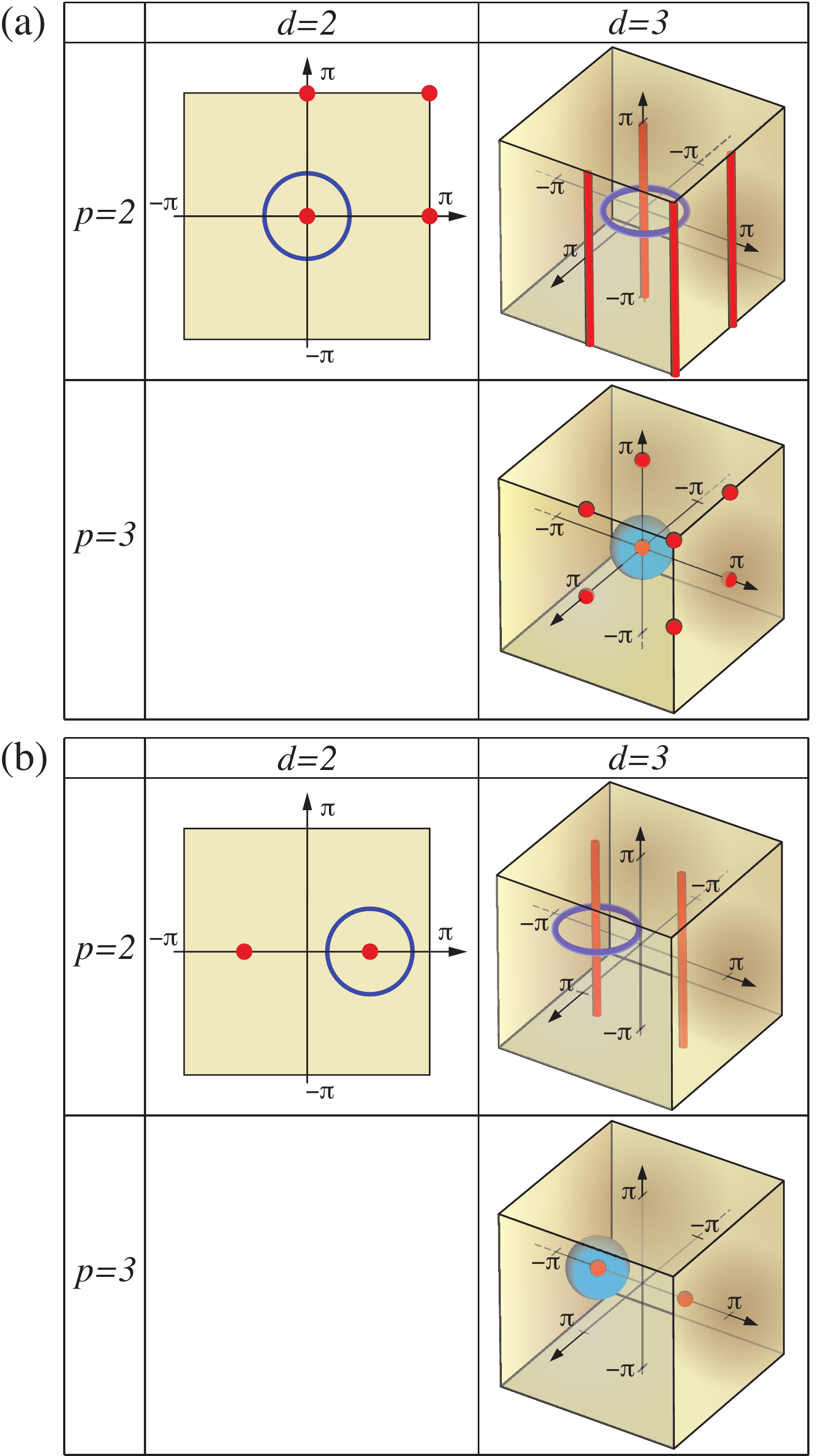}
\hfill
 \end{center}
 \caption{(Color online) The ten-fold classification of gapless topological materials depends on
 the location of the Fermi surfaces in the Brillouin zone, which in turn determines
how the Fermi surfaces transform under  global antiunitary symmetries, see Table~\ref{original table}. (a) Each Fermi surface (red point/line) is left invariant under
global (i.e., nonspatial) symmetries. 
The contour, on which the topological invariant is defined, is indicated by blue circles/spheres.
Here, $d$ denotes the spatial dimension and $p=d - d_{\mathrm{FS}}$ is the codimension of the Fermi surface. 
(b) Different Fermi surfaces
are pairwise related to each other by global symmetries (${\bf k} \leftrightarrow -{\bf k}$).  }
 \label{FSwithGlobalSym}
\end{figure}
%%%%%%%%%% end Figure

As it turns out, the topological classification of gapless materials  depends not only on 
the symmetry class of the Hamiltonian and the codimension $p$ of the Fermi surface  
\begin{eqnarray}
p =  d  - d_{\mathrm{FS}}, 
\end{eqnarray} 
where $d$ and $d_{\mathrm{FS}}$ denote the dimension of the Brillouin zone (BZ) and the Fermi surface, respectively, 
but also on how the Fermi surface transforms under the global symmetries.\cite{matsuuraNJP13} 
Regarding the symmetry properties of the Fermi surfaces, two different cases have to be distinguished: 
(i) each individual Fermi surface is left invariant under nonspatial symmetries, and (ii) different Fermi surfaces
are pairwise related to each other by nonspatial symmetries, see Fig.~\ref{FSwithGlobalSym}. 
While most of the recent literature has studied case (i),\cite{Sato_Crystalline_arxiv14,ZhaoWangPRL13,ZhaoWangPRB14} we emphasize that also in case
(ii) there exist topologically stable Fermi surfaces.

\subsubsection{Fermi surfaces at high-symmetry points}

As shown in Refs.~\onlinecite{matsuuraNJP13,ZhaoWangPRB14,ZhaoWangPRL13,Sato_Crystalline_arxiv14}, 
Fermi surfaces located at high-symmetry points in the BZ, can be protected by either
$\bZ$-type or $\bZ_2$-type invariants. The complete ten-fold classification of
Fermi surfaces that are left invariant under  global symmetries is shown in Table~\ref{real symmetry}, where the second row
indicates the codimension $p$ of the Fermi surface at a high-symmetry point.
This result has been obtained using a dimensional reduction procedure\cite{matsuuraNJP13} and an approach based on K-theory.\cite{ZhaoWangPRB14,ZhaoWangPRL13,Sato_Crystalline_arxiv14} In Appendix~\ref{appendixA}, we present yet another derivation
of this classification scheme by considering all possible symmetry-allowed mass terms that can be added
to a representative Dirac-matrix Hamiltonian in a given symmetry class.
It is important to note that for a given symmetry class and codimension $p$ a $\bZ$-type topological invariant guarantees the stability of the Fermi surface 
independent of the Fermi surface dimension $d_{\mathrm{FS}}$. A $\bZ_2$-type topological number, on the other hand,  only protects Fermi surfaces of dimension zero, i.e., Fermi points. 
We can see from Table~\ref{real symmetry}, that the ten-fold classification of global-symmetry invariant Fermi points (i.e., $d_{\mathrm{FS}} = 0$) is related to the
original ten-fold classification of topological insulators and superconductors by a dimensional shift, i.e., $d \to d - 1$.
Due to a bulk-boundary correspondence,\cite{matsuuraNJP13,ZhaoWangPRL13,Sato_Crystalline_arxiv14} gapless materials with nontrivial topology support protected surface states, which, depending on the case,
are either Dirac or Majorana states or are dispersionless, forming flat bands or arc surface states.

%%%%%%%%%%%%%%%%%%%%%%%%% TABLE
\begin{table}[t!]
\caption{Ten-fold classification of topological insulators and fully gapped superconductors,\cite{Kitaev,SchnyderAIP,Ryu2010ten,Schnyder2008gf} 
as well as of Fermi surfaces and nodal point/lines in
semimetals and nodal superconductors, respectively.\cite{matsuuraNJP13,ZhaoWangPRL13,ZhaoWangPRB14}
The first row indicates the spatial dimension $d$ of  topological insulators and superconductors,
whereas the second and third rows specify the codimension  $p=d - d_{\mathrm{FS}}$ of 
the Fermi surfaces (nodal lines) at high-symmetry points [Fig.~\ref{FSwithGlobalSym}(a)] and away from high-symmetry points of the Brillouin zone [Fig.~\ref{FSwithGlobalSym}(b)], respectively.
The first column gives the name of the  symmetry classes. The labels $T$, $C$, and $S$ in the last three columns indicate the presence (``$+$'', ``$-$'', and ``1'') or absence (``0'') of time-reversal, particle-hole and chiral symmetries, respectively, as well as the sign of the squared symmetry operators $T^2$ and $C^2$. 
}
\label{real symmetry}
\label{original table}
\begin{center}
\begin{threeparttable}
\begin{tabular}{|c|cccccccc|ccc|}
\toprule
            \mbox{\footnotesize{top.\ insul.\ and top.\ SC}}    & \footnotesize{$d$=1} & \footnotesize{$d$=2} & \footnotesize{$d$=3} & \footnotesize{$d$=4} & \footnotesize{$d$=5} & \footnotesize{$d$=6} & \footnotesize{$d$=7} & \footnotesize{$d$=8} & \multirow{3}{*}{T} & \multirow{3}{*}{C} & \multirow{3}{*}{S}  \\
                  \mbox{\footnotesize{FS at high-sym.\ point}}  & \footnotesize{$p$=8}  & \footnotesize{$p$=1} & \footnotesize{$p$=2} & \footnotesize{$p$=3} & \footnotesize{$p$=4} & \footnotesize{$p$=5} & \footnotesize{$p$=6} & \footnotesize{$p$=7}  &  &  &   \\
                \mbox{\footnotesize{FS off high-sym.\ point}}    & \footnotesize{$p$=2} & \footnotesize{$p$=3} & \footnotesize{$p$=4} & \footnotesize{$p$=5} & \footnotesize{$p$=6} & \footnotesize{$p$=7} & \footnotesize{$p$=8} & \footnotesize{$p$=1}  &  &  & \\
\hline  \hline
     A   &   0 & $\mathbb{Z}$ & 0 & $\mathbb{Z}$ & 0 & $\mathbb{Z}$ & 0    & $\bZ$         & 0 & 0 & 0    \\
  AIII    &  $\mathbb{Z}$ & 0 & $\mathbb{Z}$ & 0 & $\mathbb{Z}$ & 0 & $\mathbb{Z}$  & 0        & 0 & 0 & 1    \\ 
\hline \hline
  AI     & 0 & 0 & 0 & $2\mathbb{Z}$ & 0 & $\bZ_2^{\textrm{a,b}}$ & $\bZ_2^{\textrm{a,b}}$   &  $\mathbb{Z}$  & $+$ & 0 & 0     \\
  BDI    & $\mathbb{Z}$ & 0 & 0 & 0 & $2\mathbb{Z}$ & 0 & $\bZ_2^{\textrm{a,b}}$   & $\bZ_2^{\textrm{a,b}}$  & $+$ & $+$ & 1    \\
  D   & $\bZ_2^{\textrm{a,b}}$ & $\mathbb{Z}$ & 0 & 0 & 0 & $2\mathbb{Z}$ & 0  & $\bZ_2^{\textrm{a,b}}$      & 0 & $+$ & 0     \\
  DIII      & $\bZ_2^{\textrm{a,b}}$ & $\bZ_2^{\textrm{a,b}}$ & $\mathbb{Z}$ & 0 & 0 & 0 & $2\mathbb{Z}$ & 0    & $-$ & $+$ & 1   \\
  AII      & 0 & $\bZ_2^{\textrm{a,b}}$ & $\bZ_2^{\textrm{a,b}}$ & $\mathbb{Z}$ & 0 & 0 & 0   & $2\mathbb{Z}$   & $-$ & 0 & 0   \\
  CII      & $2\mathbb{Z}$ & 0 & $\bZ_2^{\textrm{a,b}}$ & $\bZ_2^{\textrm{a,b}}$ & $\mathbb{Z}$ & 0 & 0   & 0  & $-$ & $-$ & 1    \\
  C   &   0 & $2\mathbb{Z}$ & 0 & $\bZ_2^{\textrm{a,b}}$ & $\bZ_2^{\textrm{a,b}}$ & $\mathbb{Z}$ & 0  & 0      & 0 & $-$ & 0    \\
  CI    &  0 & 0 & $2\mathbb{Z}$ & 0 & $\bZ_2^{\textrm{a,b}}$ & $\bZ_2^{\textrm{a,b}}$ & $\mathbb{Z}$   & 0  & $+$ & $-$ & 1\\
\hline \hline
\end{tabular}
 \begin{tablenotes}
    \item[${}^{\textrm{a}}$]  $\bZ_2$ numbers only protect Fermi surfaces of dimension zero ($d_{\mathrm{FS}}=0$) at high-symmetry points of the Brillouin zone.
        \item[${}^{\textrm{b}}$] Fermi surfaces located away from high symmetry points of the Brillouin zone cannot be protected by a $\bZ_2$ topological number. Nevertheless, the system can exhibit gapless  surface states (at time-reversal invariant momenta of the surface Brillouin zone) that are protected by a $\bZ_2$ topological invariant.
    \end{tablenotes}
    \end{threeparttable}
\end{center}
\end{table}
%%%%%%%%%%%%%%%%%%%%%%%%% TABLE

\begin{figure*}[thb]
 \begin{center}
\includegraphics[clip,width=1.98\columnwidth]{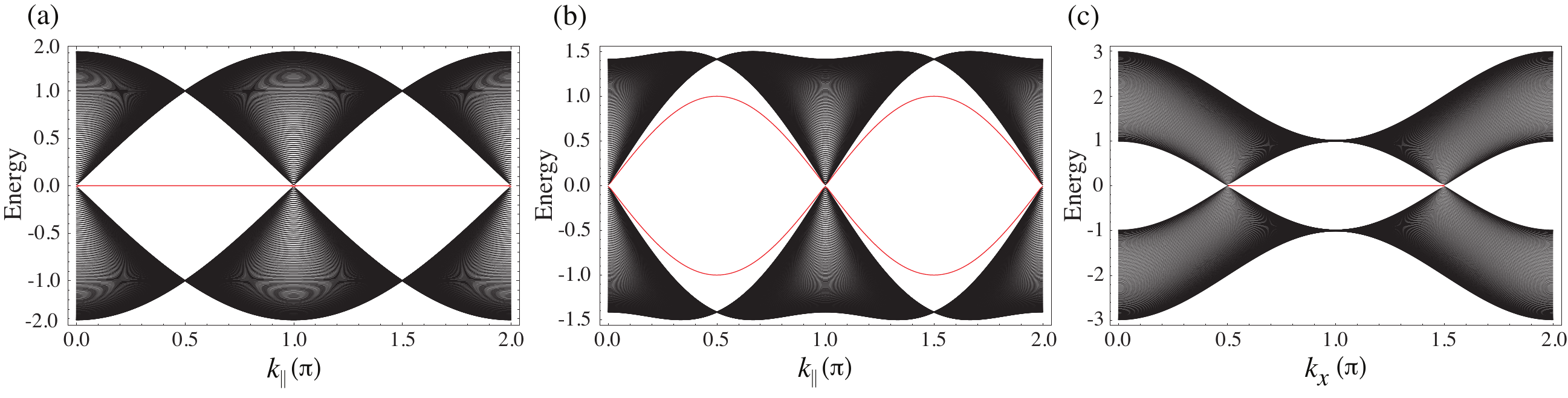}
\\
 \end{center}
 \caption{ (Color online)
 (a) Edge band structure of the nodal topological superconductor~\eqref{exampDIII} (class DIII) for the (11) face as a function
 of edge momentum $k_{\parallel} =( k_x + k_y)  / \sqrt{2}$.
 The flat-band edge states (red traces) are protected by time reversal symmetry and particle-hole symmetry. 
 (b)~Band structure of the time-reversal invariant semimetal~\eqref{exampClassAII} (class AII) at the (11) edge as a function 
 of edge momentum $k_{\parallel}$. Linearly dispersing edge states (red traces) connect the projected Fermi points in the edge BZ.
 (c)
 Edge spectrum of the sublattice-symmetric (chiral-symmetric) semimetal~\eqref{AIII_Fermi} with $A=B=0.7$ at the (01) face as a function of edge momentum $k_x$. 
 The flat-band edge states (red trace) are protected by sublattice (chiral) symmetry. 
 }
 \label{spectrumExamps}
\end{figure*}

Let us illustrate some of the gapless topological states  listed in Table~\ref{real symmetry} by considering specific lattice models. 
\paragraph{Nodal superconductor with TRS (class DIII)}
To demonstrate that $\bZ$-type invariants protect Fermi surfaces (nodal lines) of arbitrary dimension $d_{\mathrm{FS}}$
we study the following  two-dimensional tight-binding Hamiltonian on the square lattice
\bee \label{exampDIII}
H^{\rm DIII}_{\rm s}=\sin k_x \sigma_x + \sin k_y \sigma_y, 
\ee
which describes a nodal superconductor with point nodes ($d_{\mathrm{FS}}=0$) at the four time-reversal invariant momenta $(0,0),\ (0,\pi),\ (\pi,0)$, and $(\pi,\pi)$.
Hamiltonian \eqref{exampDIII} preserves time reversal symmetry, with $T=\sigma_y \mK$, and particle-hole symmetry, with $C=\sigma_x \mK$.
Here, $\mK$ denotes the complex conjugation operator. 
Since $T^2=- \bI$ and $C^2 = + \bI$, the Hamiltonian
belongs to symmetry class DIII, where $\bI$ is the $2\times 2$ identity matrix.
According to Table~\ref{real symmetry}, superconducting nodes with codimension $p=2$ in class DIII are protected by 
a $\bZ$-type topological invariant.
Indeed, we find that the winding number 
\bee
\nu =\frac{i}{2\pi}\int_{\mathcal{C}}q^*dq, \label{DIIIWno}
\ee
where $q=(\sin k_x -i \sin k_y)/\sqrt{\sin^2 k_x+\sin^2 k_y } $, is quantized to $\pm 1$ for closed contours $\mathcal{C}$ encircling one of the four nodal points. 
Specifically, for an anticlockwise-oriented contour we obtain $\nu=+1$ for the nodes at $(0,0)$ and $(\pi,\pi)$, whereas $\nu=-1$ for 
the nodes at $(0,\pi)$ and $(\pi,0)$. The topological nature of these point nodes results in the appearance of protected
flat-band edge states for all edge orientations, except the (10) and (01) faces. As demonstrated in Fig.~\ref{spectrumExamps}(a), these  flat-band states
connect two projected nodal points with different topological charge (i.e., different winding number $\nu$) in the edge BZ.
The BdG Hamiltonian \eqref{exampDIII} can be converted in a straightforward manner to a three-dimensional topological  superconductor 
with protected line nodes ($d_{\mathrm{FS}} = 1$) by including an extra momentum-space coordinate.  Similar to the two-dimensional example, Eq.~\eqref{exampDIII}, 
the stability of these nodal lines is guaranteed by  the quantized winding number $\nu$, Eq.~\eqref{DIIIWno}.

 %%%%%%%%%%% begin Figure
%%%%%%%%%% end Figure

\paragraph{Semimetal with TRS (class AII)}
As stated above, $\bZ_2$-type invariants only protect Fermi surfaces of dimension zero ($d_{\mathrm{FS}}$=0) at high-symmetry points of the BZ
and cannot give rise to topologically stable Fermi surfaces with $d_{\mathrm{FS}}>0$.
To exemplify this, we consider
the following two-dimensional Bloch Hamiltonian on the square lattice
\begin{eqnarray} \label{exampClassAII}
H^{\rm AII}_{\rm s}= \sin k_x \sigma_x + \sin k_y \sigma_y+ \sin(k_x+k_y)\sigma_z 
\end{eqnarray}
that describes a semimetal with Fermi points at the four time-reversal invariant momenta of the two-dimensional BZ.
Hamiltonian~\eqref{exampClassAII} preserves  time-reversal symmetry, with $T=\sigma_y \mK$, but breaks
particle-hole symmetry, thus belonging to symmetry class AII. 
The four Fermi points are protected by a binary $\bZ_2$ invariant,
which can be defined in terms of  an extension of $H^{\rm AII}_{\rm s}$ to three dimensions\cite{ZhaoWangPRL13,wangZhangPRL10}
\begin{eqnarray}  \label{extendHamAA}
&&
\widetilde{H}^{\rm AII}_{\rm s}(k,\theta)=  \big[ \sin k_x \sigma_x + \sin k_y \sigma_y
\\
&& \qquad  \qquad \quad
+ \sin(k_x+k_y)\sigma_z \big] \sin \theta
+ \sigma_z \cos \theta ,
\nonumber
\end{eqnarray}
where $\theta \in [0, \pi]$ is the parameter for the extension in the third direction. The extended Hamiltonian~\eqref{extendHamAA} is required to preserve TRS 
\bee
T^{-1}\widetilde{H}^{\rm AII}_{\rm s}(-k,\pi - \theta)T=\widetilde{H}^{\rm AII}_{\rm s}(k, \theta).
\ee
Performing a small-momentum expansion around a given Fermi point, we find that
the $\bZ_2$ invariant is expressed as
\bee \label{Z2 extension}
n_{\bZ_2} = \frac{1}{4\pi} \int_0^{2\pi}d\phi \int^{\pi}_0 
d\theta \, \hat{\bf g}\cdot (\partial_\theta \hat{\bf g}\times \partial_\phi\hat{\bf g}) \text{ mod } 2 , 
\ee
with $\hat{\bf g} = {\bf g} / | {\bf g} |$ and
\begin{align}
{\bf g} &=(\pm \tilde{k}_x, \pm \tilde{k}_y, \tilde{k}_x+\tilde{k}_y+\Delta ), &\text{ for } {\bf k}=(0,0),\ (\pi,\pi), \nonumber\\
{\bf g} &=(\pm \tilde{k}_x, \mp \tilde{k}_y, - \tilde{k}_x-\tilde{k}_y +\Delta),  &\text{ for } {\bf k}=(0,\pi),\ (\pi,0), 
\end{align} 
where $\tilde{k}_x=k \cos \phi \sin \theta$, $\tilde{k}_y=k\sin \phi \sin \theta$, $\Delta=\Delta_0 \cos \theta$, and ($k$, $\Delta_0$) are positive constants.
The integral~\eqref{Z2 extension} is evaluated along the sphere that surrounds the Fermi point and is required to preserve TRS.
We observe that the $\bZ_2$ invariant  \eqref{Z2 extension} is nontrivial (i.e., $n=1$) for all four Fermi points,
hence indicating the topological protection of these two-dimensional Dirac points.
By the bulk-boundary correspondence, the topological characteristics of these Fermi points lead to linearly dispersing edge modes, which connect two projected Dirac points in the edge BZ, see Fig.~\ref{spectrumExamps}(b).
Importantly, we find that Hamiltonian~\eqref{exampClassAII} cannot be converted to a three-dimensional semimetal with Fermi lines,
since it is possible to gap out the Fermi lines located at $(0,0,k_z),\ (0,\pi,k_z),\ (\pi,0,k_z)$, and $(\pi,\pi,k_z)$  by the symmetry preserving term $\sin k_z \sigma_z$.
That is,  in the presence of Fermi \emph{lines} along the $k_z$ direction, the topological invariant \eqref{Z2 extension} is ill-defined
for $k_z \ne 0, \pi$, since it breaks TRS.

\paragraph{Unstable semimetal with TRS and PHS (class BDI)} 
As an example of an unstable semimetal in two-dimensions we consider the square-lattice Hamiltonian
\bee \label{exampBDI}
H^{\rm BDI}_{\rm s}=\sin k_x \, \sigma_x \otimes \sigma_y + \sin k_y \,  \sigma_y \otimes \bI, 
\ee
which represents a four-band semimetal with Fermi points at the four time-reversal invariant momenta. 
Hamiltonian \eqref{exampBDI} belongs to class BDI, since it is both time-reversal and particle-hole symmetric with
 $T= \bI \otimes \bI \mK$ and $C=\sigma_z\otimes \bI \mK$, respectively. 
 In agreement with the classification of Table~\ref{real symmetry}, the four Fermi points of $H^{\rm BDI}_{\rm s}$ are unstable, as they 
 can be gapped out by the symmetry-preserving mass   $\sigma_x\otimes \sigma_z$. 
 This is in accordance with the fact that the winding number 
\bee \label{BDIwind1d}
\nu=\frac{i}{2\pi}\int_{\mathcal{C}}\tr ({\bf q}^\dag d {\bf q}),
\ee 
where 
\bee
{\bf q}=\frac{-i}{\sqrt{\sin^2 k_x +\sin^2 k_y }}
\bma 
\sin k_y & \sin k_x \\
-\sin k_x & \sin k_y \\
\ema,
\ee
vanishes identically for \emph{any} closed contour $\mathcal{C}$.

\subsubsection{Fermi surfaces off high-symmetry points}
\label{FermiSurfOffHighSym}

Second, we discuss the topological classification of semimetals and nodal superconductors
with Fermi surfaces (or superconducting nodes) that are located away from high-symmetry points 
of the BZ. In this case, global antiunitary  symmetries pairwise relate different Fermi surfaces with each other, see Fig.~\ref{FSwithGlobalSym}(b). 
Interestingly, only $\bZ$-type invariants can guarantee the stability of Fermi surfaces off high-symmetry points. $\bZ_2$-type numbers, 
on the other hand, cannot protect these Fermi surfaces, but  may nevertheless lead to the appearance of zero-energy surface states at 
time-reversal invariant momenta of the surface BZ.
The complete classification of Fermi surfaces that are pairwise related by global symmetries
is shown in Table~\ref{real symmetry}, where the third row indicates the codimension $p$ of
the Fermi surface  located away from high-symmetry points  (cf.\ Appendix~\ref{appendixA}). We observe
that the classification for the two complex symmetry classes A and AIII is identical to the one of 
Fermi surfaces that are left invariant by global symmetries, while the classification  for the eight real symmetry classes is different.
As before, we notice that this classification scheme is related to the
original ten-fold classification of topological insulators and superconductors by a dimensional shift, i.e., in this case $d \to d + 1$.

In order to exemplify some of the gapless topological states with Fermi surfaces away from high symmetry points
we consider a few specific lattice modes. 
\paragraph{Two-dimensional  semimetal with SLS (class AIII)}
To demonstrate that $\bZ$-type invariants protect Fermi surfaces at non-high-symmetry points of the BZ,
we study the following sublattice symmetric Hamiltonian on the square lattice 
\begin{align}   \label{AIII_Fermi}
H^{\textrm{AIII}}_{\rm n}= X\sigma_x +  Y\sigma_y ,
\end{align}
where $X=1+\cos k_y +A\sin k_x + B\cos k_x$ and $Y=\sin k_y$. Sublattice symmetry acts on  $H^{\textrm{AIII}}_{\rm n}$ as
$S H^{\textrm{AIII}}_{\rm n} + H^{\textrm{AIII}}_{\rm n} S =0$, with the unitary matrix $S=\sigma_z$.
Hamiltonian~\eqref{AIII_Fermi} exhibits two Fermi points located at 
$(\delta,\pi)$ and  $(\delta-\pi,\pi)$, where $\delta=\arctan(-B/A)$  and we require that $\sqrt{A^2+B^2}<2$.
Note that, in agreement with the
fermion-doubling theorem by Nielsen and Ninomiya,\cite{Nielsen_Ninomiya_1981}
the number of Fermi points is even.
Since there exists no symmetry-allowed mass term that can be added to  Hamiltonian~\eqref{AIII_Fermi}, the
two Fermi points are stable and, according to Table~\ref{real symmetry}, 
protected by the $\bZ$ topological number Eq.~\eqref{DIIIWno},  
with $q=(X-Yi)/\sqrt{X^2+Y^2}$ and $\mathcal{C}$ a closed contour.
Choosing $\mathcal{C}$ to be parallel to the $k_y$ axis, we find that
$\nu=+1$ for $\delta-\pi<k_x<\delta$, and zero otherwise. 
Due to an index theorem,~\cite{EssinGurariePRB11}
a nonzero value of the winding number~\eqref{DIIIWno} implies the existence of flat-band edge states at zero energy. 
At the (01) edge, the zero-energy flat-band states  appear
 within the interval $k_x \in [\delta-\pi, \delta]$ of the edge BZ, 
see Fig.~\ref{spectrumExamps}(c).

\paragraph{Three-dimensional semimetal with TRS and PHS (class BDI)}
$\bZ$-type numbers can protect Fermi surfaces of arbitrary dimension $d_{\mathrm{FS}}$.
To demonstrate this for the case of Fermi surfaces located away from high-symmetry points,
we consider the following three-dimensional tight-binding model on the cubic lattice
\bee \label{BDI_Fermi}
H^{\rm BDI}_{\rm n}=(1+\cos k_y + \cos k_x)  \sigma_x + \sin k_y \sigma_y,  
\ee
which realizes a topological semimetal with two Fermi lines at  $(\pm  \pi /2, \pi,k_z)$.
Hamiltonian~\eqref{BDI_Fermi} belongs to symmetry class BDI, since it satisfies both TRS and PHS with
$T=\bI \mK$ and $C=\sigma_z \mK$, respectively.  We observe that the two Fermi lines, which are located
away from the time-reversal invariant momenta of the BZ, transform into each other under 
particle-hole and time-reversal symmetries [cf.~Fig.~\ref{FSwithGlobalSym}(b)].
As indicated in Table~\ref{real symmetry}, the Fermi lines are protected by a $\bZ$-type topological invariant, which for 
the tight-binding model~\eqref{BDI_Fermi} takes the form of Eq.~\eqref{DIIIWno}, with $q=(1+\cos k_y + \cos k_x)- i\sin k_y$. 
The integration contour in Eq.~\eqref{DIIIWno} can be chosen to be any circle enclosing the Fermi line. (The integration contour does not need to be time-reversal
or particle-hole symmetric.)
Similar to the class AIII model \eqref{AIII_Fermi}, a nonzero value of this winding number leads to zero-energy flat-band surface states that
connect the two projected Fermi lines in the surface BZ.

\paragraph{Unstable nodal superconductor with TRS (class DIII)}\label{DIIIoff}
As indicated in Table~\ref{original table}, $\bZ_2$-type topological numbers do not guarantee the topological stability of
Fermi surfaces (superconducting nodes) at non-high-symmetry points of the BZ. Nevertheless, $\bZ_2$-type invariants, which are defined on time-reversal symmetric contours, can
give rise to protected gapless surface states.
To demonstrate this, we consider an example of an unstable nodal superconductor given by the  four-band BdG Hamiltonian
\bee \label{exampDIIIoffSym}
H_{\rm n}^{\rm DIII}=(1+\cos k_x +  \cos k_y)\sigma_x \otimes \sigma_y + \sin k_x \sigma_y \otimes \bI . 
\ee	
 This superconductor belongs to symmetry class DIII, as it preserves both  time-reversal  
 and particle-hole symmetries, with
 $T=\sigma_y\otimes \bI \mK$  and $C=\sigma_x\otimes \bI \mK$, respectively. 
Hamiltonian~\eqref{exampDIIIoffSym} exhibits two point nodes at $( \pi,\pm\pi/2)$. 
These two point nodes, which are positioned away from the high-symmetry points of the BZ, are unstable, since 
the symmetry-preserving extra kinetic term $\sin k_x \, \sigma_x\otimes \sigma_x$ opens up a gap 
in the entire bulk BZ (cf.\ Table~\ref{real symmetry}).
This is corroborated by the fact that the winding number $\nu$
for model Hamiltonian~\eqref{exampDIIIoffSym} is identically zero for any closed contour $\mathcal{C}$,
which follows from a similar argument as the one given in the example of Eq.~\eqref{exampBDI}.
In contrast, the one-dimensional $\mathbbm{Z}_2$ 
number\cite{Ryu2010ten,spin_pump}
\begin{eqnarray} \label{Z2noExampDIIIccc}
n_{\bZ_2} &=& 
\prod_{{\bf K} \in \mathcal{C} }
\frac{ \rm{Pf}[\omega  ({\bf K})] }
{ \sqrt{ \rm{det}Ê[\omega  ({\bf K})] }}
\end{eqnarray}
for Hamiltonian \eqref{exampDIIIoffSym} can take on nontrivial values, which however 
does not lead to a protection of the point nodes
of the superconductor (cf.\ Table~\ref{real symmetry} and {Appendix~\ref{appendixA}). 
In Eq.~\eqref{Z2noExampDIIIccc} the product is over the two time-reversal invariant momenta {\bf K} (high-symmetry points) of
the contour $\mathcal{C}$ and
$\omega  ({\bf K})$ denotes the $2Ê\times 2$ sewing matrix
\begin{eqnarray} \label{sewingMatrixDIII}
\omega_{\hat{a}\hat{b}} ( {\bf k} )
&=&
 \langle u_{\hat{a}}^- (-{\bf k} ) | \, T \,  u_{\hat{b}}^-({\bf k}) \rangle ,
\end{eqnarray}
with $\ket{u_{\hat{a}}^-({\bf k})}$ the negative-energy BdG wave functions of Hamiltonian~\eqref{exampDIIIoffSym}.
Even though  $\mathbbm{Z}_2$ number \eqref{Z2noExampDIIIccc} does not stabilize point nodes in the bulk,
it nevertheless leads to protected zero-energy surface states at time-reversal invariant momenta of the surface BZ. 
To exemplify this, we consider two time-reversal invariant contours $\mathcal{C}$ oriented along the $k_x$ axis with
$k_y$ held fixed at $k_y = 0$ or $k_y = \pi$. With these contours, the $\mathbbm{Z}_2$ number~\label{Z2noExampDIIInS}
takes on the values $n=+1$ and $n=-1$  at $k_y =0$ and $k_y = \pi$, respectively, indicating 
the existence of a zero-energy edge state at $k_y = \pi$ of the $(10)$ edge BZ of the superconductor. 
We observe that
the unstable nodal superconductor~\eqref{exampDIIIoffSym} can be  connected to a fully gapped topological superconductor without removing the
zero-energy edge-states. That is, the edge-states of Hamiltonian~\eqref{exampDIIIoffSym} are inherited from the fully gapped topological phase.\cite{poYao14}

%%%%%%%%%%%%%%%%%%%%%%%%% TABLE
\begin{table*}[t!]
\caption{Classification of reflection symmetry protected topological insulators and fully gapped superconductors,\cite{chiuPRB13,morimotoPRB13,Sato_Crystalline_arxiv14}
as well as of Fermi surfaces and nodal points/lines in reflection symmetry protected semimetals and nodal superconductors, 
respectively. 
The first row specifies the spatial dimension $d$ of reflection symmetry protected topological insulators and fully gapped superconductors,
while the second and third rows indicate the codimension $p=d-d_{\mathrm{FS}}$ of the reflection symmetric
Fermi surfaces (nodal lines) at high-symmetry points [Fig.~\ref{FigLocFermiWMirror}(a)]
and away from high-symmetry points of the Brillouin zone [Fig.~\ref{FigLocFermiWMirror}(b)], respectively. 
}
\label{reflection table}
\label{reflection_table_full}
\begin{center}
\begin{threeparttable}
\begin{tabular}{|c|c|cccccccc|}
\toprule
& top.\ insul.\ and top.\ SC   & $d$=1 & $d$=2 & $d$=3 & $d$=4 & $d$=5 & $d$=6 & $d$=7 & $d$=8  \\
  \cline{2-10}
  $\begin{array}{c}
\mbox{ \phantom{a} } \\
\mbox{ Reflection }
\end{array}$   &  
  $\begin{array}{c}
\mbox{FS within mirror plane} \\
\mbox{ at high-sym.\ point }
\end{array}$
  & $p$=8 & $p$=1 & $p$=2 & $p$=3 & $p$=4 & $p$=5 & $p$=6 & $p$=7    \\
  \cline{2-10}
   &  
     $\begin{array}{c}
\mbox{FS within mirror plane} \\
\mbox{ off high-sym.\ point }
\end{array}$  
  & $p$=2 & $p$=3 & $p$=4 & $p$=5 & $p$=6 & $p$=7 & $p$=8 & $p$=1     \\
\hline\hline
 $R$ &  A  & $M\bZ$ & 0 & $M\bZ$ & 0 & $M\bZ$ & 0 & $M\bZ$ & 0                \\
 $R_+$ &  AIII  & 0 & $M\bZ$ & 0 & $M\bZ$ & 0 & $M\bZ$ & 0 & $M\bZ$            \\ 
$R_-$ &  AIII   &  $M\bZ\oplus\mathbb{Z}$ & 0 & $M\bZ\oplus\mathbb{Z}$ & 0 & $M\bZ\oplus\mathbb{Z}$ & 0 & $M\bZ\oplus\mathbb{Z}$ & 0            \\ 
\hline
\multirow{8}{*}{$R_+$,$R_{++}$} & AI  & $M\bZ$ & 0 & 0 & 0 & $2M\bZ$ & 0 & $M\bZ_2^{\textrm{a,b}}$ & $M\bZ_2^{\textrm{a,b}}$        \\
  & BDI  & $M\bZ_2^{\textrm{a,b}}$ & $M\bZ$ & 0 & 0 & 0 & $2M\bZ$ & 0 & $M\bZ_2^{\textrm{a,b}}$        \\
  & D  & $M\bZ_2^{\textrm{a,b}}$ & $M\bZ_2^{\textrm{a,b}}$ & $M\bZ$ & 0 & 0 & 0 & $2M\bZ$   & 0       \\
  & DIII  & 0 & $M\bZ_2^{\textrm{a,b}}$ & $M\bZ_2^{\textrm{a,b}}$ & $M\bZ$ & 0 & 0 & 0  & $2M\bZ$       \\
  & AII   & $2M\bZ$ & 0 & $M\bZ_2^{\textrm{a,b}}$ & $M\bZ_2^{\textrm{a,b}}$ & $M\bZ$ & 0 & 0   & 0  \\
  & CII  & 0 & $2M\bZ$ & 0 & $M\bZ_2^{\textrm{a,b}}$ & $M\bZ_2^{\textrm{a,b}}$ & $M\bZ$ & 0     & 0   \\
  & C & 0 & 0 & $2M\bZ$ & 0 & $M\bZ_2^{\textrm{a,b}}$ & $M\bZ_2^{\textrm{a,b}}$ & $M\bZ$  & 0        \\
  & CI & 0 & 0 & 0 & $2M\bZ$ & 0 & $M\bZ_2^{\textrm{a,b}}$ & $M\bZ_2^{\textrm{a,b}}$ & $M\bZ$    
   \\
\hline
\multirow{8}{*}{$R_-$,$R_{--}$} & AI  & 0 & 0 & $2M\bZ$ & 0 & $T\bZ_2^{\textrm{a,b,c}}$ &  $\bZ_2^{\textrm{a,b}}$ & $M\bZ$ & 0      \\
  & BDI   & 0  & 0 & 0 & $2M\bZ$ & 0 & $T\bZ_2^{\textrm{a,b,c}}$ &  $\bZ_2^{\textrm{a,b}}$ & $M\bZ$        \\
  & D &  $M\bZ$   &  0  & 0 & 0 & $2M\bZ$ & 0 & $T\bZ_2^{\textrm{a,b,c}}$ &  $\bZ_2^{\textrm{a,b}}$        \\
  & DIII  &  $\bZ_2^{\textrm{a,b}}$ & $M\bZ$   &  0  & 0 & 0 & $2M\bZ$ & 0 & $T\bZ_2^{\textrm{a,b,c}}$       \\
  & AII  & $T\bZ_2^{\textrm{a,b,c}}$ &  $\bZ_2^{\textrm{a,b}}$ & $M\bZ$   &  0  & 0 & 0 & $2M\bZ$ & 0    \\
  & CII  & 0 & $T\bZ_2^{\textrm{a,b,c}}$ &  $\bZ_2^{\textrm{a,b}}$ & $M\bZ$   &  0  & 0 & 0 & $2M\bZ$        \\
  & C   & $2M\bZ$ & 0 & $T\bZ_2^{\textrm{a,b,c}}$ &  $\bZ_2^{\textrm{a,b}}$ & $M\bZ$   &  0  & 0 & 0          \\
  & CI  & 0 & $2M\bZ$ & 0 & $T\bZ_2^{\textrm{a,b,c}}$ &  $\bZ_2^{\textrm{a,b}}$ & $M\bZ$   &  0  & 0     
   \\
\hline
$R_{-+}$ & BDI, CII  & $2\bZ$ & 0 & $2M\bZ$ & 0 & $2\bZ$ & 0 & $2M\bZ$ & 0       \\
$R_{+-}$  & DIII, CI  & $2M\bZ$ & 0 & $2\bZ$ & 0 & $2M\bZ$ & 0 & $2\bZ$ & 0        \\
\hline
$R_{+-}$  & BDI  & $M\bZ\oplus\bZ$ & 0 & 0 & 0 &  $2M\bZ\oplus 2\bZ$ & 0 & $M\bZ_2\oplus \bZ_2^{\textrm{a,b}}$ & $M\bZ_2\oplus \bZ_2^{\textrm{a,b}}$        \\
$R_{-+}$ & DIII  & $M\bZ_2\oplus \bZ_2^{\textrm{a,b}}$ & $M\bZ_2\oplus \bZ_2^{\textrm{a,b}}$    & $M\bZ\oplus\bZ$ & 0 & 0 & 0 &  $2M\bZ\oplus 2\bZ$ & 0     \\
$R_{+-}$  & CII &   $2M\bZ\oplus 2\bZ$ & 0 & $M\bZ_2\oplus \bZ_2^{\textrm{a,b}}$ & $M\bZ_2\oplus \bZ_2^{\textrm{a,b}}$    & $M\bZ\oplus\bZ$ & 0 & 0 & 0       \\
$R_{-+}$  & CI   & 0 & 0    & $2M\bZ\oplus 2\bZ$ & 0 & $M\bZ_2\oplus \bZ_2^{\textrm{a,b}}$ & $M\bZ_2\oplus \bZ_2^{\textrm{a,b}}$    & $M\bZ\oplus\bZ$ & 0        \\
\hline
\hline
\end{tabular}
 \begin{tablenotes}
  \item[${}^{\textrm{a}}$]  $\bZ_2$ and $M\bZ_2$ invariants only protect Fermi surfaces of dimension zero ($d_{\mathrm{FS}}=0$) at high-symmetry points of the Brillouin zone.
        \item[${}^{\textrm{b}}$] Fermi surfaces located within the mirror plane but away from high symmetry points  cannot be protected by a $\bZ_2$ or $M\bZ_2$  topological invariant. Nevertheless, the system can exhibit gapless surface states 
        that are protected by a $\bZ_2$ or $M\bZ_2$ topological invariant. 
                \item[${}^{\textrm{c}}$] For gapless topological materials the presence of translation symmetry is always  assumed. 
Hence, there is no distinction between    $T\bZ_2$ and $\bZ_2$ for gapless topological materials.    
    \end{tablenotes}
    \end{threeparttable}
\end{center}
\end{table*}
%%%%%%%%%%%%%%%%%%%%%%%%% TABLE

\section{Classification of reflection symmetry protected gapless topological materials}
\label{classReflecGapless}

Having discussed the classification of
gapless topological materials in terms of global symmetries,
we are now ready to classify reflection symmetric topological semimetals and nodal superconductors. 
Reflection symmetries lead to an enrichment of the ten-fold classification of
topological semimetals (nodal superconductors) with new topological phases.  
The classification depends on the codimension $p = d - d_{\mathrm{FS}}$ of the Fermi surface (nodal line/point)
and on whether the reflection operator $R$ commutes or anticommutes with the nonspatial symmetries.
Moreover, we need to distinguish how the Fermi surface (nodal line/point)  transforms
under the mirror reflection and nonspatial symmetries. There are three different cases to be considered:
(i) The Fermi surface is invariant under both reflection and global symmetries [Fig.~\ref{FigLocFermiWMirror}(a) and Table~\ref{reflection_table_full}],
(ii) Fermi surfaces are  invariant under reflection, but transform pairwise into each other
by the global antiunitary symmetries [Fig.~\ref{FigLocFermiWMirror}(b) and Table~\ref{reflection_table_full}],
and (iii) different Fermi surfaces are pairwise related to each other by both reflection
and nonspatial symmetries [Fig.~\ref{FigLocFermiWMirror}(c) and Table~\ref{table_reflection_off_off}].

Our derivation of these classification schemes, which are presented in Tables~\ref{reflection_table_full} and~\ref{table_reflection_off_off}, relies primarily on
the so-called minimal Dirac-matrix Hamiltonian method.\cite{stoneMathTheo11,chiuPRB13,Chiu_nontrivial_surface}
This method is based on considering reflection symmetric Dirac-matrix Hamiltonians with the smallest possible
matrix dimension for a given symmetry class of the ten-fold way.
The topological properties of the Fermi surfaces (nodal lines) described by these Dirac-matrix Hamiltonians is
then determined by the existence or nonexistence of  
symmetry-preserving gap-opening terms (SPGTs), i.e., symmetry-allowed terms that
fully gap out the bulk Fermi surfaces. The existence of such an SPGT  indicates that the 
Fermi surface is topologically trivial and hence unstable.  This is denoted by the label ``0" in Tables~\ref{reflection_table_full} and~\ref{table_reflection_off_off}.
On the other hand, if no SPGT exists, then the Fermi surface is topologically stable and protected by a topological invariant (for more details see Appendix~\ref{appendixA} and Ref.~\onlinecite{chiuPRB13}).
The minimal Dirac-matrix Hamiltonian approach is complemented by a discussion of different types of topological invariants (i.e., $\mathbb{Z}$-,  $\mathbb{Z}_2$-, 
$M \mathbb{Z}$-,   $M \mathbb{Z}_2$-, and   $C \mathbb{Z}_2$-type invariants) that guarantee the  stability of these Fermi surfaces. 
For some concrete examples we derive explicit expressions for these topological numbers in Sec.~\ref{sec:examples}.  
The classification of reflection symmetric gapless  materials
in terms of topological invariants is  consistent with the  Dirac-matrix Hamiltonian method. 

Before discussing in detail the classification of reflection symmetric topological semimetals and nodal superconductors, let us  first examine how reflection symmetry acts on the
Hamiltonian and how it is related to the global symmetries. 

\subsection{Reflection symmetries}
\label{sec:ReflectSym}

Crystal reflection is a spatial symmetry,  which acts nonlocally in position space. 
For concreteness, let us consider a $d$-dimensional Bloch or BdG Hamiltonian $H( {\bf k} )$ in momentum space which is invariant  under reflection in the first direction.
The invariance of $H( {\bf k} )$ under this mirror symmetry implies 
\begin{eqnarray} \label{refSym}
R^{-1} H( -k_1, \tilde{\bf k}) R = H (k_1,\tilde{\bf k} ),
\end{eqnarray}
where $\tilde{\bf k} = (k_2, \ldots, k_d)$ and the reflection operator $R$ is a unitary matrix.
Due to a phase ambiguity in the definition of the reflection operator $R$,\cite{chiuPRB13} we can assume without loss of generality that $R$ is Hermitian (at least for electronic insulators), i.e.,
\bee
R^\dagger=R. \label{hermitian}
\ee 
With this assumption, the commutation or anticommutation relations
between $R$ and the global nonspatial symmetry operators $T$, $C$, and $S$,
\begin{equation} \label{indicesComAnti}
SRS^{-1} = \eta_S R,
\;
TRT^{-1} = \eta_T R,
\; 
CRC^{-1} = \eta_C R,
\end{equation}
can be determined in an unambiguous way,
which in turn simplifies the classification of reflection symmetry protected insulators and superconductors. 
The three indices $\eta_{S}$, $\eta_{T}$, and $\eta_{C}$ in Eq.~\eqref{indicesComAnti} take
values $+1$ or $-1$ and specify whether $R$ commutes ($+1$) or anticommutes ($-1$) with the
corresponding global symmetry operator.
These different possibilities are labeled by $R_{\eta_T}$, $R_{\eta_{S}}$, and $R_{\eta_{C}}$ for the five symmetry classes
AI, AII, AIII, C, and D, respectively, which contain only one global symmetry operation. For the remaining four symmetry classes
BDI, CI, CII, and DIII, which contain two nonspatial symmetries, the four different possible (anti)commutation relations are denoted by $R_{\eta_T \eta_C}$.
Hence, there are a total of 27 different symmetry classes for reflection symmetry protected topological insulators and fully gapped superconductors, see
Table~\ref{reflection_table_full}. We observe that since the reflection operator $R$ is both Hermitian and unitary, $R^2= \mathds{1}$ and all  eigenvalues of $R$ are either $+1$ or~$-1$. Here,  $\mathds{1}$ denotes the identity matrix with unspecified matrix dimension.

\subsection{Fermi surfaces at high-symmetry points within mirror planes}
\label{TopCrystAtMirrorHighSym}

Fermi points that are invariant under both reflection and global symmetries [red points in Fig.~\ref{FigLocFermiWMirror}(a)],
can be protected by $\mathbbm{Z}$-, $M \mathbbm{Z}$-, $\mathbbm{Z}_2$-, or $M \mathbbm{Z}_2$-type topological numbers.
The topological classification of these Fermi points ($d_{\rm FS}=0$) in $d$ dimensions is related to the classification of reflection symmetric
fully gapped systems in $d+1$ dimensions.\cite{chiuPRB13,morimotoPRB13,Sato_Crystalline_arxiv14}
(For a brief review  of the classification of fully gapped reflection symmetric topological materials see Appendix~\ref{sec:III}).
To demonstrate this relation, let us consider  a  $d$-dimensional  Dirac Hamiltonian
of a reflection symmetric insulator (or fully gapped superconductor) in a given symmetry class
\bee  \label{TI Dirac}
H^{\rm TI}_{\rm Dirac}=\sum_{i=1}^{d} k_i\gamma_i+m\tilde{\gamma}_0 .
\ee
Reflection symmetry $R$ is implemented by $R^{-1} H^{\rm TI}_{\rm Dirac} (- k_1, \tilde{\bf k} ) R = H^{\rm TI}_{\rm Dirac} ( k_1, \tilde{\bf k} ) $.
Here and in the following, $\gamma_i$ denote Dirac matrices which anticommute (commute) with the time-reversal operator $T$ (particle-hole operator $C$) 
of the given symmetry class, whereas $\tilde{\gamma}_i$ are Dirac matrices that  commute (anticommute) with $T$ ($C$), see Appendix~\ref{appendixA}. 
By considering the reflection symmetric surface states of $H^{\rm TI}_{\rm Dirac}$,
 we can derive from Eq.~\eqref{TI Dirac} a Dirac Hamiltonian describing 
a reflection symmetric Fermi point in the same symmetry class as Eq.~\eqref{TI Dirac} but in one dimension lower
\bee \label{DiracFermiaaa}
H_{\rm s}^{R}=\sum_{i=1}^{d-1} k_i {\bf P} \gamma_i  {\bf P},
\ee
with the projection operator ${\bf P} = ( \mathds{1} - i \tilde{\gamma_0} \gamma_d) / 2 $.
The topological property of $H^{\rm TI}_{\rm Dirac}$ is signaled by the existence or nonexistence of
an \emph{extra} symmetry-allowed mass term $\tilde{\Gamma}$,\cite{chiuPRB13} i.e., a symmetry preserving Dirac matrix that anticommutes with all Dirac matrices $\gamma_i$ and $\tilde{\gamma_0}$ of Eq.~\eqref{TI Dirac}. Whenever such an extra mass term $\tilde{\Gamma}$ exists, it is possible to construct an
SPGT for $H_{\rm s}^{R}$, Eq.~\eqref{DiracFermiaaa}, by
$\tilde{\Gamma}_{\bf P}Ê= {\bf P}Ê\tilde{\Gamma} {\bf P}$, which is nonzero since $\tilde{\Gamma}$ anticommutes  with both $\tilde{\gamma_0}$ and $\gamma_d$.
Vice versa, one can show that whenever there exists an SPGT for $H_{\rm s}^{R}$, i.e., a symmetry-allowed Dirac matrix $\tilde{\gamma}$ that anticommutes with
$H_{\rm s}^{R}$,
there is a corresponding extra symmetry-allowed mass term for $H^{\rm TI}_{\rm Dirac}$.\cite{chiuPRB13,altlandZirnbauerPRB10}
Hence,  the classification of Fermi points (i.e., $d_{\mathrm{FS}} = 0$) at high-symmetry positions within the mirror plane follows from the
classification of reflection symmetric fully gapped systems by the dimensional shift $d \to d - 1$ (Table~\ref{reflection_table_full}). 
We observe that this finding is in agreement with the classification of Fermi points reported by Shiozaki and Sato in Ref.~\onlinecite{Sato_Crystalline_arxiv14}
(see Eq.~(9.5) in their work).

For Fermi surfaces with $d_{\rm FS}>0$, on the other hand, the classification differs from the one of Fermi points ($d_{\rm FS}=0$).
That is, only $\bZ$-type invariants (i.e., $\bZ,\ M\bZ,$ and $M\bZ\oplus \bZ$ topological numbers) can protect Fermi surfaces
with \mbox{$d_{\rm FS}>0$}.
This is because for a gapless $d$-dimensional system with, e.g., Fermi lines along the $k_d$ direction [described by Eq.~\eqref{DiracFermiaaa}],
we can add to the Hamiltonian the additional symmetry-preserving kinetic term $k_d\gamma_{d}$, which gaps out the Fermi lines (except at high-symmetry points).
For gapless systems with a $\bZ_2$-type invariant such an extra kinetic term always exists, whereas for Fermi surfaces 
with a $\bZ$-type topological number this extra kinetic term is absent (cf.\ Appendix~\ref{appendixA} for more details and Sec.~\ref{exampFShighhigh} for some  examples).

The classification of Fermi surfaces that are located within the mirror plane at high-symmetry positions  is
summarized in Table~\ref{reflection_table_full}, where the second row indicates the codimension $p$ of the Fermi surface.
The prefix ``$M$" in Table~\ref{reflection_table_full} indicates that the corresponding topological invariant  is defined 
on a $(p-2)$-dimensional contour within the reflection plane [blue points/lines in Fig.~\ref{FigLocFermiWMirror}(a)]. The topological invariants 
labeled by $\bZ$ and $\bZ_2$, on the other hand, are defined on $(p-1)$-dimensional contours that intersect with the mirror plane (same invariants as in the absence of reflection symmetry, cf.\ Table~\ref{original table}).

%%%%%%%%%%% begin Figure
\begin{figure*}[thb]
 \begin{center} 
\includegraphics[clip,width=1.98\columnwidth]{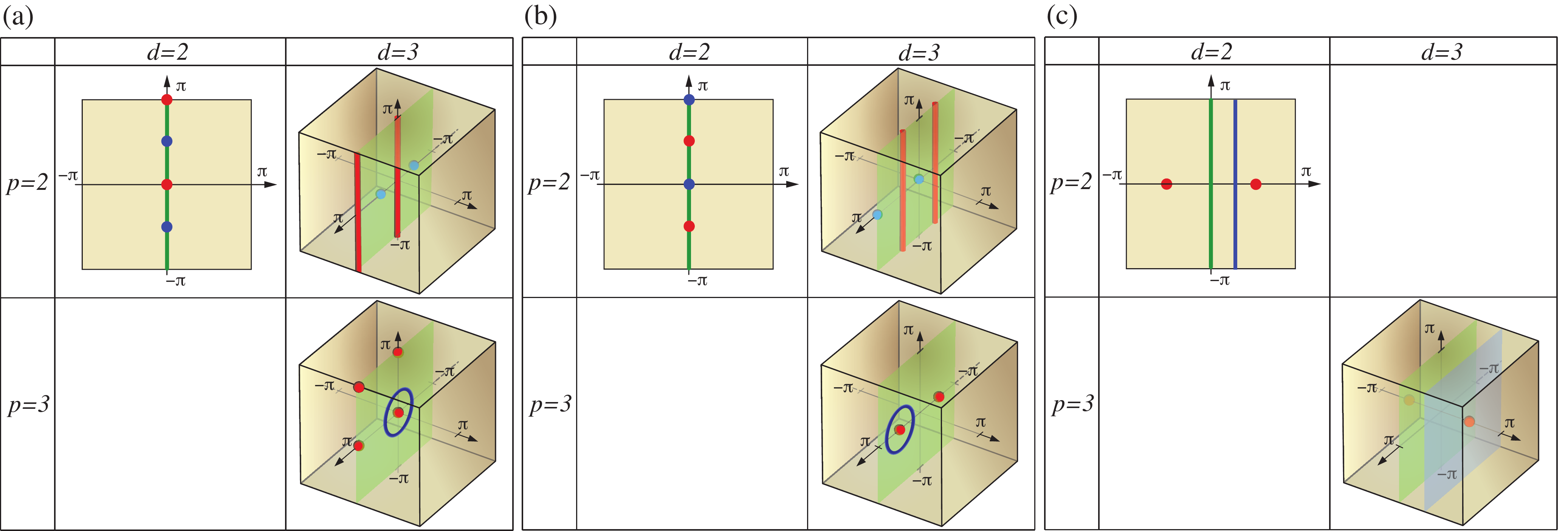}
 \end{center}
 \caption{(Color online) 
The classification of reflection symmetry protected topological semimetals and nodal superconductors depends on the location of the
Fermi surfaces with respect to the reflection plane (highlighted in green) in the Brillouin zone, which in turn determines how the Fermi surfaces transform
under reflection and global antiunitary  symmetries, see Tables~\ref{reflection_table_full} and \ref{table_reflection_off_off}.
(a)  Each Fermi surface (red points/lines) is left invariant under reflection and global antiunitary symmetries.
(b) The Fermi surfaces are left invariant by reflection, but transform pairwise into each other
by the global antiunitary symmetries.
The contours on which the $M\bZ$- and $M\bZ_2$-type invariants are defined are indicated by blue points/circles in panels (a) and (b).
 (c) Different Fermi points are pairwise related to each other by both reflection
and global symmetries. 
The contours on which the $C\bZ_2$-type invariants are defined are indicated by blue lines/planes (see Table~\ref{table_reflection_off_off}).}
 \label{FigLocFermiWMirror}
\end{figure*}
%%%%%%%%%% end Figure

\subsection{Fermi surfaces within mirror planes but off high-symmetry points}

Second, we classify Fermi surfaces that are located within the mirror plane but away from high-symmetry points [Fig.~\ref{FigLocFermiWMirror}(b)]. 
These Fermi surfaces are  invariant under reflection, but transform pairwise into each other
by the nonspatial antiunitary symmetries.
We discuss this classification by considering the following reflection symmetric Dirac-matrix Hamiltonian
\bee  \label{R off high-symmetry}
H_{n}^R=\sum_{i=1}^{p-1}\sin k_i \gamma_i + (1-p+\sum_{i=1}^{p} \cos k_i )\tilde{\gamma}_0,
\ee
which describes a semimetal (nodal superconductor) with a $(d-p)$-dimensional Fermi surface (superconducting node) located at
\bee \label{FSmirrorOFF}
{\bf k}=(0,\ldots,0,\pm \pi/2, k_{p+1},\ldots,k_{d} ) .
\ee
Reflection symmetry acts on Hamiltonian~\eqref{R off high-symmetry} as $R^{-1} H_{n}^R ( -k_1, \tilde{\bf k} ) R = H_{n}^R (k_1, \tilde{\bf k} )$.
We observe that Fermi surface~\eqref{FSmirrorOFF} lies within the mirror plane $k_1=0$, but away from the high-symmetry points $( 0, 0, 0, \ldots, 0)$, $( \pi, 0, 0, \ldots, 0)$,
$( 0, \pi, 0, \ldots, 0)$, etc.\ of the BZ.
Comparing Eq.~\eqref{R off high-symmetry} to Eq.~\eqref{TI Dirac} we find that $H_{n}^R$, with $k_p \ne \pm \pi/2$ and $k_{p-1}, \ldots, k_d$ held fixed,
can be interpreted as a reflection symmetric insulator
(fully gapped superconductor) in $d=p-1$ dimensions. 
Hence, the existence (or nonexistence) of an extra symmetry-allowed mass term $\tilde{\Gamma}$ for $H^{\rm TI}_{\rm Dirac}$, Eq.~\eqref{TI Dirac},
implies the existence (or nonexistence) of a \emph{momentum-independent} SPGT for $H_{n}^R$, Eq.~\eqref{R off high-symmetry}.
However, Fermi surface~\eqref{FSmirrorOFF} can also be gapped out by an additional symmetry-allowed kinetic term, i.e., by the momentum-dependent SPGT $\sin k_p\gamma_p$. 
It turns out that for symmetry classes with a $\bZ_2$- or $M\bZ_2$-type invariant this extra kinetic term is always allowed by symmetry, whereas for
classes with a $\bZ$- or $M\bZ$-type number this term is symmetry forbidden (cf.\ Appendix~\ref{appendixA}).
With this, it follows that the classification of $p$-dimensional Fermi surfaces (superconducting nodes) within the reflection plane but off high-symmetry points is given by
the classification of reflection symmetric topological insulators (fully gapped superconductors) in $d=p-1$ dimensions which
are protected by a $\bZ$- or $M\bZ$-type invariant (cf.\ Table~\ref{reflection_table_full}). 
We note that while $\bZ_2$- or $M\bZ_2$-type invariants cannot protect Fermi surfaces that are located within the mirror plane but away from high-symmetry points, they nevertheless might give rise to protected gapless surface states (see Sec.~\ref{Z2 like comparison} for an example).

\subsection{Fermi surfaces outside mirror planes}
\label{outside mirror planes}

Finally, we discuss the classification of Fermi surfaces (superconducting nodes) that are located outside the mirror plane. These
Fermi surfaces are pairwise related to each other by both reflection and nonspatial antiunitary symmetries, see Fig.~\ref{FigLocFermiWMirror}(c).
Reflection symmetry alone cannot protect Fermi surfaces that lie outside the reflection plane, since the reflection symmetry does not restrict the form
of the mass term at the position of the Fermi surface. However, a combination of reflection and global antiunitary symmetries can give rise to
topologically stable Fermi points (or point nodes in the superconducting gap).\cite{morimotoFurusakiPRB14,Ryu2010ten} 
In order to study this possibility we introduce the combined symmetry operators 
\begin{subequations} \label{combineSYMdef}
\bee 
\tilde{T}=RT \quad \textrm{and} \quad \tilde{C}=RC,
\ee	
which are antiunitary. 
These combined symmetry operators act on the $d$-dimensional Bloch or BdG Hamiltonian as follows
\begin{eqnarray} \label{combineSYMdefB}
\tilde{T}^{-1} H(k_1,-{\bf \tilde{k}})\tilde{T} =&+ H(k_1,{\bf \tilde{k}}) 
\end{eqnarray}
and
\begin{eqnarray} \label{combineSYMdefC}
\tilde{C}^{-1} H(k_1,-{\bf \tilde{k}})\tilde{C} =&- H(k_1,{\bf \tilde{k}}).
\end{eqnarray}
\end{subequations}
Hence, $\tilde{T}$ ($\tilde{C}$) can be viewed as an effective time-reversal (particle-hole) symmetry acting within $(d-1)$-dimensional planes that
are perpendicular to the $k_1$ direction [blue lines/planes in Fig.~\ref{FigLocFermiWMirror}(c)].
For each of these planes  it is possible to define a topological number and study its evolution as a function of the parameter $k_1$.\cite{Brydon10} 
These $k_1$-dependent topological numbers can only change across gap closing points. Hence, the stability of Fermi points or
superconducting point nodes (i.e., gap closing points) can be discussed in terms of these topological 
invariants which are defined in the presence of the combined symmetry $\tilde{T}$ and/or $\tilde{C}$, Eq.~\eqref{combineSYMdef}.
Moreover, at surfaces that are parallel to the $k_1$ direction, these $k_1$-dependent topological numbers give rise to arc surface states that connect
two projected Fermi points in the surface BZ. 

In this section, we derive the classification of Fermi surfaces outside the mirror plane, by examining which types of
topological invariants can be defined within the ($d-1$)-dimensional planes perpendicular to the $k_1$ axis.
For this, we have to distinguish between two different kinds of invariants: (i) mirror invariants that are defined
within the mirror plane for a given eigenspace of the reflection operator $R$ and (ii)  
invariants which are defined for any given plane perpendicular to the $k_1$ axis [green and blue lines/planes in Fig.~\ref{FigLocFermiWMirror}(c), respectively]. Since 
these two kinds of invariants are constrained differently by symmetry, they can in principle give rise 
 to different classifications. However, it turns out that the Fermi points are only protected by the ``weaker" 
 of these two invariants. That is, e.g., if one invariant is of $\bZ$-type whereas the other one is of $\bZ_2$-type,
 then the Fermi points only exhibit a $\bZ_2$-type topological characteristic. 
This follows from the fact that the topological invariant cannot change as a function of $k_1$ as long as the bulk gap does not close. Hence,
 the invariant defined in the mirror plane must equal the invariant defined in a plane that is perpendicular to $k_1$ and infinitesimally close to the mirror plane.
 This condition can only be satisfied if the ``stronger" of the two invariants reduces to the ``weaker" one.
In Appendix~\ref{appendixB} we present a complementary derivation of the classification scheme of Table~\ref{table_reflection_off_off} using the 
  Dirac-matrix  Hamiltonian approach.

Le us now discuss in detail for which of the 27 symmetry classes listed in Tables~\ref{reflection_table_full} and~\ref{table_reflection_off_off} there
exist topologically stable Fermi points (point nodes) protected by the combined symmetry $\tilde{T}$ and/or $\tilde{C}$.

\subsubsection{$R_{+}$ and $R_{++}$}

First, we study the situation where the reflection symmetry operator $R$ commutes with all global antiunitary symmetries, which is 
denoted by $R_{+}$ and $R_{++}$ in Table~\ref{table_reflection_off_off}.
Since $[R,T]=0$ and $[R,C]=0$, we have $\tilde{T}^2=T^2$ and $\tilde{C}^2=C^2$, from which it follows that the ten-fold symmetry class
defined in terms of $T$ and $C$ is the same as the one defined in terms of the combined symmetries $\tilde{T}$ and $\tilde{C}$.
Hence, the classification of $R_{+}$ ($R_{++}$) reflection symmetric systems with Fermi points outside the reflection plane is
almost the same as the classification of Fermi points off high-symmetry momenta  in the absence of reflection symmetry (compare Table~\ref{original table} with Table~\ref{table_reflection_off_off} and see Appendix~\ref{appBRpp}). 
The only difference is that the $C \bZ_2$-type invariants of Table~\ref{table_reflection_off_off}, which are defined in terms of the
combined symmetries~\eqref{combineSYMdef}, lead to stable Fermi points outside the reflection plane, whereas
the $\bZ_2$-type invariants of Table~\ref{original table} do not protect Fermi points that are located away from high symmetry momenta (cf.\ Sec.~\ref{FermiSurfOffHighSym}).
We observe that for systems with $R_{+}$ ($R_{++}$) reflection symmetry in Table~\ref{table_reflection_off_off}
the mirror invariants which are defined in the mirror planes for a given eigenspace of $R$ yield the same classification 
as the invariants which are defined in the planes perpendicular to $k_1$ with $k_1 \ne 0, \pi$.

%%%%%%%%%%%%%%%%%%%%%%%%% TABLE
\begin{table*}[t!]
\caption{Classification of  Fermi points and superconducting point nodes of reflection symmetric
semimetals and nodal superconductors, respectively, where the Fermi points 
(point nodes) are located outside the mirror plane
[see Fig.~\ref{FigLocFermiWMirror}(c)].
The first row indicates the spatial dimension $d$ of the semimetal (nodal superconductor).
The prefix ``$C$" indicates that the corresponding topological invariant  is defined in terms of the combined symmetries $\tilde{T}$ and/or $\tilde{C}$ [see Eq.~\eqref{combineSYMdef}] on a $(d-1)$-dimensional plane which is perpendicular to the $k_1$ axis [blue line/plane in Fig.~\ref{FigLocFermiWMirror}(c)].
The $\bZ$- and $\bZ_2$-type invariants, on the other hand, 
are identical to the   ones of the original ten-fold classification in the absence of mirror symmetry (cf.\ Table~\ref{original table}) and are defined on $(d-1)$-dimensional hyperspheres surrounding the Fermi point.}
\label{table_reflection_off_off}
\begin{center}
\begin{threeparttable}
\begin{tabular}{|c|c|cccccccc|}
\toprule
  Reflection &   
       $\begin{array}{c}
\mbox{FS off mirror plane} \\
\mbox{ and off high-sym.\ point }
\end{array}$  
  &  $d$=1 & $d$=2 & $d$=3 & $d$=4 & $d$=5 & $d$=6 & $d$=7 & $d$=8  \\
\hline\hline
 $R$ &  A & $\bZ$ & 0 & $\bZ$ & 0 & $\bZ$ & 0 & $\bZ$ & 0                \\
 $R_+$ &  AIII & 0 & $\bZ$ & 0 & $\bZ$ & 0 & $\bZ$ & 0 & $\bZ$            \\ 
\hline
\multirow{8}{*}{$R_+$,$R_{++}$}  & AI & $\mathbb{Z}$ & 0 & 0 & 0 & $2\mathbb{Z}$ & 0 & $C\bZ_2$ & $C\bZ_2$      \\
 & BDI & $C\bZ_2$ & $\mathbb{Z}$ & 0 & 0 & 0 & $2\mathbb{Z}$ & 0 & $C\bZ_2$         \\
 & D & $C\bZ_2$ & $C\bZ_2$ & $\mathbb{Z}$ & 0 & 0 & 0 & $2\mathbb{Z}$ & 0          \\
 & DIII & 0 & $C\bZ_2$ & $C\bZ_2$ & $\mathbb{Z}$ & 0 & 0 & 0 & $2\mathbb{Z}$       \\
 & AII & $2\mathbb{Z}$ & 0 & $C\bZ_2$ & $C\bZ_2$ & $\mathbb{Z}$ & 0 & 0 & 0       \\
 & CII & 0 & $2\mathbb{Z}$ & 0 & $C\bZ_2$ & $C\bZ_2$ & $\mathbb{Z}$ & 0 & 0       \\
 & C & 0 & 0 & $2\mathbb{Z}$ & 0 & $C\bZ_2$ & $C\bZ_2$ & $\mathbb{Z}$ & 0          \\
&  CI & 0 & 0 & 0 & $2\mathbb{Z}$ & 0 & $C\bZ_2$ & $C\bZ_2$ & $\mathbb{Z}$      
   \\
\hline
\multirow{8}{*}{$R_-$,$R_{--}$} & AI & $2\mathbb{Z}$ & 0 & $C\bZ_2$ & 0 & $2\mathbb{Z}$ & 0 & 0 & 0  \\
  & BDI & 0 & $2\mathbb{Z}$ & 0 & $C\bZ_2$ & 0 & $2\mathbb{Z}$ & 0 & 0        \\
  & D & 0 & 0 & $2\mathbb{Z}$ & 0 & $C\bZ_2$ & 0 & $2\mathbb{Z}$ & 0      \\
  & DIII &  0 & 0 & 0 & $2\mathbb{Z}$ & 0 & $C\bZ_2$ & 0 & $2\mathbb{Z}$       \\
  & AII & $2\mathbb{Z}$ & 0 & 0 & 0 & $2\mathbb{Z}$ & 0 & $C\bZ_2$ & 0   \\
  & CII & 0 & $2\mathbb{Z}$ & 0 & 0 & 0 & $2\mathbb{Z}$ & 0 & $C\bZ_2$       \\
  & C & $C\bZ_2$ & 0 & $2\mathbb{Z}$ & 0 & 0 & 0 & $2\mathbb{Z}$ & 0         \\
  & CI & 0 & $C\bZ_2$ & 0 & $2\mathbb{Z}$ & 0 & 0 & 0 & $2\mathbb{Z}$      
   \\
\hline
$R_{+-}$ & CI & $C\bZ_2$ & 0 & 0 & 0 & 0 & 0 & 0 & $C\bZ_2$ \\
$R_{-+}$ & BDI & 0 & $C\bZ_2$ & $C\bZ_2$ & 0 & 0 & 0 & 0 & 0  \\
$R_{+-}$ & DIII & 0 & 0   & 0 & $C\bZ_2$ & $C\bZ_2$ & 0 & 0 & 0  \\
$R_{-+}$ & CII & 0 & 0 & 0 & 0 & 0 & $C\bZ_2$ & $C\bZ_2$ & 0  \\

\hline
$R_-$ & AIII  & 0 & 0 & 0 & 0 & 0 & 0 & 0 & 0   \\
$R_{-+}$ & DIII,\ CI & 0 & 0 & 0 & 0 & 0 & 0 & 0 & 0   \\
$R_{+-}$ & BDI,\ CII   & 0 & 0 & 0 & 0 & 0 & 0 & 0 & 0   \\
\hline
\hline
\end{tabular}
\end{threeparttable}
\end{center}
\end{table*}
%%%%%%%%%%%%%%%%%%%%%%%%% TABLE

\subsubsection{$R_{-}$ and $R_{--}$}

Second, we study the case where the reflection operator $R$ anticommutes with the nonspatial symmetries $T$ and $C$, which 
is labeled by $R_{-}$ and $R_{--}$ in Table~\ref{table_reflection_off_off}. Here, we find that $\tilde{T}^2=-T^2$ and $\tilde{C}^2=-C^2$
which implies that the symmetry class defined in terms of $\tilde{T}$ and $\tilde{C}$ is shifted by four positions on the ``Bott clock"\cite{stoneMathTheo11} with respect to the symmetry class defined in terms of $T$ and $C$. 
Note that since the ``Bott clock" has periodicity eight,  the direction of the shift is ir\-relevant.
Therefore, the types of invariants that can be defined in ($d-1$)-dimensional planes with fixed $k_1 \ne 0, \pi$ can be inferred from
column $p=d+4$ of the classification of Fermi surfaces that are away from high-symmetry points (Table~\ref{original table}). This, however, is inconsistent with the invariants that can be defined within the mirror planes $k_1 = 0, \pi$. That is, 
 since $[ H (k_1 = 0, \pi ;  \tilde{\bf k} ),  R  ] = 0$ and $[ S= TC, R  ] = 0$, 
it is possible to block-diagonalized $H$ within the mirror plane 
with respect to $R$,
and for each block one can define a Chern number (class BDI, DIII, CII, and CI) or a winding number (class AI, D, AII, and C).
For example, for three-dimensional systems, there are the following invariants that can be defined within the mirror planes  (fixed $k_1=0, \pi$)
or within planes with fixed $k_1 \ne 0, \pi$
\begin{eqnarray}
\begin{array}{c | c c c c c c c c c}
 d=3 &  \textrm{AI}  &  \textrm{BDI}  &  \textrm{D}  &  \textrm{DIII}  &  \textrm{AII}  &  \textrm{CII}  &  \textrm{C}  &  \textrm{CI} \cr
 \hline
\textrm{mirror plane} & \bZ & 0 & \bZ & 0 & \bZ & 0 & \bZ & 0 \cr
\textrm{($k_1 \ne 0, \pi$) - plane} &  \bZ_2 & 0 & \bZ & 0 & 0 & 0 & \bZ & \bZ_2 \cr
\end{array}
\nonumber
\end{eqnarray}
As discussed above, the Fermi points are only protected  by the ``weaker" of these two invariants.\cite{weaker_invariant}
Extending these arguments to other dimensions yields the classification shown in Table~\ref{table_reflection_off_off}.\cite{footnoteVier}
The derivation of this result using the Dirac-matrix Hamiltonian approach is given in Appendix~\ref{R--}.
We observe that the classification for classes with  $\bZ$-type invariants almost agrees with the classification
of Fermi points located away from high-symmetry momenta  in the absence of reflection symmetry (Table~\ref{original table}). 
The only difference is that reflection symmetry requires that the $\bZ$ invariants are even (indicated by ``$2\bZ$" in Table~\ref{table_reflection_off_off}),
whereas in the absence of reflection symmetry the $\bZ$ numbers can also take on odd values.

\subsubsection{DIII \& CI with $R_{+-}$ and BDI \& CII with $R_{-+}$}

Third, we discuss the case where the reflection operator $R$ commutes with one of the global antiunitary symmetries but anticommutes with the other one, i.e.,
class DIII \& CI with $R_{+-}$-type reflection symmetry and class BDI \& CII with $R_{-+}$-type reflection symmetry. 
From the (anti-)commutation relations of $R$ with the nonspatial symmetries we find that the symmetry class defined in terms of $\tilde{T}$ and $\tilde{C}$
(symmetry class for plane with fixed $k_1 \ne 0, \pi$) is
shifted with respect to the symmetry class defined in terms of $T$ and $C$ (symmetry class of entire system) as follows
\begin{subequations} \label{shiftsRmppm}
\begin{eqnarray}
\textrm{DIII} \to \textrm{CII}, \; 
\textrm{CII} \to \textrm{CI} ,  \; 
\textrm{CI} \to \textrm{BDI},  \;
\textrm{BDI} \to \textrm{DIII}.
\end{eqnarray}
On the other hand, since only one global symmetry commutes with the reflection operator $R$,  the symmetry class within the mirror plane is reduced in the following way
\begin{eqnarray}
\textrm{DIII} \to \textrm{AII}, \; 
\textrm{CI} \to \textrm{AI},  \;
\textrm{BDI} \to \textrm{D}, \; 
\textrm{CII} \to \textrm{C}  .
\end{eqnarray}
\end{subequations}
By a similar logic as above, we find by use of Eq.~\eqref{shiftsRmppm} and Table~\ref{original table} that, e.g., for 
three-dimensional systems,  the following invariants can be defined within the mirror planes  (fixed $k_1=0, \pi$)
or within planes with fixed $k_1 \ne 0, \pi$
\begin{eqnarray}
\begin{array}{c | c c c c c c c c c}
 d=3 &  \textrm{DIII}  &  \textrm{CI}  &  \textrm{BDI}  &  \textrm{CII}  \cr
 \hline
\textrm{mirror plane} & \bZ_2 & 0 & \bZ & \bZ  \cr
\textrm{($k_1 \ne 0, \pi$) - plane} & 0 & 0 & \bZ_2 & 0  \cr
\end{array}
\nonumber
\end{eqnarray}
As before we find that only the ``weaker" of these two types of invariants leads to a protection of the Fermi point (cf.\ Appendix~\ref{R-+}).
Extending these arguments to other dimensions gives the classification of Table~\ref{table_reflection_off_off}.

\subsubsection{AIII with $R_-$, DIII \& CI with $R_{-+}$, and  BDI \& CII with $R_{+-}$} 
\label{ClassifyOffOffRmpRpm}

Finally, we consider class AIII with $R_-$-type reflection symmetry, class DIII \& CI with $R_{-+}$-type reflection symmetry, and class BDI \& CII with $R_{+-}$-type reflection symmetry.
Repeating the steps of the previous subsection, we find that for, e.g., three-dimensional systems  the following invariants
can be defined within the mirror plane and within planes with fixed $k_1 \ne 0, \pi$  
\begin{eqnarray}
\begin{array}{c | c c c c c c c c c c}
 d=3 &  \textrm{AIII}  &  \textrm{DIII}  &  \textrm{CI}  &  \textrm{BDI}  &  \textrm{CII}  \cr
 \hline
\textrm{mirror plane} & \bZ  &  \bZ  &  0  &   2\bZ  &  \bZ_2   \cr
\textrm{($k_1 \ne 0, \pi$) - plane} &  0  &  0  &  0  &   0   & \bZ_2   \cr
\end{array} , 
\nonumber
\end{eqnarray}
which suggests that Fermi points in three-dimensional systems with class CII symmetries are protected by a $\bZ_2$-type invariant.
However, this is in contradiction with the result obtained from the Dirac-matrix Hamiltonian approach, 
which shows that all Fermi points have trivial topology (Appendix~\ref{R+-}).
It turns out that even though some nontrivial $\bZ_2$-type invariants can in principle be defined,
these invariants do not protect Fermi points outside the mirror plane.
We conclude that Fermi points outside the mirror plane in 
class AIII with $R_-$-type reflection symmetry,
class DIII \& CI with $R_{-+}$-type reflection symmetry, and class BDI \& CII with $R_{+-}$-type reflection symmetry have trivial topology in all spatial dimensions (Table~\ref{table_reflection_off_off}).

\section{Examples of reflection symmetry protected topological semimetals and nodal superconductors}
\label{sec:examples}

In this section we present several examples of  gapless topological phases protected by reflection symmetry.
As in Sec.~\ref{classReflecGapless} we consider three different types of Fermi surface positions,
which are defined by how the Fermi surface transforms under the
mirror reflection and nonspatial symmetries (see Fig.~\ref{FigLocFermiWMirror}).

\subsection{Fermi surfaces at high-symmetry points within mirror planes}
\label{exampFShighhigh}

We start by discussing four examples of reflection symmetry protected Fermi surfaces (superconducting nodes)
that are left invariant under both reflection and global symmetries. These Fermi surfaces
are located at high symmetry points within the reflection plane, see Fig.~\ref{FigLocFermiWMirror}(a).

\setcounter{paragraph}{0}

\subsubsection{Reflection symmetric nodal spin-triplet superconductor with~TRS (class~DIII with $R_{-+}$ and $p=2$)}
\label{high-symmetry DIII}

As indicated in Table~\ref{reflection_table_full}, 
point nodes ($d_{\rm FS}=0$) in two-dimensional spin-triplet superconductors with TRS and \mbox{$R_{-+}$-type} reflection symmetry
(class DIII with $R_{-+}$) are protected by an $M\bZ\oplus \bZ$ invariant. 
That is, the number of protected point nodes at high symmetry points within the mirror plane is given 
by $\mathrm{max} \left\{ \left| n_{\bZ} \right|,  \left| n_{M \bZ} \right| \right\}$, where $n_{\bZ}$ denotes 
the one-dimensional winding number, whereas $n_{M \bZ}$ is the mirror invariant.
Let us illustrate this type of reflection symmetric nodal superconductor by considering the following
continuum model
\bee \label{symmetry point DIII}
H^{\rm DIII}_{s}=k_x \sigma_x+ k_y \sigma_y. 
\ee 
This superconductor has a point node at ${\bf k}=(0, 0)$ and  is invariant under reflection $k_x \to - k_x$ with  $R= \sigma_y$. 
Time-reversal and particle-hole symmetry operators are given by 
$T=\sigma_y \mK$ and $C=\sigma_x \mK$, respectively.  
Since $\left\{ T, R \right\}  =0$ and $\left[ C, R \right]=0$, Hamiltonian~\eqref{symmetry point DIII} exhibits an $R_{-+}$-type reflection symmetry. 
The global invariant $n_{\bZ}$ of this nodal superconductor is given by the one-dimensional winding number, Eq.~\eqref{DIIIWno}, 
with $q = (  k_x - i k_y ) / \sqrt{ k_x^2 + k_y^2}$ and an integration contour $\mathcal{C}$ that surrounds the point node at ${\bf k}=(0, 0)$.
We find that this winding number evaluates to $n_{\bZ} = +1$.
The mirror number $n_{M\bZ}$, on the other hand, is defined on the mirror line $k_x=0$ for each eigenspace of the mirror 
operator $R$ (i.e., $\sigma_y = \pm 1$).
For Eq.~\eqref{symmetry point DIII} the mirror number is given by the difference of occupied states on either side
of the point node 
\bee \label{DIIIexampMZinv}
n^{\pm}_{M \bZ}  =n^{\pm}_{\mathrm{occ}} (k_y>0)-n^{\pm}_{\mathrm{occ}} (k_y<0)= \mp 1,
\ee 
where $n^{\pm}_{\mathrm{occ}} (k_y)$ denotes the number of occupied states (i.e., the number of negative energy states) at ${\bf k} = ( 0, k_y)$ 
in the eigenspace of $R$ with eignevalue $\pm 1$. 
Hence, the nodal point at ${\bf k} = (0,0)$ is protected by both the winding number $n_{\bZ}$ and the mirror number $n^{\pm}_{M \bZ} $.
 
It is important to note, however, that gapless systems with $M\bZ \oplus \bZ$-type invariants are \emph{not} protected by the sum of 
 the $\bZ$ and $M\bZ$ invariants; rather the number of point nodes (gapless modes) is given by
$\mathrm{max} \left\{ \left| n_{\bZ} \right|,  \left| n_{M \bZ} \right| \right\}$. To exemplify this further we
consider two doubled versions of Hamiltonian~\eqref{symmetry point DIII}
\begin{subequations} \label{DIIIexmpVAdd}
\bee  \label{DIIIexmpVAddA}
H^{\rm DIII}_{s,1} =  k_x \sigma_x\otimes \sigma_z  + k_y \sigma_y\otimes \sigma_z 
\ee
and
\bee  \label{DIIIexmpVAddB}
H^{\rm DIII}_{s,2}= k_x \sigma_x \otimes \sigma_z + k_y \sigma_y \otimes \bI ,
\ee
\end{subequations}
which have the same symmetry properties as Eq.~\eqref{symmetry point DIII}
with $T=\sigma_y \otimes \bI \mathcal{K}$, $C=\sigma_x \otimes \bI \mathcal{K}$, and $R=\sigma_y \otimes \bI$.
Eqs.~\eqref{DIIIexmpVAddA} and~\eqref{DIIIexmpVAddB}
have different topological characteristics: While the topology of $H^{\rm DIII}_{s,1}$ is given by 
$n_{\bZ}=2$ and $n^{\pm}_{M \bZ} =0$, for $H^{\rm DIII}_{s,2}$ we find that $n_{\bZ}=0$ and $n^{\pm}_{M \bZ} = \mp 2$. Hence, both Hamiltonians in Eq.~\eqref{DIIIexmpVAdd} exhibit
two stable gapless modes
at ${\bf k}=0$. We now form a direct product between $H^{\rm DIII}_{s,1}$ and $H^{\rm DIII}_{s,2}$, which
yields an $8\times8$ Hamiltonian, $H^{\rm DIII}_{s,3} = \mathrm{diag} ( H^{\rm DIII}_{s,1}, H^{\rm DIII}_{s,2} )$, with 
four gapless modes. However, only two of these four modes are topologically stable, since it is possible to gap
out two states by the symmetry preserving mass term 
\begin{align}
\left(\begin{array}{cccc}
 0 & 0 & 0 & 0 \\
 0 & 0 & 0 & i m \sigma_y \\
  0 & 0 & 0 & 0 \\
   0 & - i m \sigma_y & 0 & 0 \\
 \end{array}\right) .
\end{align}
Thus, in accordance with the formula $\mathrm{max} \left\{ \left| n_{\bZ} \right|,  \left| n_{M \bZ} \right| \right\} =2$,
$H^{\rm DIII}_{s,3}$  exhibits only two stable gapless modes at ${\bf k}=0$.

In closing, we observe that by including an extra momentum-space coordinate we can convert 
Hamiltonian~\eqref{symmetry point DIII} to a three-dimensional reflection symmetric
superconductor with a protected line node ($d_{\mathrm{FS}} = 1$) located at ${\bf k}=(0,0,k_z)$.
The stability of this nodal line is guaranteed by  the quantized winding number $n_{\bZ}$, Eq.~\eqref{DIIIWno},
and the mirror invariant $n_{M \bZ}$, Eq.~\eqref{DIIIexampMZinv}.

\subsubsection{Reflection symmetric Dirac semimetal with TRS (class AII with~$R_{+}$ and $p=3$)}\label{high-symmetry AII}

 Next, we study a reflection symmetric three-dimensional Dirac semimetal with TRS, which is described by
\bee  \label{HAII}
H^{\rm AII}_{s}=k_x \sigma_x\otimes \sigma_z+ k_y \sigma_y\otimes \bI + k_z \sigma_z \otimes \bI . 
\ee
Time-reversal and reflection symmetry operators are given by $T=\sigma_y\otimes \bI \mK $ and $R =\bI\otimes \sigma_x$, respectively. 
Because $T^2 = - \mathds{1}$ and $[ T,  R ] =0$, Hamiltonian~\eqref{HAII} belongs to symmetry class AII with $R_+$. 
The semimetal of Eq.~\eqref{HAII} has a Dirac point at ${\bf k}=(0,0,0)$ which is topologically stable, since there exists no SPGT that can be added to the Hamiltonian.
Indeed, according to Table~\ref{reflection_table_full}, this Fermi point is protected by an $M\bZ_2$-type topological invariant,
which is defined on the mirror line $k_x = 0$ for each eigenspace of the reflection operator $R$. Focusing on the eigenspace $R=+1$, we
find that $H^{\rm AII}_{s}$ in this subspace on the mirror line is given by
\bee \label{projectExampHAII}
h_{R=+1}^{\rm AII}=k_y \sigma_y + k_z \sigma_z .
\ee
The  $M\bZ_2$ invariant is defined in terms of an extension of Eq.~\eqref{projectExampHAII} to three dimensions
[cf.\ Eq.~\eqref{extendHamAA}]
\bee
\widetilde{h}_{R=+1}^{\rm AII}
=
\left( k_y \sigma_y + k_z \sigma_z  \right) \cos \theta
+
\Delta\sigma_x\sin \theta   ,
\ee
where $\Delta$ is a positive constant and
 $\theta \in \left[ 0, \pi \right]$ is the parameter for the extension in the third dimension. 
With this, we find that the stability of the single Dirac point at ${\bf k} = (0,0,0)$  is guaranteed 
by the invariant~\eqref{Z2 extension} 
with ${\bf g} = (\Delta \sin \theta, k\cos \phi \cos \theta, k \sin \phi \cos \theta)$, 
which evaluates to $n_{M\bZ_2}=1$.
However, as indicated by the $M\bZ_2$-type invariant, 
a doubled version of this Dirac point is unstable.
This can be seen by considering two copies of Hamiltonian \eqref{HAII}, i.e., $H^{\rm AII}_{s} \otimes \bI$.
The doubled Dirac point of this $8\times8$ Hamiltonian can be gapped out by the momentum-independent SPGT 
$\sigma_x \otimes \sigma_x \otimes \sigma_y$,
which is in agreement with the value of the topological number $n_{M\bZ_2}=0$ for $H^{\rm AII}_{s} \otimes \bI$.

 $M\bZ_2$-type invariants only protect Fermi surfaces of dimension zero ($d_{\mathrm{FS}} = 0$) at high-symmetry points of the BZ.
 To illustrate this, we consider an extension of Hamiltonian~\eqref{HAII} to four spatial dimensions with a Fermi line along
 the fourth momentum direction $k_w$.
 This Fermi line, which is located at $(0,0,0,k_w)$, can be gapped out by the symmetry-preserving kinetic term $k_w\six\otimes \six$. Only the Fermi point at $(0,0,0,0)$
 remains gapless; it is  protected by the non-zero $M \bZ_2$ invariant which is  well-defined only for $k_w=0$.

\subsubsection{Nodal spin-singlet superconductor with TRS and $R_{+-}$-type reflection symmetry (class CII with $R_{+-}$ and $p=2$)} \label{CII R+-}

Let us now discuss  an example of a nodal superconductor with an
$M\bZ_2 \oplus \bZ_2$-type index. According to Table~\ref{reflection_table_full}, point nodes of time-reversal invariant spin-singlet superconductors with
an $R_{+-}$-type reflection symmetry are protected by an $M\bZ_2 \oplus \bZ_2$ topological invariant. A simple example of such a reflection
symmetric topological superconductor is provided by the $4 \times 4$ Hamiltonian
\bee \label{exampCIIsymRpm}
H_{\rm s}^{\rm CII}
=
k_x \siy \otimes \bI+ k_y \six \otimes \bI,
\ee
which preserves time-reversal and particle-hole symmetry with 
$T=\siy \otimes \bI \mK$ and $C=\six \otimes \siy \mK$, respectively. 
$H_{\rm s}^{\rm CII}$ is invariant under reflection $k_x \to - k_x$ with $R =\six \otimes \siy$.
Since $T^2= - \mathds{1} $, $C^2= - \mathds{1}$, $\left[ T, R \right]= 0$, and $\left\{ C, R \right\} = 0$, 
Hamiltonian~\eqref{exampCIIsymRpm} belongs to symmetry class CII with $R_{+-}$. 
The two-dimensional superconductor~\eqref{exampCIIsymRpm} exhibits a point node 
at ${\bf k}= (0,0)$ whose stability is guaranteed by a $M\bZ_2\oplus\bZ_2$ topological index.
To demonstrate this, we compute both the 
global invariant $n_{\bZ_2}$
and the mirror invariant $n_{M\bZ_2}$. 
From Table~\ref{reflection_table_full}
we find that the global invariant $n_{\bZ_2}$ in column $p=2$
is a second descendant of a $\bZ$-type invariant in column $p=4$. Hence, 
the  topological number $n_{\bZ_2}$ can be defined in terms of an extension of $H_{\rm s}^{\rm CII}$
to four dimensions\cite{ZhaoWangPRL13,ZhaoWangPRB14} 
\begin{eqnarray} \label{exampCIIsymRpmExtd}
\widetilde{H}_{\rm s}^{\rm CII}
&=&
\left[ k_x \siy \otimes \bI+ k_y \six \otimes \bI \right] \sin \theta  \sin \psi
\nonumber\\
&& \qquad 
+  \siz\otimes \siz \sin \theta \cos \psi 
+  \siz \otimes \six  \cos \theta, 
\end{eqnarray}
where $\psi,\theta \in [0,\pi]$ are the parameters for the extension to four dimensions. 
Just as Eq.~\eqref{exampCIIsymRpm}, Hamiltonian~\eqref{exampCIIsymRpmExtd} satisfies both
time-reversal and particle-hole symmetry with
\begin{subequations}
\begin{eqnarray}
T^{-1}\tilde{H}^{\rm{CII}}_{\rm{s}}(- {\bf k},  \pi -\psi, \pi -\theta)T=&\tilde{H}^{\rm{CII}}_{\rm{s}}( {\bf k}, \psi, \theta), 
\end{eqnarray}
and
\begin{eqnarray}
C^{-1}\tilde{H}^{\rm{CII}}_{\rm{s}}(- {\bf k}, \pi -\psi, \pi -\theta)C=&-\tilde{H}^{\rm{CII}}_{\rm{s}}( {\bf k}, \psi, \theta),
\end{eqnarray}
\end{subequations}
respectively. 
We note that for the definition of the \emph{global} invariant $n_{\bZ_2}$ we do not need to consider the
restrictions imposed by reflection symmetry.
Using the extension~\eqref{exampCIIsymRpmExtd}, the  $n_{\bZ_2}$ invariant is expressed as
\begin{eqnarray} \label{defNZ2expCII}
n_{\bZ_2}
=
\frac{1}{48 \pi^2}\oint_{\mathcal{C}} \tr{ \Big[S \left( \widetilde{H}_{\rm s}^{\rm CII}  {\bf d}  [ \widetilde{H}_{\rm s}^{\rm CII}  ]^{-1} \right)^3 \Big]} \text{ mod }2 , 
\end{eqnarray}
with the chiral symmetry operator $S=\siz \otimes \siy$ and $\mathcal{C}$ a three-dimensional contour which encloses the point node and which is mapped onto itself by
both TRS and PHS [see Fig.~\ref{FSwithGlobalSym}(a)].
Choosing $\mathcal{C}$ to be the unit three-sphere $S^3$, we parametrize the momenta as 
$k_x= \cos \phi$ and $k_y= \sin \phi$, which yields
\begin{eqnarray}
&&
n_{\bZ_2}
=
 \frac{1}{8\pi^2}\int^{2\pi}_0 d\phi \int^\pi_0 d\psi \int^\pi_0 d\theta \, \tr \Big[S \left( \widetilde{H}_{\rm s}^{\rm CII} \partial_\phi  [ \widetilde{H}_{\rm s}^{\rm CII} ]^{-1} \right)
\nonumber \\
 && \quad
\times \left( \widetilde{H}_{\rm s}^{\rm CII} \partial_\psi  [ \widetilde{H}_{\rm s}^{\rm CII} ]^{-1} \right)
 \left( \widetilde{H}_{\rm s}^{\rm CII} \partial_\theta  [ \widetilde{H}_{\rm s}^{\rm CII} ]^{-1} \right) \Big]
  \text{ mod }2 = 1,
  \nonumber\\  
\end{eqnarray}
indicating that the point node at ${\bf k}=(0,0)$ is protected by the nontrivial value
of $n_{\bZ_2}$.

As opposed to the global invariant $n_{\bZ_2}$, the mirror invariant $n_{M\bZ_2}$ is defined in the 
reflection plane $k_x=0$ for a given eigenspace of the reflection operator $R$. Focusing on the eigenspace
$R=+1$, we find that the extended Hamiltonian~\eqref{exampCIIsymRpmExtd} in this eigenspace within the mirror plane is given by
\begin{equation} \label{exampCIIsymRpmBlock}
\widetilde{h}_{R=+1}^{\rm CII}
= 
k_y \sigma_y  \sin \psi  \sin \theta 
 - \siz  \cos \psi \sin \theta +  \six  \cos \theta , 
\end{equation}
where $\psi \in [0,  \pi]$ and $\theta \in [0, \pi]$.
Hamiltonian~\eqref{exampCIIsymRpmBlock} is invariant under TRS
\bee
T^{-1}_R \widetilde{h}_{R=+1}^{\rm CII} (- {\bf k} ,\pi-\psi,\pi-\theta)T_R =  \widetilde{h}_{R=+1}^{\rm CII}  ( {\bf k} , \psi,\theta),
\ee
with $T_R= i \sigma_y \mathcal{K}$. 
The mirror invariant $n_{M\bZ_2}$  is of the same form as Eq.~\eqref{Z2 extension}
with an integration contour that preserves TRS, that lies within the mirror plane, and that surrounds the nodal point [see Fig.~\ref{FigLocFermiWMirror}(a)].
As the integration contour we choose a two-sphere $S^2$ which intersects the $(k_x, k_y)$-plane at ${\bf k }=(0, \pm a)$, such that the
Fermi point at ${\bf k} = (0,0)$ on the mirror line is enclosed by $k_y = \pm a$, see Fig.~\ref{FigLocFermiWMirror}(a).
That is, to perform the contour integration $k_y=0$ in $\widetilde{h}_{R=+1}^{\rm CII}$ is replaced by $a$ and $\psi$ is integrated over the interval $[0,2\pi]$,
whereas $\theta$ is integrated over $[0, \pi]$.
With this integration contour we find that $n_{M\bZ_2}$  is given by Eq.~\eqref{Z2 extension}
with ${\bf g}=( \cos \theta, a \sin \psi \sin \theta, - \cos \psi \sin \theta)$, which evaluates to $n_{M\bZ_2}=1$. Hence,
the point node at ${\bf k}=(0,0)$ is protected also by the mirror invariant  $n_{M\bZ_2}$.

As indicated in Table~\ref{reflection_table_full}, $M\bZ_2 \oplus \bZ_2$-type indices only protect Fermi surfaces (superconducting nodes) of
dimension zero, i.e., $d_{\mathrm{FS}} =  0$. To exemplify this, we consider a trivial extension of Hamiltonian~\eqref{exampCIIsymRpm} to three spatial dimensions by
including the extra momentum component $k_z$, which yields a three-dimensional superconductor with a line node at  $(0,0,k_z)$. However, 
this line node is unstable, since it can be gapped out by the symmetry-preserving kinetic term $k_z\siz \otimes \six$. Only the point node
at ${\bf k} = (0,0,0)$ is topologically stable.  Moreover, we find that  the global invariant $n_{\bZ_2}$, Eq.~\eqref{defNZ2expCII}, as well as the mirror invariant $n_{M \bZ_2}$
cannot be defined for the three-dimensional superconductor with a line node along the $k_z$ direction, since it is impossible to choose a time-reversal invariant
integration contour that surrounds this nodal line (except for $k_z=0$ and $k_z = \pi$).

\subsubsection{Reflection symmetric nodal spin-singlet superconductor (\mbox{class C with $R_{-}$ and $p=2$})}

As a fourth example we consider a two-dimensional nodal spin-singlet superconductor with reflection symmetry,
which is described by the $4 \times 4$ Hamiltonian
\bee \label{hamDefExpCmmm}
H_{\rm s}^{\rm C}=k_x \sigma_x \otimes \sigma_y + k_y \sigma_y \otimes \sigma_y.
\ee
Eq.~\eqref{hamDefExpCmmm} satisfies PHS with $C=\sigma_y \otimes \bI \mK$ and is invariant under
reflection $k_x \to - k_x$ with $R=\sigma_y \otimes \bI$.
Because $C^2=- \dI$ and $\{ C, R \} = 0$, Hamiltonian~\eqref{hamDefExpCmmm} belongs to symmetry class C with an $R_-$-type reflection symmetry.
This superconductor has a point node at ${\bf k}=(0,0)$, which, according to Table~\ref{reflection_table_full}, is protected by a $T\bZ_2$ invariant.
Indeed, there exists no SPGT that can gap out this point node. 
To demonstrate the $\bZ_2$-type property of Eq.~\eqref{hamDefExpCmmm},
we consider different doubled versions of the Hamiltonian.  
Using $H_{\rm s}^{\rm C}$, there are four possibilities to construct an $8 \times 8$ Hamiltonian in the symmetry class C with $R_-$\cite{chiuPRB13}
\begin{subequations}
\begin{eqnarray}
H^{\rm C}_{++}
&=&
H_{\rm s}^{\rm C}\otimes \bI,
\qquad
H^{\rm C}_{--}
=
H_{\rm s}^{\rm C}\otimes \sigma_z,
\\
H^{\rm C}_{-+}
&=&
k_x \sigma_x \otimes \sigma_y\otimes \bI + k_y \sigma_y \otimes \sigma_y \otimes \sigma_z,
\end{eqnarray}
and
\begin{eqnarray}
H^{\rm C}_{+-}
&=&
k_x \sigma_x \otimes \sigma_y\otimes \sigma_z + k_y \sigma_y \otimes \sigma_y \otimes \bI .
\end{eqnarray}
\end{subequations}
We find that  the first three Hamiltonians can be fully gapped out by the momentum-independent SPGTs $\bI\otimes \sigma_z \otimes \sigma_y$, $\bI\otimes\bI \otimes \sigma_y$, and $\sigma_y\otimes \sigma_y \otimes \sigma_y$, respectively. 
Interestingly, the fourth Hamiltonian $H^{\rm C}_{+-}$ has a stable point node at ${\bf k}=0$, i.e., there exists no SPGT for $H^{\rm C}_{+-}$.
However, if we consider quadrupled versions of $H_{\rm s}^{\rm C}$, Eq.~\eqref{hamDefExpCmmm}, we find that for
each quadrupled Hamiltonian there exists at least one SPGT which gaps out all the point nodes.  
(In a sense, the Hamiltonian has a $\bZ_4$-property rather than a $\bZ_2$-property.)

\subsection{Fermi surfaces within mirror planes but off high-symmetry points}
	
Second, we present some examples of Fermi surfaces (superconducting nodes) that
are left invariant by the mirror symmetry but transform pairwise into each other under the global symmetries. 
These Fermi surfaces are located within the mirror plane but away from the time-reversal invariant momenta, see Fig.~\ref{FigLocFermiWMirror}(b).

\subsubsection{Reflection symmetric Dirac semimetal with TRS (\mbox{class AII with $R_+$ and $p=2$})}

We begin by considering the  following two-orbital tight-binding Hamiltonian
$\mathcal{H}^{\rm AII}_{\rm n} = \sum_{\bf k} \Psi^{\dag}_{\bf k}  h^{\rm AII}_{\rm n} ( {\bf k} ) \Psi^{\ }_{\bf k}$,
with the spinor $\Psi_{\bf k}  = [ \psi_{\uparrow  1} ({\bf k}),  \psi_{\uparrow  2}  ({\bf k}),      \psi_{\downarrow  1} ({\bf k}) ,  \psi_{\downarrow  2}  ({\bf k}) ]^{\mathrm{T}}$ 
and
\begin{eqnarray} \label{AIImodelA}
h^{\rm AII}_{\rm n} ( {\bf k} )
&=& 
t_x \sin k_x \, \sigma_z \otimes \tau_x + [1 - t_y \cos k_y]  \sigma_0 \otimes \tau_z , \;  \; 
\end{eqnarray}
where  $\sigma_i$ operates in  spin grading and $\tau_i$  in  orbital grading.\cite{footnoteABC}
This Hamiltonian satisfies TRS,
with $T= \sigma_y \otimes \tau_0 \mathcal{K} $, and reflection symmetry $k_x \to - k_x$, with $R= \sigma_0 \otimes \tau_z $.
Because $T^2 = - {\dI}$ and $[ R, T]=0$, semimetal~\eqref{AIImodelA} belongs to symmetry class AII with $R_+$.
The spectrum of the Hamiltonian is given by
\begin{eqnarray}
E 
=
\pm \sqrt{ t_x^2  \sin^2 k_x  + (1 - t_y \cos k_y)^2 } .
\end{eqnarray}
For $t_y > 1$ Hamiltonian~\eqref{AIImodelA} has four Dirac points at $(k_x, k_y ) = ( 0, \pm \arccos [ 1 / t_y ])$ and
$( \pi,  \pm \arccos [ 1 / t_y ])$,
for $t_y=1$ there are two Dirac points at $(k_x, k_y )=(0,0)$ and  $(\pi, 0)$,
and for  $t_y < 1$ there is a full gap in the BZ.  The reflection symmetry $R$ maps each Dirac point onto itself, i.e., the Fermi points
are located within the mirror lines $k_x = 0$ and $k_x = \pi$, see Fig.~\ref{FigLocFermiWMirror}(b).
Since there does not exist any SPGT that can be added to Eq.~\eqref{AIImodelA}, the four Dirac points of  Hamiltonian~\eqref{AIImodelA} with $t_y > 1$ are topologically stable and protected
against gap opening by TRS and reflection symmetry. This is in agreement with the classification of Table~\ref{reflection_table_full} (column $p=2$), 
which shows that the Fermi points are protected by a mirror invariant of type $2 M \bZ$, where 
the prefix ``$2$" indicates that the mirror invariant only takes on even values.
To exemplify this for semimetal~\eqref{AIImodelA}, we evaluate the mirror number $n_{2 M \bZ}$ 
for the reflection line $k_x = 0$.
We find that $h^{\rm AII}_{\rm n} $ in the eigenspace $R = \pm 1$ for $k_x = 0$ reads
\begin{eqnarray}
h^{\rm AII}_{R = \pm 1} =
\pm (1-t_y \cos k_y)\bI .
\end{eqnarray}
The mirror index $n^{\pm}_{2 M \bZ}$  for the eigenspace $R=\pm 1$ is given by the difference of occupied states (i.e., states with $E < 0$)
of Hamiltonian $h^{\rm AII}_{R = \pm 1}$
on either side of the Dirac point, i.e.,  
\begin{eqnarray} \label{AIImodelAINV}
n^{\pm}_{2 M \bZ}
=
 n^{\pm}_{\rm occ} ( | k_y | < k_0 ) - n^{\pm}_{\rm occ} ( | k_y | > k_0 ) =  \pm 2,
\end{eqnarray}
where $k_0 =  \arccos [ 1 / t_y ] $ and  
\bee 
n^{+}_{\rm occ}Ê(k_y)=\left\{ 
  \begin{array}{l l}
    2,& \, \left| k_y \right| < k_0 \\
    0,&   \,  \left| k_y \right| > k_0
  \end{array} \right. ,
  \;
  n^{-}_{\rm occ}Ê(k_y)=\left\{ 
    \begin{array}{l l}
    0,&  \, \left| k_y \right| < k_0 \\
    2,&   \,  \left| k_y \right| > k_0
  \end{array} \right.
\ee	 
denotes the number of occupied states  at ${\bf k} = ( 0, k_y)$ in the eigenspace or $R$
with eigenvalue $+1$ and $-1$, respectively. Hence, the two 
Dirac points at $(0,\pm k_0)$ are protected by the invariant~\eqref{AIImodelAINV}.
The index $n_{2 M \bZ}$ for the $k_x = \pi$ line, which guarantees the stability
of the Fermi points at  $(\pi,\pm k_0)$,   can be computed in a similar fashion.

\subsubsection{Reflection symmetric tight-binding model on the honeycomb lattice (class AI with $R_+$ and $p=2$)}
\label{secGraphene}

 As a second example we discuss a tight-binding model of spinless fermions on the honeycomb lattice, which describes
the electronic properties of graphene\cite{castroNetoRMP09} (ignoring any spin-dependent terms).
Considering both first- and second-neighbor hopping the tight-binding Hamiltonian can be written as
$\mathcal{H}^{\rm AI}_{\rm n}  = 
\sum_{\bf k} 
\Psi^{\dag}_{\bf k} h^{\rm AI}_{\rm n} ( {\bf k} ) \Psi^{\ }_{\bf k}$
with the spinor $\Psi^{\ }_{\bf k} = \left( a^{\ }_{\bf k},
b^{\ }_{\bf k} \right)^{\mathrm{T}}$ and
\begin{eqnarray} \label{hamGraphene}
h^{\rm AI}_{\rm n} ( {\bf k} )
=
\begin{pmatrix}
\Theta_{\bf k} & \Phi_{\bf k} \cr
\Phi^{\ast}_{\bf k} & \Theta_{\bf k} \cr
\end{pmatrix} ,
\end{eqnarray}
where $a_{\bf k}$ and $b_{\bf k}$ denote the fermion annihilation operators with momentum ${\bf k}$ on sublattice $A$ and $B$, respectively. The 
hopping terms are given by $\Phi_{\bf k} = t_1 \sum_{i=1}^3 e^{+ i {\bf k} \cdot {\bf s}_i }$ and
$\Theta_{\bf k} = t_2 \sum_{i=1}^6 e^{+ i {\bf k} \cdot {\bf d}_i }$, where ${\bf s}_i$ and 
${\bf d}_i$ denote the nearest- and second-neighbor bond vectors, respectively [Fig.~\ref{Honeycomb}(a)].
The hopping integrals  $t_1$ and $t_2$ are assumed to be positive. 

%%%%%%%%%%% begin Figure
\begin{figure}[t]
 \begin{center} 
\includegraphics[width=0.48\textwidth,angle=-0]{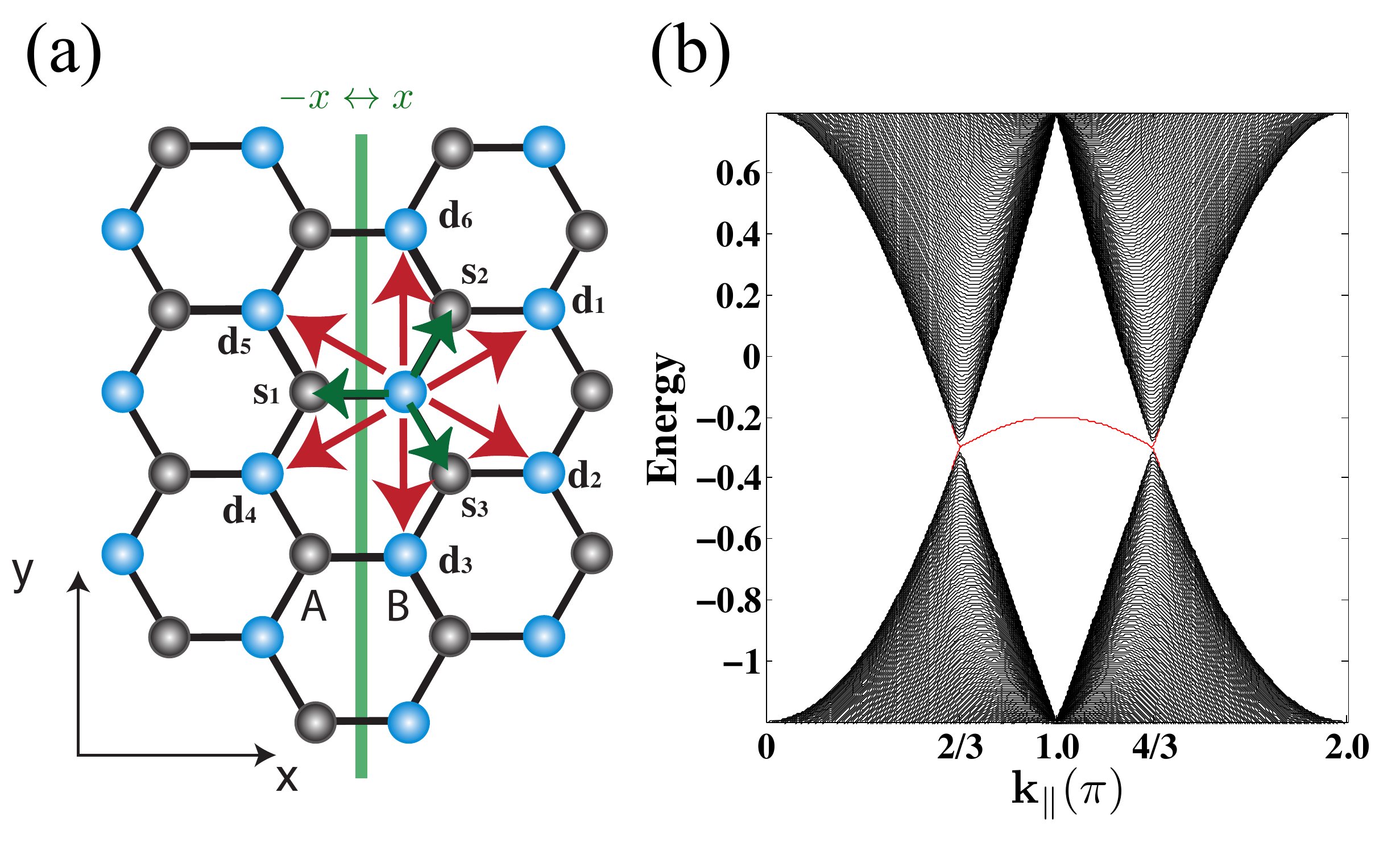}
\end{center}
\caption{
(a) 
The honeycomb lattice of graphene is a bipartite lattice composed of two interpenetrating triangular sublattices. 
The two sublattices are marked ``A" (black dots) and ``B" (blue dots). 
The nearest-neighbor bond vectors (green arrows) are given by ${\bf s}_1=(-1,0)$, ${\bf s}_2=\frac{1}{2}(1,\sqrt{3})$, and ${\bf s}_3=\frac{1}{2}(1, -\sqrt{3})$.
The second-neighbor bond vectors (red arrows) are 
${\bf d_1}=-{\bf d_4}=\frac{1}{2}(3,\sqrt{3})$, ${\bf d_2}=-{\bf d_5}=\frac{1}{2}(3,-\sqrt{3})$, and ${\bf d_3}=-{\bf d_6}=(0,-\sqrt{3})$. 
The mirror line $x \to - x$ is indicated by the green line.
 (b) Energy spectrum of a graphene ribbon with (10) edges (i.e., zigzag edges) and $(t_1, t_2)=(1.0, 0.1)$.
A linearly dispersing edge state (red trace) connects the Dirac points, which are located at $k_\parallel=2\pi/3$ and $k_\parallel = 4\pi/3$ in the edge BZ and are projected from the bulk Dirac points at $(0,\pm k_0)$. 
}
 \label{Honeycomb}
\end{figure}
%%%%%%%%%% end Figure 

Hamiltonian~\eqref{hamGraphene} satisfies TRS with $T= \sigma_0 \mathcal{K}$
and is invariant under the mirror symmetry $k_x \to - k_x$ with $R = \sigma_x$.
(Incidentally, Eq.~\eqref{hamGraphene} is also symmetric under $k_y \to - k_y$. However, we shall ignore this symmetry, since it does
not play any  role for the protection of the Dirac points.)
Because $T^2=+\dI$ and $[R, T] = 0$ we find that Hamiltonian~\eqref{hamGraphene} belongs to symmetry class AI with $R_+$.
The energy spectrum 
\begin{eqnarray}
&&
E_{\bf k}^{\pm}
=
+ 2 t_2  \left[ 2 \cos \left( \frac{ 3 k_x }{2}  \right) \cos \left( \frac{ \sqrt{3} k_y }{2}  \right) +  \cos ( \sqrt{3} k_y ) \right]
\nonumber\\
&&
\;
\pm t_1 \left[  3 + 4 \cos \left( \frac{ 3 k_x }{2}  \right) \cos \left( \frac{ \sqrt{3} k_y }{2}  \right) +  2 \cos ( \sqrt{3} k_y ) \right]^{\frac{1}{2}}
\; \;
\end{eqnarray} 
exhibits two Dirac points, which are located on the mirror line $k_x =0$, i.e., at $(k_x, k_y) = (0, \pm k_0 )$ in the BZ, with $k_0=4 \pi /( 3 \sqrt{3}  )$.
These two Dirac points transform pairwise into each other under TRS.
Because there does not exist any SPGT that can be added to Eq.~\eqref{hamGraphene}, we find that the Dirac points are topologically stable
and protected against gap opening by TRS, reflection symmetry, and $SU(2)$ spin-rotation symmetry. In particular, we note
that the TRS preserving mass term $\sigma_3$ is forbidden by reflection symmetry $R$.
This finding is confirmed by the classification of Table~\ref{reflection_table_full}, which indicates that the stability of the Dirac points 
is guaranteed by  an $M \bZ$-type invariant. 

To compute this mirror invariant $n_{M \bZ}$ we
determine the eigenstates $\psi^{\pm}_{\bf k}$ of $h^{\rm AI}_{\rm n} (  {\bf k} )$ with energy $E^{\pm}_{\bf k}$
\begin{eqnarray}
\psi^{-}_{\bf k}
=
\frac{1}{\sqrt{2}} 
\begin{pmatrix}
- e^{i \varphi_{\bf k}} \cr
1 \cr
\end{pmatrix},
\qquad
\psi^{+}_{\bf k}
=
\frac{1}{\sqrt{2}} 
\begin{pmatrix}
e^{i \varphi_{\bf k}} \cr
1 \cr
\end{pmatrix},
\end{eqnarray}
where $\varphi_{\bf k} = \arg [ \Phi_{\bf k}] $.
On the mirror line $k_x = 0$ we have
\begin{eqnarray} \label{phaseFacGraph}
e^{i \varphi_{(0, k_y) }}  
=
  \left\{  \begin{array}{l l}
    +1 ,& \, \left| k_y \right| < k_0 \\
    - 1,&   \,  \left| k_y \right| >  k_0
  \end{array} \right. .
\end{eqnarray}
Hence, $ \psi^{\pm}_{(0, k_y ) }$ are simultaneous eigenstates of the reflection operator $R = \sigma_x$ with opposite eigenvalue ($+1$ or $-1$),
which prohibits the hybridization between them. 
The mirror invariant $n^{\pm}_{M \bZ}$ is given by the difference of the number of states with energy $E^-_{\bf k}$ and
reflection eigenvalue $R = \pm 1$ on either side of the Dirac point, i.e.,
\begin{eqnarray} \label{mirroInvGraph}
n^{\pm}_{M \bZ} 
=
 n^{\pm}_{\rm neg} ( | k_y | > k_0) - n^{\pm}_{\rm neg} ( | k_y | < k_0),
\end{eqnarray}
where $n^{\pm}_{\rm neg} ( k_y) $ denotes the number of states with energy $E^-_{\bf k}$ and reflection eigenvalue $R = \pm 1$.
Using Eq.~\eqref{phaseFacGraph}Ê we find that $n^{\pm}_{M \bZ} = \pm 1$, and hence the Dirac points are 
protected by the mirror invariant~\eqref{mirroInvGraph}.
By the bulk-boundary correspondence, the nontrivial topology of the Dirac points leads to a linearly dispersing edge mode, which connects the projected Dirac points in the (10) edge BZ, see Fig.~\ref{Honeycomb}(b).

%%%%%%%%%%% begin Figure
\begin{figure*}[t]
 \begin{center} 
\includegraphics[clip,width=1.95\columnwidth]{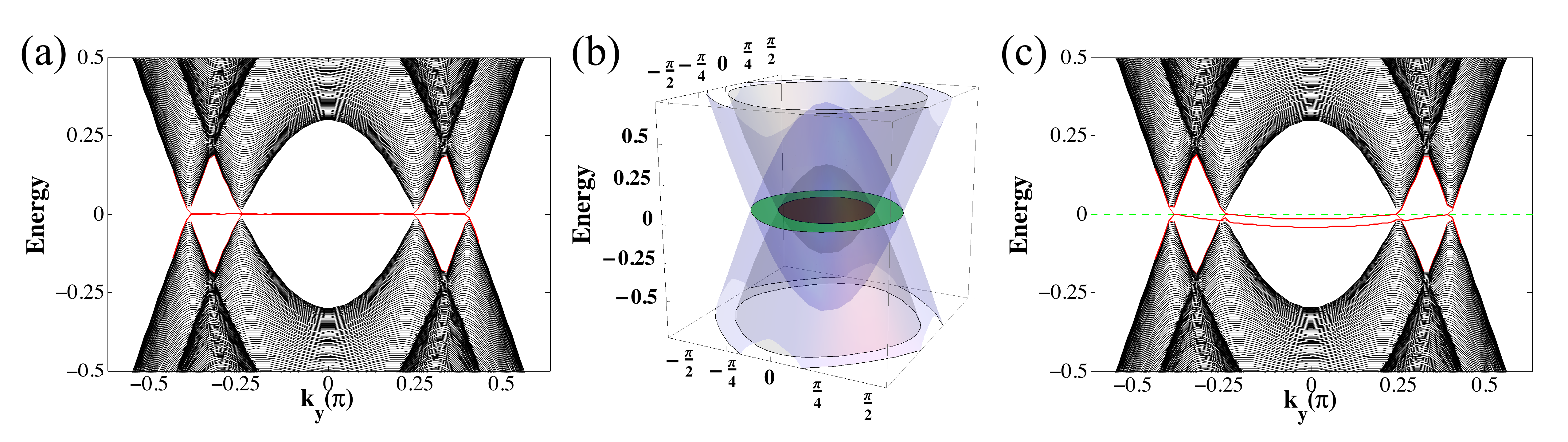}
 \end{center}
 \caption{(Color online) {
(a) Surface band structure of semimetal (\ref{Fermi rings}) for the (100) face with $\mu_s=0$, $k_z=0$, $m_1=2.5$, and $m_2=0.2$
 as a function of surface momentum $k_y$. Note that the (100) surface is not symmetric under $k_x \to - k_x$.
 Zero-energy surface flat bands (red traces) appear within regions of the surface BZ that are bounded by the projected bulk Fermi lines.
 (b) Surface spectrum on the (100) face as a function of both $k_y$ and $k_z$ for the same parameters as in panel (a). 
Nondegenerate zero-energy flat bands protected by the winding number $n_{\bZ} = 1$ [see Eq.~\eqref{windinNoExmpAR}]
appear within the region  $1.3 < \cos k_y+ \cos k_z < 1.7$ of the surface BZ (green area).
Doubly degenerate flat bands  protected by $n_{\bZ}=2$ exist within the region $\cos k_y + \cos k_z >1.7$ (brown area).
(c) Surface spectrum in the presence of a  staggered chemical potential~\eqref{stagCemPot} with $\mu_s = 0.05$.  
Linearly dispersing surface states (red traces) connect the projected Fermi rings in the surface BZ. }
 }
 \label{Ring figures}
\end{figure*}
%%%%%%%%%% end Figure

\subsubsection{Reflection symmetric semimetal with Fermi rings \mbox{(class A with $R$ and $p=3$)}}

To exemplify that $M\bZ$-type invariants can give rise to topologically stable Fermi surfaces with $d_{\mathrm{FS}} > 0$,
we consider the following three-dimensional semimetal 
on the square lattice
$\mathcal{H}^{\rm A}_{\rm n} = \sum_{\bf k} \Psi^{\dag}_{\bf k} h_{\rm n}^{\rm A}({\bf k}) \Psi^{\ }_{\bf k}$,
with the spinor $ \Psi^{\ }_{\bf k} = \left[ c_1 ( {\bf k} ), c_2 ( {\bf k} ), c_3 ( {\bf k} ), c_4 ( {\bf k} ) \right]^{\mathrm{T}}$ and
\bee  \label{Fermi rings}
h_{\rm n}^{\rm A}({\bf k})=M({\bf k})\tau_0\otimes \siz+ m_2 \tau_z \otimes \siz + \sin k_x \tau_0 \otimes \six .
\ee
Here, $M({\bf k})=m_1 -\cos k_x -\cos k_y - \cos k_z$ is a momentum dependent mass term, and $m_1$ and $m_2$ are positive constants.
Eq.~\eqref{Fermi rings} breaks both TRS and PHS, but is symmetric under $k_x \to - k_x$ with $R= \tau_0 \otimes \sigma_z $.
Incidentally, Eq.~\eqref{Fermi rings} also exhibits a chiral symmetry with $S = \mathbbm{1} \otimes \sigma_y$ and $\{ R, S \} =0$,
which corresponds to class AIII with $R_-$ in Table~\ref{reflection_table_full}. However, chiral symmetry can be broken by
including a staggered chemical potential 
\begin{eqnarray} \label{stagCemPot}
V_s
=
 \mu_s \sum_{i=1}^N  (-1)^i \Psi^\dagger (x_i)  \bI  \otimes \siy \Psi^{\ }_\nu (x_i) ,
\end{eqnarray}
with $N$ the number of lattice sites in the $x$ direction. 
For simplicity we assume that $N$ is an even number.
The Hamiltonian with the staggered chemical potential, i.e., $\mathcal{H}^{\rm A}_{\rm n}  + V_s$, is still reflection
symmetric about the mirror plane $x=(x_1+x_N)/2$, and hence belongs to
class A with $R$ in Table~\ref{reflection_table_full}.

The energy spectrum of $\mathcal{H}^{\rm A}_{\rm n}$ in the absence of $V_s$ is given by
\begin{eqnarray}
E_{\pm, \mu } = \pm \sqrt{ \left[ M + (-1)^\mu  m_2 \right]^2+\sin^2 k_x },
\end{eqnarray}
with $\mu \in \{1,2\}$.
Assuming that $m_2 > 0$ and $m_1 - m_2 >1$, 
we find that Hamiltonian~\eqref{Fermi rings}
exhibits two Fermi rings  (i.e., two Fermi surfaces with $d_{\rm FS}=1$) located within the mirror plane $k_x = 0$,
which are described by
\bee  \label{location Fermi rings}
 \cos k_y + \cos k_z = m_1 -1 \pm   m_2 .
\ee
These Fermi rings are topologically stable, since there does not exist any reflection symmetric mass term nor any reflection symmetric kinetic term
that can be added to Eq.~\eqref{Fermi rings} (cf.\ Appendix \ref{appendixA}).
This finding is in agreement with Table~\ref{reflection_table_full}, which shows that the Fermi rings~\eqref{location Fermi rings} are protected by an 
$M\bZ$-type invariant (in the presence of $V_s$)
or an
$M\bZ \oplus \bZ$-type invaraint
(in the absence of~$V_s$).  
To demonstrate this, let us compute the corresponding mirror and winding numbers.

The mirror number $n_{M \bZ}$ is defined within the mirror plane \mbox{$k_x = 0$} for a given eigenspace of the reflection operator $R$.
Focusing on the eigenspace $R = +1$, we find that $h_{\rm n}^{\rm A} (0, k_y, k_z)$ in this subspace reads
\begin{eqnarray}
h_{R=+1}^{\rm A} 
= (m-1 - \cos k_y - \cos k_z) \bI - m_2 \sigma_z. 
\end{eqnarray}
The mirror topological number $n_{M \bZ}$ is given by the 
difference of occupied states (i.e., states with negative energy) on either side of the Fermi ring
\begin{eqnarray}
n^+_{M \bZ }
=
n^+_{\rm occ} ( k^>_y, k^>_z) - n^+_{\rm occ} ( k^<_y, k^<_z) ,
\end{eqnarray}
where $( k^>_y, k^>_z)$ and $( k^<_y, k^<_z)$ are two momenta on either side of the Fermi ring
and 
\bee
n^+_{\rm occ} ( k_y, k_z)=
\left\{ 
  \begin{array}{l l}
    2,&  \; \;   \widetilde{m} (k_y,k_z) < -   m_2  \\
    1,&  \; \;  -m_2 < \widetilde{m} (k_y,k_z) <  + m_2  \\
    0,&  \; \;  \widetilde{m} (k_y,k_z) > + m_2 
 \end{array} \right. ,
\ee
with $ \widetilde{m} (k_y, k_z) =m_1 -1 -\cos k_y-\cos k_z$,
represents the number of occupied states in the eigenspace with $R=+1$. }

In the absence of the staggered chemical potential $V_s$,  Hamiltonian \eqref{Fermi rings} satisfies chiral symmetry and the Fermi rings are also protected
by a winding number $n_{\bZ}$, which takes the form of Eq.~\eqref{BDIwind1d} with 
\bee
{\bf q} = 
\bma
\frac{\sin k_x -i (M({\bf k})+m_2)}{r_+} & 0 \\
0 & \frac{\sin k_x -i(M({\bf k})-m_2)}{r_-} \\
\ema, 
\ee
where $r_\pm=\sqrt{(M({\bf k})\pm m_2)^2+\sin ^2 k_x}$, and an integration contour $\mathcal{C}$ that encircles the Fermi ring. 
Choosing the contour along the $k_x$ direction we find
\bee \label{windinNoExmpAR}
n_{\bZ} ( k_y, k_z)=
\left\{ 
  \begin{array}{l l}
    2,&  \; \;   \widetilde{m} (k_y,k_z) < -   m_2  \\
    1,&  \; \;  -m_2 < \widetilde{m} (k_y,k_z) <  + m_2  \\
    0,&  \; \;  \widetilde{m} (k_y,k_z) > + m_2 
 \end{array} \right. .
\ee
By the bulk-boundary correspondence, a nontrivial value of $n_{\bZ} $, Eq.~\eqref{windinNoExmpAR}, leads to zero-energy flat bands
at the surface of the semimetal. These zero-energy states appear within regions of the surface BZ that are bounded by the projection of the bulk Fermi rings, see Figs.~\ref{Ring figures}(a) and~\ref{Ring figures}(b).
When chiral symmetry is broken, for example by a finite staggered chemical potential $V_s$, the surface flat bands acquire a finite dispersion,
see Fig.~\ref{Ring figures}(c).

\subsubsection{Unstable reflection symmetric nodal superconductors (class~DIII with $R_{-+}$ and $p=2$, class~D~with~$R_+$ and $p=2$)}
 \label{Z2 like comparison}

As shown in Table~\ref{reflection_table_full}, $\bZ_2$-type topological invariants (i.e., $\bZ_2$, $M\bZ_2$, and $M\bZ_2\oplus \bZ_2$) do not protect Fermi surfaces
(superconducting nodes) that are located within the mirror planes but away from high-symmetry points (cf.\ Sec.~\ref{DIIIoff}).
However, these $\bZ_2$-type invariants can lead to protected gapless surface states. To exemplify this behavior we study
in this subsection two-dimensional unstable nodal superconductors belonging to class DIII with $R_{-+}$-type reflection and class D with $R_+$-type reflection, 
 which are classified as $M\bZ_2\oplus \bZ_2$ and $M\bZ_2$, respectively, in Talbe~\ref{reflection_table_full}.
For this purpose, we  borrow an example from Sec.~\ref{DIIIoff}, i.e., 
$\mathcal{H}^{\rm DIII}_{\rm n} = \sum_{\bf k} \Psi^{\dag}_{\bf k} h_{\rm n}^{\rm DIII} \Psi^{\ }_{\bf k} $ with the Nambu spinor
$\Psi^{\ }_{\bf k}  = ( a^\dag_{\bf k}, b^\dag_{\bf k}, a_{- {\bf k}}^{\phantom{\dag}}, b^{\phantom{dag}}_{- {\bf k}} )^{\rm T}$
and
\bee  \label{exampDIIIoffTwo}
h_{\rm n}^{\rm DIII}=(1+\cos k_x +  \cos k_y)\sigma_x \otimes \sigma_y + \sin k_x \sigma_y \otimes \bI  ,
\ee	
which describes a time-reversal symmetric  superconductor
with point nodes located at  $(\pi, \pm\pi/2)$. 
Here, $a^{\dag}_{\bf k}$  and $b^{\dag}_{\bf k}$ represent fermionic creation operators with momentum ${\bf k}$.
Hamiltonian~\eqref{exampDIIIoffTwo} preserves TRS and PHS with $T=\siy\otimes \bI \mK$ and $C=\six \otimes \bI \mK$, respectively, and
is invariant under $k_x \to - k_x$ with $R =\six \otimes \bI$.
Because $T^2=-\dI$, $C^2=+\dI$, $\{ R, T \}Ê= 0$, and $[ R, C ]=0$, Eq.~\eqref{exampDIIIoffTwo} belongs to class DIII with $R_{-+}$.
According to Table~\ref{reflection_table_full}
the point nodes of Hamiltonian \eqref{exampDIIIoffTwo}, which transform pairwise into each other by TRS and PHS, are topologically unstable,
even though the topological numbers $n_{\bZ_2}$ [cf.\ Eq.~\eqref{Z2noExampDIIIccc}] and $n_{M\bZ_2}$ for Hamiltonian~\eqref{exampDIIIoffTwo} take on nontrivial values.
Indeed, we find that 
the symmetry-preserving extra kinetic term $\delta t \sin k_y \six\otimes \six$ gaps out the Fermi points at $(\pi, \pm\pi/2)$
and turns Eq.~\eqref{exampDIIIoffTwo} into a fully gapped reflection symmetric topological superconductor
\begin{eqnarray} \label{exampDIIIoffTwoFullG}
\mathcal{H}^{\rm DIII}_{\rm fg} 
=
\mathcal{H}^{\rm DIII}_{\rm n} 
+
\delta t
\sum_{\bf k} \Psi^{\dag}_{\bf k} 
 \sin k_y \six\otimes \six
\Psi^{\phantom{\dag}}_{\bf k}  .
\end{eqnarray}
That is, the unstable nodal superconductor \eqref{exampDIIIoffTwo} is connected to the fully gapped reflection symmetric topological superconductor~\eqref{exampDIIIoffTwoFullG} and
inherits topological edge states from the fully gapped phase.\cite{poYao14} 

To demonstrate this, let us compute the global $n_{\bZ_2}$ invariant
and the mirror invariant $n_{M \bZ_2}$ for Hamiltonian~\eqref{exampDIIIoffTwo} and~\eqref{exampDIIIoffTwoFullG}. The computation of the global invariant $n_{\bZ_2}$, which is given by Eq.~\eqref{Z2noExampDIIIccc}, follows along similar lines as in the example of Sec.~\ref{DIIIoff}.
(Note that for the definition of a $\bZ_2$-type invariant the reflection symmetry does not play any role; the $\bZ_2$ number $n_{\bZ_2}$ is  defined
solely in terms of the global symmetries.)
We find that 
for a contour $\mathcal{C}$ oriented along the $k_x$ axis with
$k_y$ held fixed at $k_y =0$ (or $k_y = \pi$), the topological index evaluates to
$n_{\bZ_2}=+1$ (or $n_{\bZ_2}=-1$) both for the nodal superconductor
$\mathcal{H}^{\rm DIII}_{\rm n}$  and the fully gapped superconductor $\mathcal{H}^{\rm DIII}_{\rm fg} $.
This indicates that 
there appear zero-energy edge states at $k_y = \pi$ of the (10) edge BZ of both the fully gapped and the nodal  system.

To calculate the mirror number $n_{M \bZ_2}$ we focus on the eigenspace of the reflection operator with eigenvalue $R=+1$
and transform Hamiltonian~\eqref{exampDIIIoffTwoFullG} to a Majorana basis.\cite{Kitaev1D} 
On the mirror lines $k_x=0$ and $k_x=\pi$, $\mathcal{H}^{\rm DIII}_{\rm fg}$  in the eigenspace $R=+1$ can be expressed as
\begin{eqnarray} \label{exampDIIIoffTwoTrafo1}
&&
\mathcal{H}^{\rm DIII, \nu}_{R=+1} =
\\
&& \quad
\sum_{k_y} M_{\nu} (k_y)
\bma
d_{\nu, k_y}^\dag & d^{\phantom{\dag}}_{\nu, -k_y} \\
\ema
\bma
1 & - i \delta T \\
i \delta T  & -1
\ema
\bma
d_{\nu, k_y} \\
d_{\nu, -k_y}^\dag
\ema,
\nonumber
\end{eqnarray}
with $\nu \in \{ 0, \pi \}$ and
 where $M_{\nu} (k_y)=1+ (-1)^{\nu / \pi } + \cos k_y$ and $\delta T(k_y)= \delta t \sin k_y$.
In Eq.~\eqref{exampDIIIoffTwoTrafo1} the transformed fermion operators $d_{\nu, k_y}$ are given by
\bee \label{transformedOps1}
d_{\nu, k_y}
=\frac{1}{2} \left[ a^\dag_{\nu, k_y}+a^{\phantom{\dag}}_{\nu, -k_y}-i(b^\dag_{\nu, k_y}+b^{\phantom{\dag}}_{\nu, -k_y}) \right].
\ee
Using Eq.~\eqref{transformedOps1} we can construct real Majorana operators
$\Lambda_{\nu, k_y}= (\lambda_{\nu, k_y}, \lambda'_{\nu, k_y})^{\mathrm{T}}$, with
\bee  
\lambda_{\nu, k_y}:=d^\dag_{\nu, k_y}+d^{\phantom{\dag}}_{\nu, k_y},
\; \;
 \lambda'_{\nu, k_y}:=i(d_{\nu, -k_y}^\dag-d^{\phantom{\dag}}_{\nu, -k_y}),\ 
\ee
and rewrite the Hamiltonian in the $R=+1$ eigenspace as
\begin{subequations}
\bee
\mathcal{H}^{\rm DIII, \nu}_{R=+1}
=
\frac{i}{ 2}\sum_{k_y}
\Lambda^{\mathrm{T}}_{\nu, -k_y} 
B _{\nu }(k_y)
\Lambda^{\phantom{T}}_{\nu, k_y} ,
\ee
with 
\bee
B_{\nu }(k_y)=
\bma
\delta T(k_y) & M_{\nu }(k_y) \\
-M_{\nu }(k_y) & \delta T(k_y)
\ema .
\ee
\end{subequations} 
It follows that the mirror invariant $n_{M \bZ_2}$ on the two mirror lines $k_y=0$ and $k_y=\pi$ is given by
\begin{equation}
n^{\nu}_{M \bZ_2}
=
{\rm sgn}  [{\rm Pf} B_\nu (0)]{\rm sgn}  [{\rm Pf} B_\nu (\pi)] 
=
\left\{ 
  \begin{array}{l l}
 + 1, & \nu = 0  \\
  - 1, & \nu = \pi 
  \end{array}
  \right. . \; 
\end{equation}
Interestingly, the value of  $n^{\nu}_{M \bZ_2}$ does not depend on the extra kinetic term $\delta t \sin k_y \six\otimes \six$.
Hence, we conclude that the unstable nodal superconductor $\mathcal{H}^{\rm DIII}_{\rm n}$ 
 can be connected to the fully gapped topological superconductor $\mathcal{H}^{\rm DIII}_{\rm fg}$
 (whose bulk topology is described by $n^{0}_{M \bZ_2} n^{\pi}_{M \bZ_2}$)
 without changing the values of the invariants $n_{\bZ_2}$ and $n^{\nu}_{M \bZ_2}$.
Both $n_{\bZ_2}$ and $n^{\nu}_{M \bZ_2}$ lead to protected zero-energy states
at the edge of the nodal (or fully gapped) superconductor. 

We observe that in systems that are classified as $M\bZ_2\oplus \bZ_2$ in Table~\ref{reflection_table_full}
the two invariants $n_{M \bZ_2}$ and $n_{\bZ_2}$ always take on the same values. 
This is in contrast to topological materials with an $M \bZ \oplus \bZ$ classification,
where the two invariants $n_{M \bZ}$ and $n_{\bZ}$ can be distinct, see example in Sec.~\ref{high-symmetry DIII}.
That is, the presence of reflection symmetry in $M\bZ_2\oplus \bZ_2$-type systems does not lead to any new topological characteristics,
but it simplifies the calculation of the topological index. I.e., the topological characteristics can be inferred from
the wavefunctions at reflection planes alone.
(This situation is in a  sense similar to the $\bZ_2$ time reversal symmetric topological insulator with inversion symmetry of Ref.~\onlinecite{InversionFu},
where the inversion symmetry  does not lead to new topological features, but simplifies the formula for the topological index.)

A similar analysis as above can be preformed for a two-dimensional unstable nodal superconductor in class $D$ with $R_{+}$-type reflection symmetry.
In the absence of TRS the global $\bZ_2$ number $n_{\bZ_2}$ is ill defined, however the mirror invariant
$n_{M \bZ_2}$ is still well defined and takes on nontrivial values (cf.\ Table~\ref{reflection_table_full}). 
This mirror index leads to stable zero-energy modes at edges that are invariant under reflection.
As before, we find that a reflection symmetric nodal superconductor in class D with $R_+$
can be connected to a fully gapped topological superconductor without removing the zero-energy edge-states.

\subsubsection{Reflection symmetric nodal spin-triplet superconductor with TRS (\mbox{class DIII with $R_{--}$ and $p=3$})}
\label{exampDIIInodalRmm}

%%%%% begin Figure
\begin{figure*}[t!]
\begin{center}
\includegraphics[width=0.95\textwidth,angle=-0]{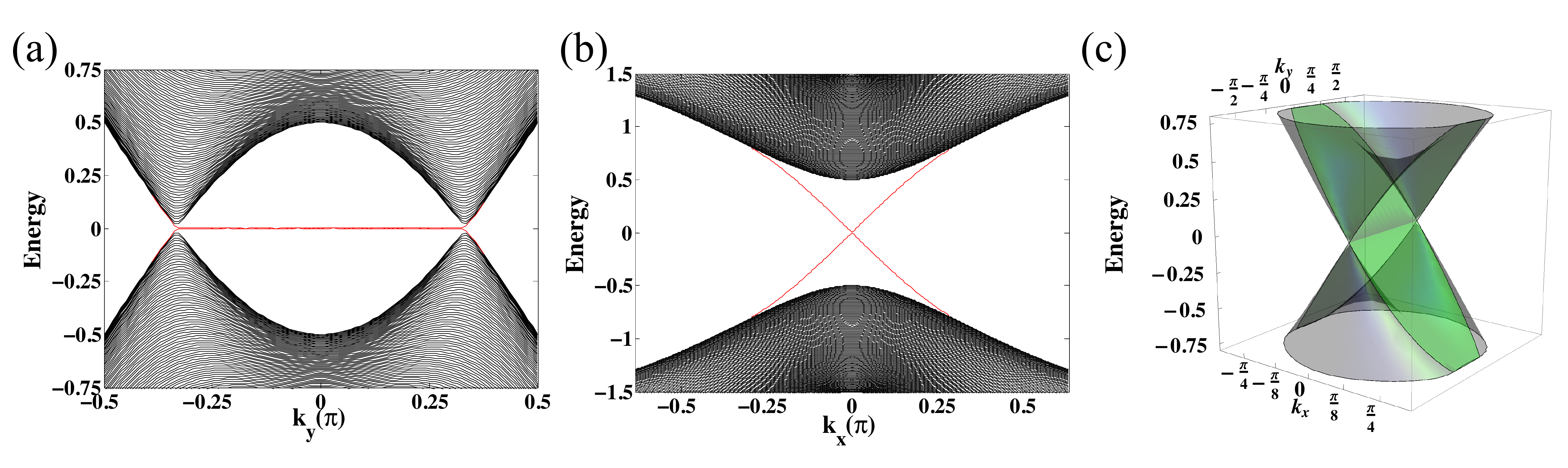}
  \caption{(Color online) 
Surface band structure of 
the reflection symmetric nodal superconductor~\eqref{exampDIII3DsymO} (class DIII with $R_{--}$) for the (001) face as a function
 of (a) surface momentum $k_y$ with $k_x=0$ and  (b) surface momentum $k_x$ with $k_y=0$.
A zero-energy arc surface state (red trace) connects the projected point nodes in the surface BZ.
(c)  Surface spectrum on the (001) face as a function of both $k_x$ and $k_y$.
The surface states and bulk states are indicated in green and grey, respectively.}
\label{DIII_reflection}
\end{center}
\end{figure*}
%%%%% end Figure

As the last example of this subsection, we study a three-dimensional reflection symmetric superconductor in class~DIII 
\begin{equation} \label{exampDIII3DsymO}
h^{\mathrm{DIII}}_{\rm 3D}
=
M ({\bf k}) \sigma_z \otimes \mathbbm{1} 
+
\sin k_x \sigma_x \otimes \sigma_x
+
\sin k_z \sigma_x \otimes \sigma_z , \; \; 
\end{equation}
which exhibits point nodes at  ${\bf k} = (0, \pm \pi /3, 0)$.
The k-dependent mass $M ({\bf k} )$ is given by $M ({\bf k} ) = -2.5 + \cos k_x + \cos k_y + \cos k_z$.
Hamiltonian~\eqref{exampDIII3DsymO} satisfies 
TRS and PHS with $T =  \mathbbm{1} \otimes \sigma_y \mathcal{K}$ and $C = \sigma_x \otimes \mathbbm{1} \mathcal{K}$,
respectively, and is reflection symmetric under $k_x \to - k_x$ with $R = \sigma_z \otimes \sigma_x$.
Because $T^2 = -  \dI $, $C^2 =+ \dI$, $\{ T, R \}=0$, and $\{ C, R \}=0$, Eq.~\eqref{exampDIII3DsymO}
is classified as DIII with $R_{--}$. 
The two point nodes, which are located within the mirror plane at ${\bf k} = (0, \pm \pi /3, 0)$ are 
protected by TRS, PHS, and reflection symmetry, since there does not exist any SPGT  that can be added to Eq.~\eqref{exampDIII3DsymO}. 
We note that the gap opening term $\sin k_y \sigma_y \otimes \mathbbm{1}$ is symmetric under TRS and PHS but breaks
mirror symmetry, which shows that the reflection symmetry $R$ is crucial for the protection of the point nodes. 
Indeed, as indicated by Table~\ref{original table},  the point nodes are unstable in the absence of reflection symmetry.

Let us now compute the mirror invariant $n_{M \bZ}$ which, as listed in Table~\ref{reflection_table_full}, protects the  point nodes. 
Since the chiral symmetry operator $S=T C = \sigma_x \otimes \sigma_y $ commutes with $R$, the mirror number 
$n_{M \bZ}$ can be expressed as a one-dimensional winding number, i.e., for the eigenspace $R=+1$ it takes the form
of Eq.~\eqref{BDIwind1d} with 
\bee
{\bf q} = \frac{M({\bf k})-\sin k_z i}{\sqrt{M({\bf k})^2+\sin^2 k_z}},
\ee 
and a contour $\mathcal{C}$ that lies within the mirror plane
and encloses one of the point nodes [see Fig.~\ref{FigLocFermiWMirror}(b)].
Choosing the contour along the $k_z$ axis with $k_x =0$ and $k_y$ a fixed parameter, we find that the mirror number
evaluates to
\begin{equation} \label{mZclassDIII3D}
n^+_{M \bZ} (k_y)
=
\left\{ 
  \begin{array}{l l}
  1, & \; \;  0 \leq  | k_y | <  \frac{ \pi }{ 3}  \\
   0, &  \; \;     \frac{ \pi }{  3} < | k_y | \leq \pi 
  \end{array}
  \right. . \; 
\end{equation}
By the bulk boundary correspondence, the nontrivial value of Eq.~\eqref{mZclassDIII3D} leads to zero-energy arc states on surfaces that
are perpendicular to the mirror plane. As shown in Fig.~\ref{DIII_reflection},
these zero-energy arc states connect  two projected point nodes in the surface BZ.

\subsection{Fermi surfaces outside mirror planes}

Third, {we discuss three examples of Fermi surfaces (superconducting nodes) that lie outside the mirror plane.
These Fermi surfaces are pairwise related to each other by both reflection
and nonspatial symmetries, see Fig.~\ref{FigLocFermiWMirror}(c).
Their topological properties are classified by Table~\ref{table_reflection_off_off}.}

\subsubsection{Reflection symmetric Dirac semimetal with TRS (class~AII~with~$R_{+}$ and $p=3$)}

We start by studying an example of a three-dimensional Dirac semimetal with an $R_{+}$-type reflection symmetry,
which is described by\cite{OjanenPRB13,morimotoFurusakiPRB14,footnoteABC}
\begin{align} \label{exampAIIoff_off}
H_{\rm off}^{\rm AII}=
 \sin k_y \tau_x\otimes \sigma_z +  \sin k_z \tau_y \otimes \bI + \mathcal{M} ({\bf k}) \tau_z\otimes \bI .
\end{align}
Here, $\mathcal{M} ({\bf k} ) =M-\cos k_x - \cos k_y - \cos k_z$ and $M$ is a positive constant, which we set to $M=2.0$.
The Pauli matrices $\sigma_i$ and $\tau_i$ operate in  spin and orbital grading, respectively.
Hamiltonian~\eqref{exampAIIoff_off} preserves TRS with $T=\bI\otimes i\sigma_y \mathcal{K}$
and is symmetric under $k_x \to - k_x$ with $R =\bI \otimes \bI$.
Since $T^2 = -  \dI $ and $[T, R] =0$, the Hamiltonian belongs to  class AII with $R_+$.
By computing the energy spectrum we find that the semimetal
exhibits two doubly degenerate Dirac points that are located outside the reflection plane $k_x =0$, i.e., at
${\bf k}=(\pm\pi/2,0,0)$.
These Fermi points are protected by a combination of time-reversal and reflection symmetry,
because there does not exist any SPGT
that can be added to Eq.~\eqref{exampAIIoff_off}.
We note, however, that in the absence of reflection symmetry, the Dirac points
can be gapped out by the time-reversal invariant term $\sin k_x \tau_x \otimes \sigma_x$,
which turns Hamiltonian~\eqref{exampAIIoff_off} into a class AII topological insulator. This finding
is in agreement with the ten-fold classification of gapless topological materials shown in Table~\ref{original table}.
To determine whether the Dirac points have a  $\bZ$- or  $\bZ_2$-type character, we consider a doubled version of  
$H_{\rm off}^{\rm AII}$, i.e., $H_{\rm off}^{\rm AII}\otimes\bI $.
For the doubled Hamiltonian there exist a momentum-independent SPGT (i.e., $\tau_x \otimes \sigma_x \otimes \sigma_y$),
demonstrating that the Dirac points are protected by a $\bZ_2$-type invariant, which is denoted
as ``$C \bZ_2$" in Table~\ref{table_reflection_off_off}.

The $C \bZ_2$ invariant  $n_{C \bZ_2}$ is defined in terms of the combined symmetry~\eqref{combineSYMdefB}, i.e.,
$ \tilde{T}^{-1} H_{\rm off}^{\rm AII} ( k_x , - {\bf \tilde{k}} ) \tilde{T}
= 
H_{\rm off}^{\rm AII}( k_x, {\bf \tilde{k}} ) $.
Since each plane perpendicular to the $k_x$ axis is left invariant by the combined symmetry~\eqref{combineSYMdefB},
we can define the topological number $n_{C \bZ_2}$ for any given plane $E_{k_x}$ with fixed $k_x$ [see Fig.~\ref{FigLocFermiWMirror}(c)].
We find that
\begin{equation}
n_{C \bZ_2} (k_x) 
=
\left\{ 
  \begin{array}{l l}
    +1,&  \; \, \frac{\pi}{2} < | k_x|  \leq \pi \\
    -1,&  \; \, 0 \leq | k_x |< \frac{\pi}{2} 
  \end{array} \right. \; .
\end{equation} 
Due to the bulk-boundary correspondence, the nontrivial value of $n_{C \bZ_2} (k_x)$ in the interval $[-Ê\pi /2 , + \pi /2 ]$ gives rise to 
helical Fermi arcs on surfaces that are perpendicular to the reflection plane.\cite{OjanenPRB13,morimotoFurusakiPRB14} These helical arc states connect the project bulk Dirac points in the surface BZ.

\subsubsection{Reflection symmetric nodal spin-triplet superconductor  (class~D~with~$R_-$ and $d=3$)}

Next, we consider a reflection symmetric nodal spin-triplet superconductor, which is described by the BdG Hamiltonian
\bee \label{exmpHDoffoff}
H_{\rm off}^{\rm D}=  \sin k_y \tau_y \otimes \sigma_z+ \sin k_z \tau_x \otimes \sigma_z + \mathcal{M} ( {\bf k} ) \tau_z \otimes\bI, 
\ee
where $\mathcal{M} ( {\bf k} ) =2-\cos k_x - \cos k_y - \cos k_z$. Here, the Pauli matrices $\sigma_i$ and $\tau_i$ act in spin and particle-hole space, respectively.
 $H_{\rm off}^{\rm D}$ satisfies PHS with $C=\tau_x \otimes \bI \mathcal{K}$ and is invariant under $k_x \to - k_x$ with
$R_-=\tau_z\otimes \sigma_x \label{R D}$.
Because $C^2 = + \dI$ and $\{ R, C \} =0$, the BdG Hamiltonian belongs to class D with $R_-$. As an aside, we note that
reflection symmetry $k_x \to - k_x$ for spin-$\frac{1}{2}$ systems is usually implemented by the operator $R'_{\rm p} =+ i\sigma_x$ 
($R'_{\rm h} =- i\sigma_x$)
for  particle-like (hole-like) degrees of freedom, i.e., by
the operator $R' = i\tau_z\otimes \sigma_x$ in particle-hole space. However, in order to correctly categorize the Hamiltonian
with respect to the 27 symmetry classes of Table~\ref{table_reflection_off_off}, we need to ensure that the reflection operator $R$
is Hermitian (cf.~Eq.~\ref{hermitian}).
Therefore we have dropped the factor $i$ in the above definition of $R$.

The spectrum of Hamiltonian~\eqref{exmpHDoffoff} exhibits two doubly degenerate point nodes, which are located outside the mirror plane at 
${\bf k}=(\pm\pi/2,0,0)$. These point nodes are topologically stable, since there does not exist any SPGT that can be added to Eq.~\eqref{exmpHDoffoff}. According to Table~\ref{table_reflection_off_off} the point nodes of $H_{\rm off}^{\rm D}$ are protected by
an invariant of type ``$2 \bZ$" (i.e., a Chern number),  
where the prefix ``$2$" indicates that the topological number only takes on even values.
Choosing the two-dimensional integration contour to be a plane perpendicular to the $k_x$ axis, we find
that the Chern number for Hamiltonian~\eqref{exmpHDoffoff} is given by
\begin{small}
\begin{align}
n_{\bZ}(k_x) =& \frac{i}{2\pi}\int \sum_{i=1}^2 d\braket{ u^-_i}{d u^-_i} \nonumber \\
=& -\int \frac{1}{2\pi R^3} {\big (} Z dX \wedge dY + X dY \wedge dZ + Y dZ \wedge dX \big ),
\end{align}
\end{small}
where $X=\sin k_z,\ Y=\sin k_y,\ Z=\mathcal{M}  ( {\bf k}) $, and $R=\sqrt{X^2+Y^2+Z^2}$. 
Evaluating the integral, we obtain
\bee  \label{ChernNoDoffoff}
n_{\bZ} (k_x)=\left\{ 
  \begin{array}{r l}
    0,& \; \;  \frac{\pi}{2}  < | k_x|  \leq \pi, \\
     -2,& \; \;   0 \leq | k_x |<  \frac{\pi}{2} 
  \end{array} \right. .
\ee	
Note that for the definition of the Chern number~\eqref{ChernNoDoffoff}, the combined symmetry $\tilde{C} = R C$, Eq.~\eqref{combineSYMdefC}, does not  
not play any role, except to ensure that there are an even number of point nodes on either side of the reflection planes.
By the bulk-boundary correspondence,  the nontrivial value of $n_{\bZ}$, Eq.~\eqref{ChernNoDoffoff}, gives rise to 
arc surface states which connect the projected point nodes in the surface BZ.\cite{Brydon10}

\subsubsection{Unstable reflection symmetric nodal superconductor with TRS (class DIII with $R_{-+}$ and $d=2$)}

As stated in Section~\ref{ClassifyOffOffRmpRpm}, superconducting nodes outside the mirror plane in systems of class DIII with $R_{-+}$-type reflection symmetry
are unstable, 
even though a nontrivial   $M\bZ_2$-type invariant can be defined for these systems.
To illustrate this, we consider the following BdG Hamiltonian 
\bee  \label{exampOffOffDIIIunstable}
H^{\rm DIII}_{\rm off}=\sin k_y \six \otimes \bI +(1+\cos k_x +\cos k_y)\siz\otimes \siy,
\ee
which describes a superconductor with unstable point nodes. 
Eq.~\eqref{exampOffOffDIIIunstable} preserves TRS and PHS with $T=\siy \otimes \bI \mK$ and $C=\six\otimes \siz \mK$, respectively, and
is symmetric under $k_x \to - k_x$ with $R=\six\otimes \siz$.
Because $T^2=- \dI $, $C^2=+  \dI $, $\{ T, R \}=0$, and $[ C, R ] =0$,
Hamiltonian~\eqref{exampOffOffDIIIunstable} is classified as DIII with $R_{-+}$.
We find that the spectrum of Eq.~\eqref{exampOffOffDIIIunstable} exhibits 
point nodes located away from the mirror lines $k_x =0$ and $k_x =\pi$, i.e., at ${\bf k} =(\pm \pi/2,0)$.
These point nodes are topologically unstable, since there exists a momentum-dependent SPGT
(i.e., $\sin k_x\siy \otimes \bI)$, which opens up a full gap.

Let us now examine topological invariants for Hamiltonian~\eqref{exampOffOffDIIIunstable}. First, we consider a winding number 
$\nu_{\bZ}$, which is defined by chiral symmetry with $S=TC=-i \siz \otimes \siz $ on a line perpendicular to the $k_x$ direction. Since chiral symmetry is momentum independent, combining reflection and chiral symmetries is not required to define the winding number $\nu_{\bZ}$.
We find that this one-dimensional winding number is given by Eq.~\eqref{DIIIWno} with 
\bee
q=\frac{1}{\sqrt{\sin^2 k_y+M^2}}
\bma 
\sin k_y & -iM \\
-iM & \sin k_y \\
\ema ,
\ee
where $M=1+\cos k_x + \cos k_y$.
Evaluating the integral, one obtains that
$\nu_{\bZ}$ is trivial for any fixed $k_x$ (i.e., $\nu_{\bZ}=0$), in agreement with the fact that
the point nodes are unstable. 
Second, we consider the mirror invariant, which is defined within the mirror lines $k_x = 0$ and $k_x = \pi$ for a given eigenspace of $R$.
Since   $H^{\rm DIII}_{\rm off}$ restricted to the
mirror lines satisfies PHS,
a mirror invariant of type $M \bZ_2$ can be defined.
By a similar calculation as in example~\ref{Z2 like comparison}, we find that the mirror invariant $n_{M\bZ_2}$ is given by $n_{M\bZ_2}=1$ for $k_x=0$ and $n_{M\bZ_2}=-1$ for $k_x=\pi$. 
However, even though $n_{M\bZ_2}$ takes on a nontrivial value, this $M\bZ_2$-type invariant does not protect the point nodes that are located at ${\bf k} =(\pm \pi/2,0)$
(see Appendix~\ref{R+-}).

\section{Summary and Conclusions}
\label{sec:Conclu}

In this paper we have performed an exhaustive classification of reflection symmetry protected topological semimetals and nodal superconductors. We have shown that the classification depends on (i) the codimension $p=d-d_{\mathrm{FS}}$ of the Fermi surface (nodal line) of the semimetal (nodal superconductor), (ii) how the Fermi surface (nodal line) transforms under the crystal reflection and the global symmetries, and (iii) whether the reflection symmetry operator $R$ commutes or anticommutes with the global (i.e., nonspatial) symmetries. 
The result of this classification scheme is summarized in Tables~\ref{reflection_table_full} and~\ref{table_reflection_off_off}, which show that the presence of reflection symmetries leads to an enrichment of the ten-fold classification of gapless topological materials (cf.\ Table~\ref{original table}) with additional topological states.
The reflection symmetry $R$ together with the three nonspatial symmetries, time-reversal, particle-hole, and chiral symmetry, define a total of 27 different symmetry classes. 
For Fermi surfaces with even (odd) codimension $p$ located within the mirror plane,
17 (10) out of these 27 classes allow for  nontrivial topological characteristics of the Fermi surface (Table~\ref{reflection_table_full}).
For Fermi surfaces located outside the mirror plane, on the other hand, there are 9 symmetry classes
which permit the existence of nontrivial topological properties (Table~\ref{table_reflection_off_off}).

To illustrate the general principles of the classification schemes, we have discussed in Sec.~\ref{sec:examples} concrete examples of reflection symmetry protected topological semimetals and nodal superconductors. 
The topological properties of these gapless  materials manifest themselves at the surface in the form of
linearly dispersing Dirac or Majorana modes, or dispersionless states, which form two-dimensional flat-bands 
or one-dimensional arcs (see Figs.~\ref{Honeycomb},~\ref{Ring figures}, and \ref{DIII_reflection}). These different types of surface states are protected by different types
of topological invariants. For the examples of Sec.~\ref{sec:examples} we have derived
 explicit expressions for these  topological numbers.

Probably, the most prominent example of a reflection symmetric topological semimetal is graphene,\cite{castroNetoRMP09} whose Dirac points are protected against gap opening by time-reversal symmetry together with reflection and $SU(2)$ spin-rotation symmetry.
In the classification scheme of Table~\ref{reflection_table_full}, graphene belongs to class AI with $R_+$-type reflection symmetry. 
Hence, the Dirac points of graphene, which are located within the reflection line but away from time-reversal invariant points, 
are protected by a mirror invariant ($M \bZ$), see Sec.~\ref{secGraphene}.
The classifications of Tables~\ref{reflection_table_full} and~\ref{table_reflection_off_off} predict several new reflection symmetric topological semimetals and nodal superconductors, for which realistic physical systems have yet to be found. 
For example, a reflection symmetric topological nodal superconductor with spin-triplet pairing is predicted to exist in three spatial dimensions (class DIII
with $R_{--}$), see Sec.~\ref{exampDIIInodalRmm}. This nodal superconductor, which exhibits two point nodes within the reflection plane (but away from the time-reversal invariant momenta)
is a three-dimensional superconducting analog of graphene.

Recently, several examples of space group symmetry protected topological semimetals have been theoretically proposed.\cite{Dirac_semimetal_Kane,Steinberg:2014aa}  The surface states of Na$_3$Bi\cite{Liu21022014,DiracHasan,Dai_predition_Na3Bi} and Cd$_3$As$_2$,\cite{Yazdani_CdAs,Dirac_semimetal_Xi_Dai,Cd3As2:Chen2014} which are two
topological Dirac materials protected by rotation symmetry, have been experimentally observed using angle-resolve photoemission 
and scanning tunneling measurements.
We hope that these recent discoveries will spur the experimental search for other types of topological phases.
The results of this paper will be useful for the search and design of new gapless topological materials that are protected by reflection symmetry.

\acknowledgments
The authors thank G. Bian, M.~Franz, M.~Garcia Vergniory, P.~Horsch, S.~Ryu, and A.~Yaresko   
for useful discussions. 
The support of the Max-Planck-UBC Centre for Quantum Materials is gratefully acknowledged.
A.~P.~S.~wishes to thank the ESI (Vienna) for its hospitality.

\appendix
\section{Review of ten-fold classification scheme of gapless topological materials}
\label{appendixA}

Topological properties of gapless materials can be classified by two different methods:\cite{matsuuraNJP13,ZhaoWangPRL13,ZhaoWangPRB14,Sato_Crystalline_arxiv14} 
(i)  the minimal Dirac-matrix Hamiltonian method and
(ii) the derivation  of topological invariants.
For the former, the topological property is determined by the existence or nonexistence of a symmetry-preserving gap-opening term (SPGT).
The existence of an SPGT implies trivial topology of the gapless system, i.e., the Fermi surface (nodal line) is topologically unstable. In the absence of such an SPGT, however,
the gapless state is topologically nontrivial and exhibits topologically stable Fermi surfaces (nodal lines).
Method (i) is similar to the approach of Refs.~\onlinecite{chiuPRB13,Chiu_nontrivial_surface}, which classify fully gapped topological materials by studying symmetry preserving extra mass terms that allow to deform
different gapped states into each other without closing the bulk gap.
Method (ii), on the other hand, relies on the existence or nonexistence of nonzero topological invariants. 
A nonzero topological invariant implies nontrivial topology of the gapless quantum system.
In this Appendix and in Appendix~\ref{derivation of gapless} we use the minimal Dirac-matrix Hamiltonian approach [i.e., method (i)] to derive the topological classification
of gapless materials. These derivations should be compared to the discussions in the main text, which uses the topological invariant approach [i.e., method (ii)]; see, in particular, 
Sec.~\ref{outside mirror planes}.

\begin{table}
\begin{center}
\begin{tabular}{|c|c|c|c|}
\hline
s & AZ class ($d=0$) & Topological invariant & gamma matrix \\
\hline
0 & A & $\pi_0(\mathcal{C}_0)=\bZ$ &  \\
1 & AIII & $\pi_0(\mathcal{C}_1)=0$ & $\gamma_{d+1}$ or $\tilde{\gamma}_1$ \\
\hline
0 & AI &  $\pi_0(\mathcal{R}_0)=\bZ$ &  \\
1 & BDI & $\pi_0(\mathcal{R}_1)=\bZ_2$ & $\gamma_{d+1}$ \\
2 & D &  $\pi_0(\mathcal{R}_2)=\bZ_2$ & $\gamma_{d+1},\ \gamma_{d+2}$ \\
3 & DIII & $\pi_0(\mathcal{R}_3)=0$ & $\gamma_{d+1},\ \gamma_{d+2},\ \gamma_{d+3}$ \\
4 & AII & $\pi_0(\mathcal{R}_4)=2\bZ$ &  \\
5 & CII & $\pi_0(\mathcal{R}_5)=0$ & $\tilde{\gamma}_{1},\ \tilde{\gamma}_2,\ \tilde{\gamma}_3$ \\
6 & C & $\pi_0(\mathcal{R}_6)=0$ & $\tilde{\gamma}_1,\ \tilde{\gamma}_2$ \\
7 & CI & $\pi_0(\mathcal{R}_7)=0$ & $\tilde{\gamma}_1$ \\
\hline
\end{tabular}
\caption{
This table lists the presence or absence of symmetry-allowed kinetic terms ($\gamma_i$) or mass terms ($\tilde{\gamma}_j$) for each of the ten
Altland-Zirnbauer symmetry classes.
 Due to the periodicity of two and eight for complex and real symmetry classes, respectively, $l=0,1$ mod $2$ for $\mathcal{C}_l$ and $l=0,1,\ldots,7$ mod 8 for $\mathcal{R}_l$.}
\label{presence_gamma}
\end{center}
\end{table}

\subsection{Fully gapped materials}

Before discussing the ten-fold classification of gapless materials (cf.~Table~\ref{original table}), let us briefly state some results and definitions related to the ten-fold classification of fully gapped systems. The Dirac Hamiltonian ($H^{\rm TI}_{\rm Dirac}$) that classifies fully gapped systems (i.e., topological insulators and superconductors) is given by Eq.~\eqref{TI Dirac}, where the Dirac matrices $\gamma_i$ are kinetic terms and the Dirac matrices $\tilde{\gamma}_j$ represent mass terms. For real symmetry classes, these Dirac matrices obey 
\begin{align}
\{T,\gamma_i\}&=0,& [C,\gamma_i]&=0 , \label{kinetic symmetry} \\
[T,\tilde{\gamma}_j]&=0,& \{C,\tilde{\gamma}_j\}&=0 ,
\end{align}
to preserve TRS and PHS. 
Note that both Dirac matrices, $\gamma_i$ and $\tilde{\gamma}_j$, anticommute with the chiral symmetry operator $S=CT$. The classification of fully gapped topological 
materials follows from the homotopy groups, which are given by\cite{Kitaev,chiuPRB13,Chiu_nontrivial_surface}
\begin{subequations} \label{homotopyFullyGapped}
\begin{align}
K^{\mathbb{C}}(s,d)&=\pi_0(\mathcal{C}_{s-d}),  \\
K^{\mathbb{R}}(s,d)&=\pi_0(\mathcal{R}_{s-d}) ,
\end{align}
\end{subequations}
where $\mathcal{C}_{s}$ and $\mathcal{R}_s$ denote the complex and real classifying spaces, respectively.
Eqs.~\eqref{homotopyFullyGapped} are in line with the existence or nonexistence of symmetry allowed kinetic terms ($\gamma_i$) and mass terms ($\tilde{\gamma}_j$)
which enter in the minimal-Dirac matrix description;
see Table~\ref{presence_gamma}.
In the case, where the classification is trivial, which is labeled by  ``$0$" in Table~\ref{presence_gamma}, the symmetry preserving mass term  $m\tilde{\gamma}_1$ in $H^{\rm TI}_{\rm Dirac}$
allows to deform different gapped phases into each other without closing the bulk gap. Hence, in this case there is only one topological equivalence class, namely the trivial one. 
When there is a binary classification, which is labeled by ``$\bZ_2$"  in Table~\ref{presence_gamma}, there exists an extra symmetry allowed kinetic term $k_{j}\gamma_{d+1}$ that can be added to $H^{\rm TI}_{\rm Dirac}$.
This kinetic term allows us to deform the doubled version of $H^{\rm TI}_{\rm Dirac}$ to a trivial state without closing the bulk gap.
Finally, in the case of the $\bZ$ classification, both symmetry-allowed kinetic terms ($\gamma_i$) and mass terms ($\tilde{\gamma}_j$) are absent.

\subsection{Gapless materials}

The classification of global symmetry invariant Fermi points is related to the ten-fold classification of fully gapped systems by the dimensional shift $d \to d-1$; see Table~\ref{original table}.
In other words, the classification of gapless materials follows from the homotopy groups~\cite{Sato_Crystalline_arxiv14,matsuuraNJP13,ZhaoWangPRL13,ZhaoWangPRB14}
\begin{subequations} \label{homotopyGapless}
\begin{align}
G^{\mathbb{C}}_{\rm s}(s,d)&=\pi_0(\mathcal{C}_{s-d-1}) ,\\
G^{\mathbb{R}}_{\rm s}(s,d)&=\pi_0(\mathcal{R}_{s-d-1}).
\end{align}
\end{subequations}
Eqs.~\eqref{homotopyGapless} are 
in agreement with the results from the minimal Dirac-matrix Hamiltonian method, which we will discuss in the following. 
Let us consider a Dirac Hamiltonian describing a Fermi point at a time-reversal invariant momentum of the BZ (i.e., at $k=0$) 
\bee
H_{\rm s}^{\rm Dirac}=\sum_i^d k_i  \gamma_i . \label{Dirac symmetry}
\ee
We note that this Hamiltonian is identical to the fully gapped Dirac Hamiltonian   
 in Eq.~\eqref{TI Dirac}, except for the mass term $m\tilde{\gamma}_0$. 
 For real symmetry classes the Dirac matrices $\gamma_i$ (i.e., the kinetic term) obey Eqs.~\eqref{kinetic symmetry}. 
Furthermore, we observe that $H_{\rm s}^{\rm Dirac}$ in $d$ dimensions can be viewed as the boundary states of $H^{\rm TI}_{\rm Dirac}$ in $d+1$ dimensions; see Eq.~\eqref{DiracFermiaaa}.
In other words, the Hamiltonian $H^{\rm TI}_{\rm Dirac}$ in $d+1$ dimensions is obtained from the $d$-dimensional Hamiltonian $H_{\rm s}^{\rm Dirac}$
by adding an extra kinetic term (i.e., $k_{d+1}\gamma_{d+1}$) and a  mass term (i.e., $M\tilde{\gamma}_0$). With this, both $H^{\rm TI}_{\rm Dirac}$
and $H_{\rm s}^{\rm Dirac}$ satisfy the same global symmetries.
The extra symmetry preserving mass term $m\tilde{\gamma}_1$ that can (or cannot) be added to $H^{\rm TI}_{\rm Dirac}$ plays the role 
of an SPGT that can (or cannot) be added to $H_{\rm s}^{\rm Dirac}$.
That is, the existence of nonexistence of the term $m\tilde{\gamma}_1$  determines the topology
for both $H^{\rm TI}_{\rm Dirac}$ and $H_{\rm s}^{\rm Dirac}$.
Following, we will provide more detail and also show how the minimal Dirac-matrix Hamiltonian approach allows us
to distinguish between $\bZ_2$ and $\bZ$ classifications.

But before doing so, let us add some remarks about the classification of Fermi surfaces that are located away from high-symmetry points in the BZ.
These gapless materials are described by Hamiltonian~\eqref{R off high-symmetry} and their 
$(d-p)$-dimensional 
Fermi surfaces are located at the momenta described by Eq.~\eqref{FSmirrorOFF}.
We can interpret Eq.~\eqref{R off high-symmetry} as a $(p-1)$-dimensional fully gapped Hamiltonian
with mass term
\bee	
	(1-p+\sum_{i=1}^p \cos k_i)\tilde{\gamma}_0 .
\ee	
[This mass term corresponds to the term $m\tilde{\gamma}_0$ in $H^{\rm TI}_{\rm Dirac}$, Eq.~\eqref{TI Dirac}.]
Hence, the classification of Fermi surfaces with codimension $p$ is related to the ten-fold classification of topological
insulators and superconductor in $(p-1)$ dimensions; see Table~\ref{original table}. 
We note, however, that as opposed to $H^{\rm TI}_{\rm Dirac}$, Eq.~\eqref{R off high-symmetry} can be gapped by two different 
SPGTs, namely by the mass term $\tilde{\gamma}_{1}$  \emph{and} by the kinetic term $\sin k_p\gamma_{p}$.
For symmetry classes with a $\bZ_2$-type invariant 
the SPGT $\sin k_p\gamma_{p}$ 
is always allowed by symmetry, whereas for classes with a $\bZ$-type number this term is symmetry forbidden.
Hence,  $\bZ_2$-type invariants cannot protect Fermi surfaces located away from high symmetry points of the BZ. Nevertheless, 
because these $\bZ_2$-type numbers are well-defined in $(p-1)$-dimensional planes in the BZ that are invariant under PHS or TRS, nonzero $\bZ_2$ numbers can lead to the appearance of gapless surfaces states at high-symmetry points of the surface BZ.

\subsubsection{Topological invariant ``$0$''}

Let us now discuss in more detail the different SPGTs that can be added to the Dirac Hamiltonian~\eqref{Dirac symmetry}. 
First of all, if any SPGTs exist then $H_{\rm s}^{\rm Dirac}$ belongs to the trivial phase. That is the Fermi surface is topologically unstable,
since the spectrum can be  gapped by the SPGT without breaking any symmetries. This case is denoted by the label ``$0$'' in Table~\ref{original table}.
For example, consider the following two-dimensional Dirac Hamiltonian in class D
\bee \label{examp000}
H_{\rm s}^{\rm D}=k_x \sigma_x + k_y \sigma_y,
\ee
which describes a superconductor with a point node at ${\bf k} = (0,0)$.
Hamiltonian~\eqref{examp000} preserves PHS with $C=\sigma_x \mathcal{K}$.
The nodal point at ${\bf k}=(0,0)$ is topologically unstable, since the spectrum can 
be gapped by the SPGT $m\sigma_z$.

If there does not exist any SPGT, then $H_{\rm s}^{\rm Dirac}$ is either classified by a $\bZ_2$ or a $\bZ$ number.
To distinguish between $\bZ_2$ and $\bZ$ classifications, we need to consider doubled versions of $H_{\rm s}^{\rm Dirac}$ and then
check whether there exist any SPGTs for the doubled Hamiltonian.

\subsubsection{Topological invariant `$\bZ_2$'}\label{Z2 feature}

A doubled version of $H_{\rm s}^{\rm Dirac}$ can be obtained in several different ways. In general it can be written as
\bee
\mH_2=\sum_i k_{n_i}\gamma_{n_i}\otimes \sigma_z +\sum_{{\rm remain}} k_{n_j}\gamma_{n_j}\otimes \bI. \label{doublesize}
\ee
Here, 
the first summation is over an arbitrary set of $\gamma_{n_i}$ ($n_i \subseteq \{ 1,2,...,d-1, d \}$) and the second summation is 
over $\gamma_{n_j}$'s that are not picked up by the first summation. 
We observe that the enlarged Dirac matrices entering in the definition of $\mH_2$ all anticommute with each other and
satisfy the same global symmetries as the original Hamiltonian $H_{\rm s}^{\rm Dirac}$.
Now, if for each choice of the set $n_i$ there exists an SPGT that can be added to $\mH_2$, then the Hamiltonian exhibits a
$\bZ_2$ classification.
SPGTs for  $\mH_2$ can be constructed by considering even and odd numbers of terms in the first summation of Eq.~\eqref{doublesize} separately. 
For an odd number of terms, the SPGTs are given by $\frak{M}$ (or $i \frak{M}$), with $\frak{M}=m (\prod_{n_i}^{\rm odd} \gamma_{n_i})\otimes \sigma_u$,
where the Pauli matrix $\sigma_u \in \{\sigma_x , \sigma_y\}$  has to be chosen such that $\frak{M}$ (or $i \frak{M}$) preserves TRS and/or PHS. (The choice
between $\frak{M}$ and $i \frak{M}$ is determined by the condition that the SPGT is Hermitian.)
For an even number of terms in the first sum of Eq.~\eqref{doublesize}, the SPGTs are given by $\frak{M}$ (or $i \frak{M}$)
with $\frak{M}=m (\gamma_{d+1}\prod_{n_i}^{\rm even}\gamma_{n_i})\otimes \sigma_u$. 
As before, $\sigma_u \in \{\sigma_x , \sigma_y\}$ has to be chosen such that PHS and/or TRS is preserved.
Note that this formula is always well defined, since according to Table~\ref{presence_gamma}, there always exist a $\gamma_{d+1}$ term for systems with $\bZ_2$-type invariants.

To make this more explicit, let us consider the following example of a two-dimensional Dirac Hamiltonian with TRS
\bee \label{exampClassAII}
h^{\rm{AII}}_{\rm s}=k_x\sigma_x+k_y\sigma_y,
\ee
which describes a topological semimetal with a Fermi point at ${\bf k} = (0,0)$.
The time-reversal symmetry operator is given by  $T=i\sigma_y\mK$. Since $T^2 = - \mathbbm{1}$, Hamiltonian \eqref{exampClassAII} belongs
to symmetry class AII. We observe that $h^{\rm{AII}}_{\rm s}$ is identical to the surface Hamiltonian of a three-dimensional topological insulator with spin-orbit coupling.
 The only possible mass term, which anticommutes with $h^{\rm{AII}}_{\rm s}$, is $\sigma_z$. 
However, $\sigma_z$ breaks TRS and is therefore forbidden by symmetry. Hence, $h^{\rm{AII}}_{\rm s}$ describes a topologically stable Fermi point.
Next, we examine different doubled versions of $h^{\rm{AII}}_{\rm s}$, i.e.,
\bee
H^{\rm{AII}}_{\rm s}=
\bma
h^{\rm{AII}}_{\rm s} & 0 \\
0 & {h^{\rm{AII}}_{\rm s}}' \\
\ema,
\ee
with ${h^{\rm{AII}}_{\rm s}}' \in \left\{ h^{\rm{AII}}_{{\rm s}++}, {h^{\rm{AII}}_{{\rm s}- -}} , h^{\rm{AII}}_{{\rm s} +-} , h^{\rm{AII}}_{{\rm s} - +} \right\}$, 
where ${h^{\rm{AII}}_{{\rm s}\pm\pm}}=\pm k_x\sigma_x\pm k_y\sigma_y$ and ${h^{\rm{AII}}_{{\rm s}\pm\mp}}=\pm k_x\sigma_x\mp k_y\sigma_y$.
It is not difficult to show that for each of the four versions of $H^{\rm{AII}}_{\rm s}$ there exists at least one SPGT. For example, for $h^{\rm AII}_{\rm s ++}$ the SPGT is $\sigma_z\otimes \sigma_y$.
Thus, the Fermi point described by $H^{\rm{AII}}_{\rm s}$ is unstable. Therefore, we conclude that Eq.~\eqref{exampClassAII} exhibits 
a $\bZ_2$ topological characteristic.

\subsubsection{Topological invariant `$\bZ(2\bZ)$'}

For systems with a $\bZ$ (or $2\bZ)$ topological invariant, there does not exist any SPGT both for
$H_{\rm s}^{\rm Dirac}$ and some of its doubled versions, cf.\ Eq.~\eqref{doublesize}.
To be more specific, when the first summation in Eq.~\eqref{doublesize} includes an {\it odd\/} number of $\gamma_{n_i}$'s, there   exists  SPGTs, which open
up a gap. [I.e., $\frak{M}$ or $i \frak{M}$, with  $\frak{M}=m (\prod_{n_i}^{\rm odd} \gamma_{n_i})\otimes \sigma_u$.] 
However, when there is an {\it even \/} number of $\gamma_{n_i}$'s in the first summation of Eq.~\eqref{doublesize}, an SPGT does not exist due to the absence of an extra kinetic term ($\gamma_{d+1}$). 
It is important to note that two gapless modes are only protected if the two blocks in Eq.~\eqref{doublesize} have the same sign.
Similarly, the system can be extended to $n$ gapless modes with the same sign in each block. In the absence of an SPGT these $n$ gapless modes are protected. This behavior reveals the signature of the $\bZ$ invariant. 

For concreteness, let us consider the Hamiltonian of a Weyl semimetal\cite{WanVishwanathSavrasovPRB11,Weyl_semimetal_Fang} as an example. This two-dimensional system, which does not preserve any symmetry, belongs to class A. One of the simplest Hamiltonians, which is also a minimal Dirac-matrix Hamiltonian, can be written as 
\bee
h^{\rm{A}}_{\rm s}=k_x\sigma_x+k_y\sigma_y+k_z \sigma_z .
\ee 
It is impossible to find an extra gap term because only three Dirac matrices can be present in the $2\times 2$ matrix dimension. Therefore, the gapless mode is stable. 
To distinguish between $\bZ$ and $\bZ_2$ classification, we need to consider two copies of $h^{\rm{A}}_{\rm s}$.
One doubled version of  $h^{\rm{A}}_{\rm s}$ is given by
\bee
H^{\rm{A}}_{\rm s} =k_x\sigma_x\otimes \sigma_z+k_y\sigma_y\otimes \bI+k_z\sigma_z\otimes \bI .
\ee
We find that there are two SPGTs that can be added to  $H^{\rm{A}}_{\rm s}$, i.e., $\sigma_x\otimes \sigma_x$ and $\sigma_x\otimes\sigma_y$.
 Hence, the gapless modes of  $H^{\rm{A}}_{\rm s} $ are unstable.  
However, there exists another doubled version of $h^{\rm{A}}_{\rm s}$, namely
\bee
{H^{\rm{A}}_{\rm s}}'=k_x\sigma_x\otimes \bI+k_y\sigma_y\otimes \bI+k_z\sigma_z\otimes \bI.
\ee
There does not exist any SPGT that can be added to ${H^{\rm{A}}_{\rm s}}'$, 
so the two identical gapless modes of ${H^{\rm{A}}_{\rm s}}'$ are stable. 
Since there exists one doubled version of $h^{\rm{A}}_{\rm s}$ which has two protected gapless modes,
we conclude that the system exhibits a $\bZ$ classification.

\section{Classification of reflection symmetry protected topological insulators and fully gapped superconductors}
\label{sec:III}

As discussed in Sec.~\ref{classReflecGapless},
the classification of reflection symmetry protected semimetals (nodal superconductors) 
can be related to the classification of reflection symmetry protected insulators (fully gapped superconductors) by dimensional reduction.
To make this relation more explicit,
we briefly  survey in this appendix the classification of fully gapped topological materials protected by 
crystal reflection symmetrie.\cite{chiuPRB13,morimotoPRB13,Sato_Crystalline_arxiv14} This classification scheme crucially depends on whether the crystal reflection symmetry commutes or anticommutes with the global nonspatial symmetries.

The  classification of reflection symmetry protected topological insulators and fully gapped superconductors is summarized in Table~\ref{reflection_table_full},
where the first row indicates the dimension $d$ of the fully gapped system.\cite{chiuPRB13,morimotoPRB13,Sato_Crystalline_arxiv14}
In even (odd) spatial dimension $d$,  ten (seventeen) out of the 27 symmetry classes allow for the
existence of nontrivial topological insulators/superconductors protected by reflection symmetry.
The different topological sectors within a given class of reflection symmetry protected topological insulators/superconductors can be labeled by 
an integer  $\mathbb{Z}$ number, a binary $\mathbb{Z}_2$ quantity,  a mirror Chern or winding number $M\mathbb{Z}$, 
a mirror binary $\mathbb{Z}_2$ quantity $M\mathbb{Z}_2$, or a binary $\mathbb{Z}_2$ quantity with translation symmetry  $T\mathbb{Z}_2$.
Interestingly, reflection symmetric topological states belonging to  symmetry classes with chiral symmetry, can be protected in some cases by both
an integer $\mathbb{Z}$ number (binary $\mathbb{Z}_2$ quantity) and a mirror Chern or winding number $M\mathbb{Z}$ (mirror  $\mathbb{Z}_2$ quantity $M\mathbb{Z}_2$), as indicated by the label $M \mathbb{Z} \oplus \mathbb{Z}$ ($M \mathbb{Z}_2 \oplus \mathbb{Z}_2$) in Table~\ref{reflection_table_full}.
The nontrivial bulk topology characterized by these invariants manifests itself at the boundary  in terms of protected Dirac or Majorana  surface states,
which, depending on the type of the invariant, appear either at any surface (for $\mathbb{Z}$ and $\mathbb{Z}_2$) or only at surfaces that are left invariant
under the reflection symmetry (for $M\mathbb{Z}$ and $M\mathbb{Z}_2$).
As explained in Sec.~\ref{classReflecGapless}, by use of a dimensional reduction procedure these surface states of a $d$-dimensional fully gapped system can be interpreted as a reflection symmetry protected topological semimetal (or nodal superconductor) in $d-1$ dimensions.

Before discussing in detail the different invariants that characterize reflection symmetry protected topological materials, we remark that the recently
discovered  topological crystalline insulator SnTe is included in Table~\ref{reflection_table_full}.\cite{Tanaka:2012fk,Hsieh:2012fk,Xu2012,Dziawa2012uq_short} Specifically, SnTe belongs to symmetry class AII with $T^2=-\mathds{1}$  in $d=3$ dimensions and exhibits a reflection symmetry $R_-$ that anticommutes with the time-reversal symmetry operator $T$. As indicated by Table~\ref{reflection_table_full}, this crystalline topological insulator is described by a mirror Chern number $M \mathbbm{Z}$ and hence supports Dirac-cone states at reflection-symmetric surfaces. These  Dirac surface states have recently been observed in angle-resolved photoemission experiments.\cite{Tanaka:2012fk,Xu2012,Dziawa2012uq_short}

\subsection{$M\bZ$ and $M\bZ_2$ invariants}
	
The mirror Chern or winding numbers and mirror $\bZ_2$ invariants, denoted by $M\bZ$ and $M\bZ_2$ in Table~\ref{reflection_table_full}, respectively, are defined on the hyperplanes  in the BZ that are symmetric
under reflection $R$, i.e., the two hyperplanes $k_1=0$ and $k_1=\pi$.
 Since $R$ is Hermitian and anticommutes with 
the Hamiltonian $ H ( {\bf k} )$ restricted to the hyperplanes $k_1=0$ and $k_1= \pi$,  $\left. H ( {\bf k} ) \right|_{k_1 = 0, \pi} $ 
can be block diagonalized with respect to the two eigenspaces $R=\pm 1$ of the reflection operator.
We observe that each of the two blocks of $\left. H ( {\bf k} ) \right|_{k_1 = 0, \pi} $ is  left invariant only under those global symmetries that commute with the reflection operator $R$. Hence, depending on the nonspatial symmetries of the $R=\pm1$ blocks of $\left. H ( {\bf k} ) \right|_{k_1= 0, \pi} $, it is possible to define a 
 mirror Chern or winding invariant\cite{chiuPRB13} 
\begin{eqnarray}
\nu_{M \mathbb{Z}} = \sgn \left[ \nu^{d-1}_{k_1 = 0} - \nu^{d-1}_{k_1 = \pi} \right] \left( \left| \nu^{d-1}_{k_1 = 0} \right| - \left| \nu^{d-1}_{k_1 = \pi} \right| \right) ,
\end{eqnarray}
where $\nu^{d-1}_{k_1 = 0 (\pi)}$ denotes the Chern or winding number of the $R=+1$ block of $\left. H ( {\bf k} ) \right|_{k_1 = 0 ( \pi)} $.\cite{foonoteMirrorChern}
Similarly, the mirror $\mathbb{Z}_2$ quantity $M\mathbb{Z}_2$ is  defined by
\begin{eqnarray}
n_{M \mathbb{Z}_2} = 1 - \left|   n^{d-1}_{k_1 = 0} - n^{d-1}_{k_1 = \pi} \right| ,
 \end{eqnarray}
 with $n^{d-1}_{k_1 = 0 ( \pi )} \in \{ -1, +1 \}$ the  $\mathbb{Z}_2$ invariant of the $R=+1$ block of $\left. H ( {\bf k} ) \right|_{k_1 = 0 ( \pi)} $.
A nontrivial value of these mirror indices indicates the appearance of Dirac or Majorana states at reflection symmetric surfaces, i.e., at surfaces that are perpendicular to the reflection hyperplane $x_1=0$.
At surfaces that break reflection symmetry, however, the boundary modes are in general gapped.
Some illustrative examples of topological crystalline insulators with mirror Chern or winding numbers have been discussed in Ref.~\onlinecite{chiuPRB13}.

\subsection{$\bZ$ and $\bZ_2$ invariants}

For symmetry classes with at least one nonspatial symmetry that anticommutes with the reflection operator $R$, it is possible in certain cases to define
 a global $\bZ$ or $\bZ_2$ number even in the presence of reflection. These $\bZ$ and $\bZ_2$ indices are identical to the 
 ones of the original ten-fold classification in the absence of mirror symmetry (cf.\ Table~\ref{original table}) and
 lead to the appearance of linearly dispersing Dirac or Majorna states
at any surface, independent of the surface orientation.

\subsection{$M\bZ \oplus \bZ$  and $M\bZ_2 \oplus \bZ_2$ invariants}

Topological properties of reflection symmetric insulators (superconductors) with chiral symmetry are described in some cases by
both a global $\bZ$ or $\bZ_2$ invariant and a mirror index $M\bZ$ or $M\bZ_2$.
The global invariant and the mirror invariant are independent of each other. 
At surfaces which are perpendicular to the mirror plane the number of protected gapless states is
given by $\mathrm{max} \left\{ \left| n_{\bZ} \right|, \left| n_{M \bZ} \right| \right\}$,\cite{chiuPRB13} 
where $n_{\bZ}$   denotes the global $\bZ$ invariant, whereas 
$n_{M\bZ}$  is the mirror $\bZ$ invariant.
This should be compared to Sec.~\ref{high-symmetry DIII}, where
we provide an example of a gapless topological phases with nontrivial $M\bZ$ and  $\bZ$ invariants.
Examples of gapless topological phases with nontrivial  $M\bZ_2$ and  $\bZ_2$ invariants are given 
in Secs.~\ref{CII R+-} and~\ref{Z2 like comparison}.

\subsection{$T\bZ_2$ invariant}

In  symmetry classes where the reflection operator $R$ anticommutes with the
global antiunitary symmetries TRS and PHS ($R_-$ and $R_{--}$ in Table~\ref{reflection_table_full}) the second descendant  $\bZ_2$ invariants\cite{Ryu2010ten} are only well defined
in the presence of translation symmetry. That is, the edge or surface states of these phases can be gapped out by
density-wave type perturbations, which preserve reflection and global symmetries but break translation symmetry. Hence,  these topological states
are protected by a combination of reflection, translation, and global antiunitary symmetries. Therefore we denote their topological indices by ``$T \bZ_2$" in Table~\ref{reflection_table_full}. 

To exemplify the properties of reflection symmetric insulators (superconductors) with a $T \bZ_2$ invariant we consider 
a two-dimensional superconductor with $R_{--}$ reflection symmetry in class CII  given by the
$8 \times 8$ BdG Hamiltonian
\begin{eqnarray} \label{CIIRmmExamp}
 H^{\rm CII}_{\rm bulk}  
=M \gamma_0  + \sin k_x \gamma_1 + \sin k_y  \gamma_2 ,
\end{eqnarray}
where $M=1+ \cos k_x  +\cos k_y $, $\gamma_0=\sigma_z \otimes \bI \otimes \bI$, $\gamma_1=\sigma_x \otimes \sigma_x \otimes \bI$, and $\gamma_2=\sigma_x \otimes \sigma_y \otimes \sigma_x$.
Superconductor~\eqref{CIIRmmExamp} preserves TRS and PHS
with 
 $T_{\rm bulk}=\bI \otimes \sigma_y\otimes \bI \mK$ and
 $C_{\rm bulk}=\sigma_x \otimes \bI \otimes \sigma_y \mK$,
respectively.
Reflection symmetry is implemented as
$R^{-1}_{\rm bulk}  H^{\rm CII}_{\rm bulk}   ( -k_x, k_y ) R_{\rm bulk} 
= 
H^{\rm CII}_{\rm bulk}  (k_x, k_y) $,
with
$R_{\rm bulk} =\bI \otimes \sigma_y\otimes \bI$.
This topological crystalline superconductor is characterized by a $T\bZ_2$ invariant (cf.\ Table~\ref{reflection_table_full}), 
which indicates that the helical Majorana states at the (01) edge are only stable in the presence of translation symmetry.
We find that these Majorana-cone  edge states appear at $k_x = \pm \delta$ of the edge BZ and
are described by the following edge Hamiltonian\cite{foonoteBasisChoice}
\begin{eqnarray} \label{CIIexampSurf}
h_{\rm edge}^{\rm CII}=k_x  \, \sigma_x \otimes \sigma_x + \delta \, \sigma_z\otimes \sigma_y .
\end{eqnarray}
The edge Hamiltonian satisfies TRS, PHS, and reflection symmetry with
$T_{\rm edge}=\sigma_y\otimes \bI \mK$, $C_{\rm edge}=\sigma_y \otimes \sigma_z \mK$, and
$R_{\rm edge}=\sigma_z\otimes \bI$, respectively. 
In the absence of reflection symmetry the gap opening mass term $m \, \sigma_x\otimes\sigma_y$, which preserves both TRS and PHS,
can be added to Eq.~\eqref{CIIexampSurf}. 
Therefore, Hamiltonian \eqref{CIIRmmExamp} is topologically trivial according to the ten-fold
classification of Table~\ref{original table}. However, with reflection and translation symmetry  $h_{\rm edge}^{\rm CII}$ cannot be gapped since
$m \, \sigma_x\otimes\sigma_y$ breaks reflection symmetry $R_{\rm edge}$.
Considering two copies of the edge Hamiltonian, i.e., $H_{\rm edge}^{\rm CII}=h_{\rm edge}^{\rm CII}\otimes \bI$,
we find that the symmetry preserving  mass term  $m\siz\otimes \six \otimes \siy$ opens up a gap in the spectrum 
of the doubled Hamiltonian $H_{\rm edge}^{\rm CII}$.
Hence, BdG Hamiltonian~\eqref{CIIRmmExamp} exhibits a nontrivial $\bZ_2$-type topological characteristic (cf.\ Appendix \ref{Z2 feature}).
To demonstrate that the two Majorana edge modes, Eq.~\eqref{CIIexampSurf}, are unstable against translation symmetry breaking we consider
the density wave type mass term
\begin{eqnarray} \label{CIIexpCDWmass}
\hat{\mathfrak{M}} 
&=&
m\sum_{-\eta\leq k_x < \eta}
\Big(
i c_{k_x+\eta +\delta}^\dag \, \mathfrak{M} \, c^{\phantom{\dag} }_{k_x-\eta+\delta}
\nonumber\\
&& \qquad +i c_{-k_x +\eta -\delta}^\dag \, \mathfrak{M} \, c^{\phantom{\dag}}_{-k_x-\eta-\delta}+\mathrm{H.c.} \Big),
\end{eqnarray}
which is invariant under TRS, PHS, and reflection $\hat{R} =\sum_{k_x} c^\dag_{-k_x}\sigma_z\otimes \bI \, c^{\phantom{\dag}}_{k_x}$.
In Eq.~\eqref{CIIexpCDWmass},
$\mathfrak{M}=m \sigma_x\otimes\sigma_y$ and $\eta$  is a constant with $0<\eta < \delta$.
For $m>\eta$ the translation symmetry breaking mass term \eqref{CIIexpCDWmass} fully gaps out all  edge modes.

In closing, we remark that for the classification of gapless topological materials presented in Sec.~\ref{classReflecGapless}, the presence of translation symmetry is always  assumed. 
In particular, density-wave type mass terms are disregarded, since these can gap out the bulk by coupling 
Fermi surfaces (nodal lines) located at different parts of the BZ. Thus, the distinction between $\bZ_2$ and $T\bZ_2$ invariants is
irrelevant for the topological classification of reflection symmetric semimetals and nodal superconductors.

\section{ Classification of Fermi points outside mirror planes}
\label{derivation of gapless}
\label{appendixB}

In this appendix we derive the classification scheme of Table~\ref{table_reflection_off_off}
using the Dirac-matrix Hamiltonian approach. This should be compared to
the discussion in Sec.~\ref{outside mirror planes}, where this classification is derived by examining different types of topological invariants. 
As in the main text we assume that reflection symmetry maps $k_1 \to - k_1$.
To derive the classification we consider the following reflection symmetric Dirac-matrix Hamiltonian
\bee \label{off reflection}
H_{\rm off}=\sum_{i=2}^d\sin k_i \gamma_i +(1-d+\sum_{i=1}^d \cos k_i)\tilde{\gamma}_0 ,
\ee
which describes a $d$-dimensional gapless system with Fermi points located at 
\bee \label{FSoffoffapp}
{\bf k}=(\pm \pi/2,0,\ldots, 0) .
\ee
Reflection symmetry acts on Hamiltonian~\eqref{off reflection} as  $[R, H_{\rm off} ]=0$.
We note that the Fermi surface \eqref{FSoffoffapp} lies outside the mirror plane $k_1=0$ and away from the
high symmetry points of the BZ.
Furthermore, observe that 
%Hamiltonian~\eqref{off reflection} is similar to Hamiltonian~\eqref{R off high-symmetry}, which describes
%Fermi surfaces that are located within mirror planes but away from high-symmetry points.
%That is, $\sin k_1\gamma_1$ in Eq.~\eqref{R off high-symmetry} is replaced by $\sin k_d \gamma_d$ in Eq.~\eqref{off reflection}.
by fixing $k_1$ to  $k^0_1  \ne \pm \pi /2 $, Hamiltonian~\eqref{off reflection} can be viewed as a ($d-1$)-dimensional insulator 
\bee \label{HfullGapOffOff}
H^{d-1}_{\rm off}=\sum_{i=2}^d\sin k_i \gamma_i + \tilde{m} \tilde{\gamma}_0 ,
\ee
with mass $\tilde{m} = (1-d+   \cos k^0_1 + \sum_{i=2}^d \cos k_i) $.
 
In order to classify the Fermi surfaces described by Eq.~\eqref{off reflection}, two different types of SPGTs need to be considered, i.e., 
\bee
m\tilde{\gamma}_1  \quad \textrm{and} \quad  \sin k_1 \gamma_{1} .
\ee
The latter is a kinetic term. It will lead to a classification pattern which is quite different from the ten-fold classification.
Let us now discuss for which of the 27 symmetry classes listed in Table~\ref{table_reflection_off_off} there exist topologically
stable Fermi points. 

\begin{table}
\begin{center}
\begin{tabular}{|c|c|c|c|c|c|c|c|c|}
\hline
$s-d$ & 0 & 1 & ~2~ & 3 & ~4~ & ~5~ & 6 & 7 \\
\hline 
$G^{\mathbb{R}}_{\rm off}(R^{+},s-d)$ & $C\bZ_2$ & $C\bZ_2$ & 0 & $2\bZ$ & 0 & 0 & 0 & $\bZ$ \\
\hline
$G^{\mathbb{R}}_{\rm off}(R^{-},s-d)$ & 0 & 0 & 0 & $2\bZ$ & 0 & $C\bZ_2$ & 0 & $2\bZ$ \\
\hline
$G^{\mathbb{R}}_{\rm off}(R^{\mp \pm},s-d)$ & 0 & 0 & 0 & 0 & 0 & 0 & 0 & 0 \\
\hline
$G^{\mathbb{R}}_{\rm off}(R^{\pm \mp },s-d)$ & 0 & 0 & 0 & 0 & 0 & 0 & $C\bZ_2$ & $C\bZ_2$ \\
\hline
\end{tabular}
\caption{
Classification of Fermi points outside mirror planes; cf.\ Table~\ref{table_reflection_off_off}. The prefix ``$C$" indicates that the $\bZ_2$ invariant 
is defined in terms of the combined symmetries, see Eqs.~\eqref{combineSYMdef}.
The label ``$R^{\mp \pm}$" represents $R_{-+}$ for classes BDI and CI; and $R_{+-}$ for classes CI and DIII. Similarly, ``$R^{\pm \mp }$" represents $R_{+-}$ for classes BDI and CI; and $R_{-+}$ for classes CI and DIII.}
\label{presence gamma}
\end{center}
\end{table}
 
\subsection{$R_{+}$ and $R_{++}$}
\label{appBRpp}

 We start by considering the case where the reflection operator $R$ commutes with all global symmetries. For simplicity, we can choose  $R=\dI$.
We note that even if the global symmetries allow the kinetic mass term $\sin k_1 \gamma_1$, the reflection symmetry forbids this term due to $k_1 \rightarrow -k_1$. Therefore, 
the classification is solely determined by the presence or absence of the regular mass term $m \tilde{\gamma}_1$.   Thus, the classification of $d$-dimensional gapless modes
described by Eq.~\eqref{off reflection}
is identical to the classification of ($d-1$)-dimensional fully gapped systems [described by Eq.~\eqref{HfullGapOffOff}] in the absence
of reflection symmetry.  [I.e., we have $G_{\rm off}^{\mathbb{C}}(R^{+},s,d)=\pi_0(\mathcal{C}_{s-d+1})$.] 

\subsection{$R_{-}$ and $R_{--}$}\label{R--}

 Second, we study the case where $R$ anticommutes with all global symmetries. In this case the reflection operator $R$ can take on three different forms,
namely  $R =i\gamma_{d+1}\gamma_{d+2}$,  $R =i\tilde{\gamma}_1\tilde{\gamma}_2$, or $R =\bI\otimes \sigma_y$. 
The classification of the gapless Dirac Hamiltonian~\eqref{off reflection}, can be inferred from the 
homotopy group $\pi_0(\mathcal{R}_{l})$, where $\mathcal{R}_l$ represents the classifying space and $l=s-d+1$ mod $8$, with $s$ denoting the symmetry class and $d$ the spatial dimension.
Each symmetry class $s$ and dimension $d$
needs to be discussed separately.
Since the classification only depends on the difference $s-d$, we discuss it in terms of $l= s-d+1$.
Based on Table~\ref{presence_gamma}, we find that for $l=2,3$ and $l=5,6$ the reflection operator can be defined as follows 
\begin{align}
l=&2, 3,& R &=i\gamma_{d+1}\gamma_{d+2} , \\
l=&5, 6,& R &=i\tilde{\gamma}_{1}\tilde{\gamma}_{2}  .
\end{align}
 For $l=2, 3$ we find that there exists an SPGT, i.e., $\sin k_1\gamma_{d+1}$, which implies trivial topology. Similarly, for $l=5$, the presence of the symmetry-allowed gap opening term $i \sin k_1 \tilde{\gamma}_{1}\tilde{\gamma}_2\tilde{\gamma}_3$ signals trivial topology. For $l=6$, on the other hand, the Fermi point of Eq.~\eqref{off reflection} is stable since there does not
exist any SPGT.
To distinguish between $\bZ_2$ and  $\bZ$  classifications, we need to consider a doubled version of Hamiltonian~\eqref{off reflection} with two identical gapless modes, i.e.
\bee
H_{\rm off}'=H_{\rm off}\otimes \bI . \label{double off}
\ee
For $l=6$ there exists an SPGT ($\sin k_1 \tilde{\gamma}_1\otimes \sigma_y$) that can be added to $H_{\rm off}'$, signaling a $\bZ_2$ classification.

For $l=0,1,7$ it is not possible to implement a reflection symmetry for the minimal Dirac-matrix Hamiltonian~\eqref{off reflection}. Instead, one needs 
to consider the doubled version of $H_{\rm off}$, i.e. Eq.~\eqref{double off}, 
to study the effects of reflection symmetry. For $H_{\rm off}'$ reflection symmetry can be implemented as
$R =\bI\otimes \sigma_y$. 
For $l=1$ and $l=7$, SPGTs can be found as $m\tilde{\gamma}_1\otimes \bI$ and $m \gamma_{d+1} \otimes \sigma_y$, respectively. 
For $l=0$, however, gap opening terms are forbidden by symmetry. We find that also for the quadrupled version of $H_{\rm off}$, i.e.
\bee
H_{\rm off}''=H_{\rm off}\otimes \bI \otimes \bI ,
\ee
there do not exist any SPGTs in the case of $l=0$.
Therefore, the system exhibits a $2\bZ$ classification, due to the doubled size of the minimal Hamiltonian, Eq.~\eqref{double off}.  

Finally, for $l=4$ the system, which corresponds to $2\bZ$, can be effectively treated as two identical copies of the $\bZ$ system in the spatial dimensions 
\bee
H_{\rm off}^{2\bZ}=H_{\rm off}^{\bZ}\otimes \bI .
\ee
The relations of the global symmetry operators between $\bZ$ and $2\bZ$ are given by $T_{2\bZ}=T_{\bZ}\otimes \sigma_y$ and $C_{2\bZ}=C_{\bZ}\otimes \sigma_y$. Therefore, we can simply define $R^{-}=\bI \otimes \sigma_y$, which anticommutes with $T_{2\bZ}$ and $C_{2\bZ}$. Following the similar discussion of $l=0$, we find the system of $l=4$ inherits $\bZ$ topology.

\subsection{AIII with $R_-$, DIII \& CI with $R_{-+}$, and  BDI \& CII with $R_{+-}$}\label{R+-}

 Next, we consider
 class AIII with $R_-$-type reflection
symmetry, class DIII \& CI with $R_{-+}$-type reflection symmetry, and class BDI \& CII with $R_{+-}$-type reflection symmetry.
 That is, we have
 \begin{subequations} \label{classesOffOfftrivial}
\begin{align}
R_{-}& \text{ for class AIII,}  \\ 
R_{+-}& \text{ for class BDI and CII}, \label{BDI CII}   \\
R_{-+}& \text{ for class DIII and CI}. \label{DIII CI} 
\end{align}
\end{subequations}
 In all these cases there is a chiral symmetry operator $S$ which anticommutes with the Hamiltonian.
 Using $S$ we can construct the reflection symmetry operator $R$ as $R =i\gamma_{d+1}S$, where   $\gamma_{d+1}$ represents a kinetic term.
 Let us clarify how $S$ is related to the two global symmetry 
 operators $T = U_T \mathcal{K}$ and $C = U_C \mathcal{K}$.
 (Here, we assume that $U_T$ and $U_C$ are Hermitian and unitary.)
 In general $S$ is proportional to $T C$. 
 We choose  $S=TC$ if $[U^*_C,U_T]=0$ and $S=iTC$ if $\{ U^*_C,U_T\}=0$. 
 This choice ensures that $R$ is Hermitian and that $R$ and $T$ / $C$ 
 satisfy the commutation and anticommutation relations of Eqs.~\eqref{BDI CII} and~\eqref{DIII CI}.
 In order to verify these (anti)commutation relations one has to make use of
 Eq.~\eqref{kinetic symmetry} and the fact that
 \begin{align}
TST^{-1}&=\pm S,
\quad
CSC^{-1}=\pm S, \label{TCS}
\end{align}
where
we pick up the plus sign in front of $S$
when $T^2=\pm \dI$ and $C^2=\pm \dI$,
whereas
we pick up the minus sign
when $T^2=\pm \dI$ and $C^2=\mp \dI$.

%By using Eq.~\eqref{kinetic symmetry}, which describes the relations between the kinetic term $(\gamma_{d+1})$ and the global symmetry operator, we obtain the commutation and anticommutation relations of $R^{\mp \pm}=i\gamma_{d+1}S$ exactly shown in Eqs.~\eqref{BDI CII} and~\eqref{DIII CI}: $[T,R^{\mp \pm}]=0$ and $\{C,R^{\mp \pm}\}=0$ when $T^2=C^2=\pm \dI $ and $\{T,R^{\mp \pm}\}=0$ and $[C,R^{\mp \pm}]=0$ when $T^2=-C^2=\mp \dI $.

With these definitions, we find that the kinetic term $\sin k_1 \gamma_{d+1}$ is an SPGT for all dimensions and all the cases listed in Eq.~\eqref{classesOffOfftrivial}, i.e.,  $\sin k_1 \gamma_{d+1}$ opens up a full gap and is allowed by both the global symmetries and the reflection symmetry. Hence, for the symmetry classes \eqref{classesOffOfftrivial}
the system always has trivial topology.
Therefore, we write \mbox{$G_{\rm off}^\mathbb{R}(R^{\mp \pm},s-d)=0$}; see Table~\ref{presence gamma} and Table~\ref{table_reflection_off_off}.

\subsection{DIII \& CI with $R_{+-}$ and BDI \& CII with $R_{-+}$}\label{R-+}

Last, we discuss class DIII \& CI with $R_{+-}$-type reflection symmetry and class BDI \& CII with $R_{-+}$-type reflection symmetry. 
In a similar way as in the previous subsection, we can construct the reflection operator $R$  in the form of $R=i\tilde{\gamma}_1 S$. This ensures that  $\{T,R \}=0$ and $[C,R ]=0$ when $T^2=C^2=\pm \dI$; and $[T,R ]=0$ and $\{C,R \}=0$ when $T^2=-C^2=\pm \dI$. 
In the following we discuss the topology for each symmetry class $s$ and each spatial dimension $d$ separately.
Since the classification only depends on the difference $s-d$, we discuss it in terms of $l= s-d+1$ (cf.\ Sec.~\ref{R--}). The classification 
can also be inferred from the homotopy group $\pi_0 (\mathcal{R}_l)$; cf.\  Table~\ref{presence_gamma}.

For $l=5,\ 6$, we find that the reflection operator $R$ can be defined as
 $R =i \tilde{\gamma}_1 S$, without enlarging the matrix dimension of the minimal Hamiltonian.
According to Table~\ref{presence_gamma}, there exist at least two mass terms, i.e., $\tilde{\gamma}_1$ and $\tilde{\gamma}_2$, which preserve the global symmetries. 
The second mass term, $\tilde{\gamma}_2$, which preserves also reflection symmetry, gaps out the Fermi points. Hence, the topology is trivial and classified as ``$0$".

For $l=3$, there exist three kinetic terms, $\gamma_{d+1}$, $\gamma_{d+2}$, and $\gamma_{d+3}$, which satisfy Eq.~\eqref{kinetic symmetry}. The product of these three kinetic terms form a mass term $i\gamma_{d+1}\gamma_{d+2}\gamma_{d+3}$, which preserves global symmetries. Hence, the reflection symmetry operator can be constructed as $R =i\gamma_{d+1}\gamma_{d+2}\gamma_{d+3}S$. The kinetic term $\sin k_1 \gamma_{d+1}$, which also preserves reflection symmetry, is allowed to be added to Hamiltonian (\ref{off reflection}) as an SPGT. Hence, the case $l=3$  is classified as the trivial phase. 

For $l=7$, there is only one mass term, namely $\tilde{\gamma}_1$, which is allowed by the global symmetries (see Table.~\ref{presence_gamma}). So it is possible to construct the reflection symmetry operator $R$ 
as $R =i\tilde{\gamma}_{1}S$.  The reflection symmetry forbids $\tilde{\gamma}_1$, which is the only term that gaps the Fermi points. Although the Fermi points are stable in the minimal Hamiltonian, to distinguish $\bZ_2$ and $\bZ$ we have to consider doubled versions of the   minimal Hamiltonian. For $H_{\rm{off}}\otimes \bI $, there exists a mass term $\sin k_1 \tilde{\gamma}_1\otimes \siy$ which preserves global symmetries and reflection symmetry with $R =i\tilde{\gamma}_1S\otimes \bI$. Hence,
the case $l=7$ exhibits $\bZ_2$ characteristics. 

For $l=1,2$, the reflection operator $R$ for the minimal Hamiltonian, Eq.~\eqref{off reflection}, in the absence of the mass term $\tilde{\gamma}_1$ cannot be constructed.
In order to study the effects of reflection symmetry, we need to enlarge the matrix dimension and consider two identical copies of $H_{\rm off}$, i.e., $H_{\rm{off}}\otimes \bI$.
For $H_{\rm{off}}\otimes \bI$ a mass term can be defined as $\tilde{\gamma}_1=\gamma_{d+1}\otimes \siy $. Therefore, the reflection symmetry operator is given by $R =i\gamma_{d+1}S\otimes \siy$.  With this, we find that $\gamma_{d+1}\otimes \six$ is an SPGT that can be added to $H_{\rm{off}}\otimes \bI$.
Hence, the case $l=1,2$ is topologically trivial, i.e., classified as ``$0$".

For $l=0$ we also need to enlarge the matrix dimension in order to study the effects of reflection symmetry.
We consider the following doubled version of Eq.~\eqref{off reflection}
\bee
H_{\rm off}^{l=0}=\sum_{i=2}^d\sin k_i \gamma_i\otimes \bI  +(d-1+\sum_{i=1}^d \cos k_i)\tilde{\gamma}_0 \otimes \siz . \label{l0 enlarge}
\ee
We note that there exist several different doubled versions of $H_{\rm off}$ for which a reflection symmetry can be defined. However, all these different versions 
are unitarily equivalent, hence it is sufficient to study only one of them.
For Hamiltonian~\eqref{l0 enlarge} there exist only one mass term (i.e., $\gamma_0\otimes \six$) and one kinetic term (i.e., $\gamma_0\otimes \siy$) that
preserve the global symmetries.  
The mass term $\gamma_0\otimes \six$ can be used to define a reflection operator, i.e., $R = i\tilde{\gamma}_0S\otimes \six$.
There exist two mass terms which satisfy the global symmetries ($m\tilde{\gamma}_0\otimes \six$ and $\sin k_1 \gamma_0 \otimes \siy$). However,
these two mass terms break reflection symmetry. Hence, the Fermi points in the case $l=0$ are topologically stable.
To distinguish between $\bZ_2$ and $\bZ$,
 the Hamiltonian has to be doubled, i.e, we consider 
$H_{\rm off}^{l=0} \otimes \bI$. We find that for $H_{\rm off}^{l=0} \otimes \bI$ there exists an SPGT, 
 namely $m\tilde{\gamma}_0\otimes \siy \otimes \siy$. Thus, the system is classified as $\bZ_2$.

For $l=4$, a reflection symmetry cannot be implemented for Eq.~\eqref{off reflection} (since mass and kinetic terms are absent).
We need to consider a quadrupled version of Eq.~\eqref{off reflection}, in order to study the influence of reflection symmetry.
The quadrupled Hamiltonian can be constructed using Eq.~\eqref{l0 enlarge}. We have
\bee
H_{\rm off}^{l=4}=H_{\rm off}^{l=0}\otimes \bI.
\ee
The global symmetry operators for this Hamiltonian are given by 
\bee
T^{l=4}=T^{l=0}\otimes \siy,
\quad
C^{l=4}=C^{l=0}\otimes \siy.
\ee
The reflection symmetry operator can be constructed as $R =i\tilde{\gamma}_0S\otimes \six\otimes \bI$, where $\tilde{\gamma}_0$ and $S$ are the mass term and the chiral symmetry operator of $H_{\rm off}^{l=0}$, respectively.  
We find that the mass term $m\tilde{\gamma}_0\otimes \siy \otimes \siy$, which preserves the global symmetries and the reflection symmetry, gaps out the Fermi points
of $H_{\rm off}^{l=4}$. Thus, the system is topologically trivial and classified as ``$0$''.

\bibliography{references}

\begin{thebibliography}{92}
\bibitem[{wea()}]{weaker_invariant}
\bibinfo{howpublished}{Since the topological invariant cannot change as a
  function of $k_1$ as long as the bulk gap does not close, the invariant
  defined in the mirror plane must equal the invariant defined in a plane that
  is perpendicular to $k_1$ and infinitesimally close to the mirror plane. Due
  to this condition the ``stronger" of the two invariants reduces to the
  ``weaker" one.}
\bibitem[{foo({\natexlab{b}})}]{footnoteABC}
\bibinfo{howpublished}{Note that this model is similar to the BHZ model
  describing the quantum spin Hall effect in HgTe quantum
  wells\cite{bernevig06,QiRMP11,konigJPSJ08}.}
\bibitem[{foo({\natexlab{d}})}]{foonoteBasisChoice}
\bibinfo{howpublished}{Note that the edge Hamiltonian $h_{\rm edge}^{\rm CII}$
  and the symmetry operators $T_{\rm edge}$, $C_{\rm edge}$, and $R_{\rm edge}$
  are only defined up to a similarity transformation, since one is free to
  choose a basis among the degenerate states of the Hamiltonian. The
  topological properties, however, are independent of this basis choice.}
\end{thebibliography}

 \end{document}